\documentclass[preprint]{aastex}
\usepackage{graphicx}
\usepackage{natbib}

\begin{document}

\title{Corrugations and eccentric spirals in Saturn's D ring:\\ New insights into what happened at Saturn in 1983}
\author{M. M. Hedman$^a$, J.A. Burns$^b$ and M.R. Showalter$^c$}
\affil{ $^a$Department of Physics, University of Idaho, Moscow ID 83844-0903\\
$^b$Department of Astronomy and Department of Mechanical Engineering, \\Cornell University, Ithaca NY 14850 \\
$^c$ SETI Insitute, Mountain View CA 94043}

\maketitle

{\bf ABSTRACT:} Previous investigations of Saturn's outer D ring (73,200-74,000 km from Saturn's center) identified periodic brightness variations  whose radial wavenumber increased linearly over time. This pattern was attributed to a vertical corrugation, and its temporal variability implied that some event --possibly an impact with interplanetary debris-- caused the ring to become tilted out the planet's equatorial plane in 1983. This work examines these patterns in greater detail using a more extensive set of Cassini images in order to obtain additional insights into the 1983 event. These additional data reveal that  the D ring is not only corrugated, but also contains a  time-variable periodic modulation in its optical depth that probably represents organized eccentric motions of the D-ring's particles. This second pattern suggests that whatever event tilted the rings also disturbed the radial or azimuthal velocities of the ring particles. Furthermore, the  relative amplitudes of the two patterns indicate that the vertical motions induced by the 1983 event  were 2.3$\pm$0.5 times larger than the corresponding in-plane motions. If these structures were indeed produced by an impact, material would need to strike the ring at a steep angle ($>60^\circ$ from the ring plane) to produce such motions. Meanwhile, the corrugation wavelengths in the D ring are about 0.7\% shorter than one would predict based on extrapolations from similar structures in the nearby C ring. This could indicate that the D-ring was tilted/disturbed about 60 days before the C ring. Such a timing difference could be explained if the material that struck the ring was derived from debris released when some object broke up near Saturn some months earlier. To reproduce the observed time difference, the debris would need to have a substantial initial velocity dispersion and then have its orbital properties perturbed by some phenomenon like solar tides prior to its collision with the rings. 

\section{Introduction}

\begin{figure}
\centerline{\resizebox{3in}{!}{\includegraphics{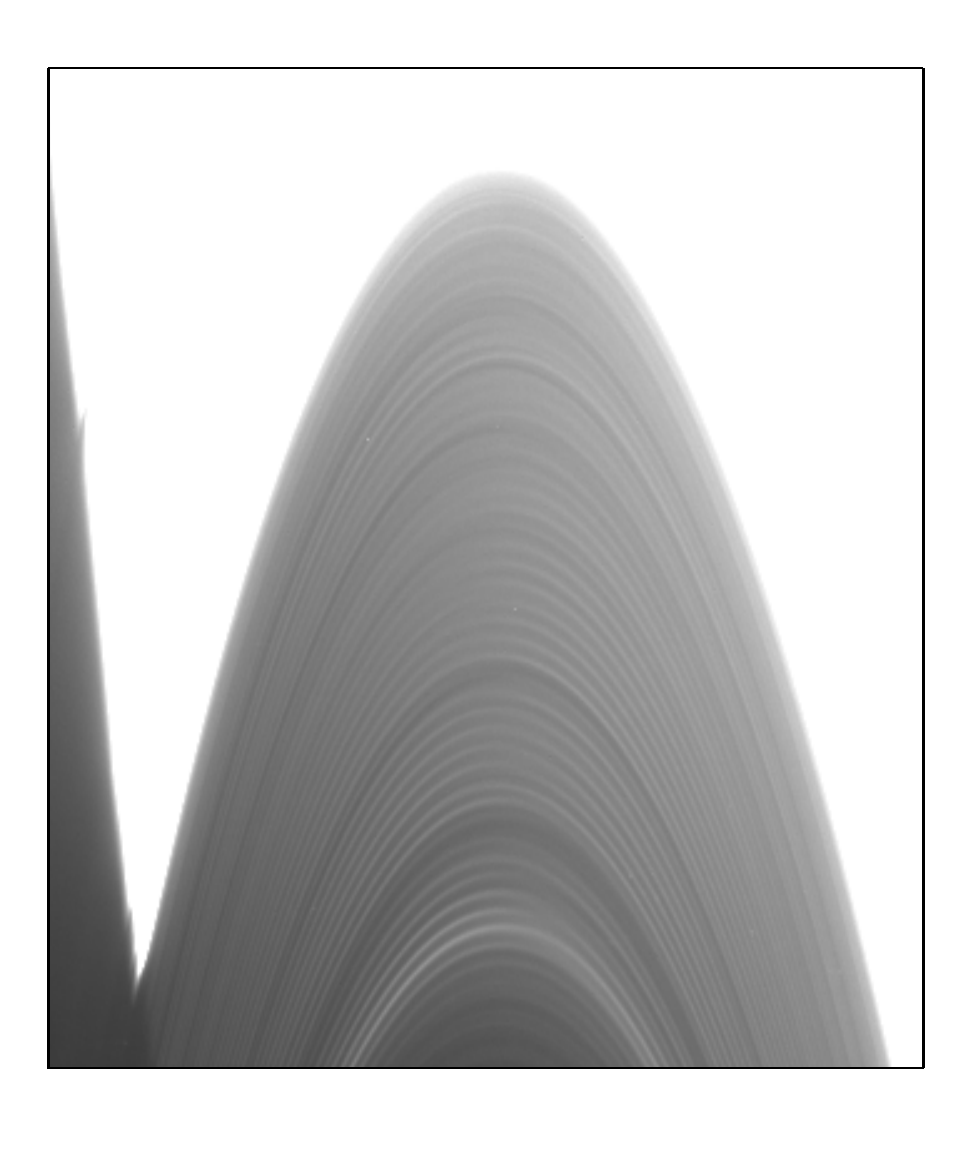}}}\caption{Image of the periodic structures in the outer D ring (Image name N1571969357 phase angle  26.7$^\circ$, ring opening angle 2.4$^\circ$, radius increases upwards, and the surrounding C ring  is overexposed). Periodic brightness variations are apparent throughout this portion of the D ring. However, one can also see that the intensity of the pattern varies with azimuth, reaching a minimum near the ring ansa. These azimuthal trends indicate that at least a fraction of these brightness variations are due to vertical structure in the ring. However, periodic patterns are also visible at the ring ansa, where the signal from vertical structures should vanish. Thus  periodic variations in the ring's opacity also seem to be present.}
\label{waveim}
\end{figure}

The D ring is the innermost component of Saturn's ring system, extending from the inner edge of the C ring towards the planet's cloud tops. One of the more intriguing structures in this region is a set of periodic brightness variations with a wavelength of $\sim$30 km covering the outermost 1500 km of this ring (see Figure~\ref{waveim}). The intensity of this periodic pattern varies with longitude around the ring, and it becomes rather indistinct near the ring ansa. Such azimuthal intensity variations are characteristic of vertical ring structures, and so \citet{Hedman07} argued that these periodic patterns were due to a vertical corrugation in the ring. Similar corrugations had previously been identified in Galileo images of Jupiter's rings \citep{OB99, Showalter11}, but the more extensive Cassini images (coupled with an earlier Hubble Space Telescope occultation) revealed that the wavelength of the D-ring pattern was steadily decreasing over time. The observed trend in the pattern's wavelength was consistent with the evolution of a corrugation under the influence of differential nodal regression. This finding not only confirmed that the D-ring pattern included a vertical corrugation, but also suggested that this structure probably arose from some event in the recent past that caused the ring to become tilted out of the planet's equatorial plane.

Later investigations revealed that both Saturn's C ring \citep{Hedman11} and Jupiter's main ring \citep{Showalter11} contained similarly evolving patterns of vertical corrugations. Furthermore, by extrapolating these trends backwards in time, we could  estimate when Jupiter's and Saturn's rings became inclined relative to their planet's equatorial plane. The event that tilted Jupiter's rings happened in the summer of 1994, when the fragments of comet Shoemaker-Levy 9 were crashing into the planet. It is therefore reasonable to conclude that this cometary debris was responsible for tilting Jupiter's rings.  The event that tilted Saturn's rings occurred in the early 1980s, but unlike the Shoemaker-Levy 9 collision with Jupiter, this earlier event was not directly observed. Fortunately, some information about the impact is encoded within the corrugations themselves. In particular, the extent of the disturbance in the rings indicates that the ring encountered a dispersed debris field rather than a single compact object. Thus \citet{Hedman11} inferred that the ring-tilting event at Saturn may have involved an object that was disrupted by a previous close encounter with Saturn, just as Shoemaker-Levy 9 broke apart during its close encounter with Jupiter in 1992. 

Here we present a more detailed analysis of the periodic patterns in Saturn's D ring that provides additional information about how the rings were disturbed and how the resulting patterns evolved over time. This investigation focuses on the D-ring structures for two reasons. First, there are extensive Cassini observations available, and suitably high-resolution Cassini images that capture the relevant periodic patterns span nearly a decade (By contrast, the corrugations in Saturn's C ring were only visible for a brief interval around Saturn's equinox in 2009). This data set yields very precise measurements of how the patterns' wavelengths vary over time, and so we can confirm that these structures are evolving at rates consistent with current models of Saturn's gravitational field. 

Second, the D ring appears to contain a second periodic structure overprinted on the corrugation. The original analysis of the D-ring patterns by \citet{Hedman07} revealed that the scatter of the wavelength measurements around the mean trend with time was larger than their individual error bars would predict. Furthermore, observations taken further from the ring ansa appeared to have systematically longer wavelengths, suggesting that another periodic structure was being revealed in certain viewing geometries. Close inspection of additional Cassini images have confirmed this supposition. For example, Figure~\ref{waveim} shows that periodic brightness variations are visible at the rings' ansa. A vertical corrugation cannot generate brightness variations at this location because the vertical slopes are all nearly orthogonal to the observer's line of sight (see below). Hence the patterns visible close to the ansa likely reflect variations in the ring's  surface density rather than its vertical structure. Since there is no obvious change in the pattern's wavelength close to the ansa, the wavelength of these opacity variations must be nearly identical to the wavelength of the vertical corrugations. This suggests that these density variations were generated by the same event that formed the corrugation, and detailed analyses of the two patterns' wavelengths confirm this hypothesis. Furthermore, the  relative amplitudes of these two patterns, along with some anomalous trends in the corrugations' wavelength with radius between the D and C rings, yield new information about how the ring was disturbed in 1983.

We begin this investigation by reviewing the theory of how corrugations in the ring are expected to evolve over time, and how similarly time-variable periodic opacity variations could be produced (Section~\ref{theory}). Section~\ref{methods} then describes the analytical procedures used to isolate opacity variations from vertical structures and to obtain estimates of the relevant patterns' wavelengths and amplitudes. Section~\ref{results} lists the images considered for this analysis and summarizes the resulting estimates of the patterns' properties and evolution over time. Based on these results, Section~\ref{spiral} demonstrates that the measured wavelengths and amplitudes are consistent with the expected evolution of patterns generated by a discrete disturbance like an impact that occurred sometime in the past. Section~\ref{event} describes how  the amplitudes and precise wavelengths of these patterns can provide new  information about the pre-impact trajectory of the debris that collided with the rings. The results and potential implications of this analysis are summarized at the end of this paper.

\section{Theoretical Background}
\label{theory}

This investigation builds upon the earlier studies of the corrugations \citep{Hedman07, Hedman11} not only by considering additional data, but also by employing image-processing techniques that can isolate signals due to vertical structure from those due to optical depth variations. In order to motivate this effort and justify some of the choices made in the analytical procedures, we first review how an inclined sheet can evolve into a vertical corrugation and the expected observable properties of such a corrugation. In addition, this section will describe how a disturbance in the ring-particles' in-plane motions can produce periodic optical-depth variations with evolving wavelengths very similar to those associated with the corrugations. 


Imagine that a portion of Saturn's ring became tilted relative to Saturn's equatorial plane at a time $t_i$. The particles in such a tilted ring all have a finite inclination $I$ and the same longitude of ascending node $\Omega$, which we can set equal to zero for the remainder of this calculation. However, if the forces exerted on the ring particles deviate from a purely central inverse-square-law, then the node positions will regress at a rate $\dot{\Omega}(r)$ that depends on the particles' mean radial distance from Saturn's spin axis $r$. Hence if the ring is observed at a time $t_f>t_i$, the node location at a given $r$ will be $(t_f-t_i)\dot{\Omega}(r)$. So long as the nodal regression is predominantly due to Saturn's oblateness parameter $J_2$, $\dot{\Omega}(r)$ will  have negative values everywhere and the absolute value of $\dot{\Omega}(r)$ will decrease monotonically with increasing $r$. The longitude of ascending node will therefore form a leading spiral that becomes progressively more tightly wound  over time. More specifically, the ring's vertical position  $z$ at a given radius $r$ and longitude $\theta$ can be expressed as the following function of the ring's inclination $I$ and the node longitude:
\begin{equation}
z=rI\sin(\theta-\Omega(r)),
\end{equation}
The vertical position of the ring at a given $\theta$ will therefore oscillate up and down as a function of radius. In the vicinity of any given radial position in the ring $r_o$, the node position can be approximated using the first two terms of the Taylor series:
\begin{equation}
\Omega(r)=\Omega(r_o)+\frac{\partial\dot{\Omega}}{\partial r}(t_f-t_i)(r-r_o).
\label{nodea}
\end{equation}
Hence
\begin{equation}
z=A_z\sin\left[\theta-\Omega(r_o)-k_z(r-r_o)\right],
\end{equation}
where $A_z=aI$ is the amplitude of the vertical corrugation  and the parameter
\begin{equation}
k_z=\left|\frac{\partial \dot{\Omega}}{\partial r}\right|(t_f-t_i).
\label{eqkz}
\end{equation}
is the corrugation's radial wavenumber. 

In practice, we do not observe $z$ directly, but instead measure variations in brightness caused by differing vertical slopes along the line of sight (see Section~\ref{methods}). For the corrugations and observations considered here, which have radial wavelengths of order 30 km, the spiral pattern is wrapped so tightly that the relevant slopes are nearly radial and so are approximately equal to the following expression:
\begin{equation}
\frac{\partial z}{\partial r}\simeq-A_zk_z\cos\left[\theta-\Omega(r_o)-k_z(r-r_o)\right],
\end{equation}
where we have assumed $k_z$ and $A_z$ are approximately constant over the scale of one wavelength. For this study, this is a reasonable approximation since including the relevant corrections has no impact on the pattern's wavelength and changes the  slope estimates by much less than 10\%.

Just as periodic vertical corrugations can arise from a ring composed of particles on aligned inclined orbits, periodic variations in the ring's surface density could arise from a ring composed of particles on aligned eccentric orbits. Say that at time $t_i$ all the particles in a ring acquired finite eccentricities $e$ with same pericenter longitude $\varpi$, which we can again set to zero for the purposes of this calculation. The finite orbital eccentricities of these ring particles cause them to move back and forth in radius, so we must take care to distinguish the ring particles' radial positions ($r$) from their orbital semi-major axes ($a$). Any force acting on these ring particles that deviates from a central inverse-square-law will cause particles with different semi-major axes to undergo different rates of pericenter precession $\dot{\varpi}(a)$, so that at any later time $t_f$ the pericenter location  will  also depend on position $\varpi(a,t_f)=(t_f-t_i)\dot{\varpi}(a)$. Provided Saturn's gravity dominates the precession rates, $\dot{\varpi}(a)$ will be a positive, monotonically-decreasing function of $a$, and so the pericenter location will become an increasingly tightly wrapped trailing spiral as time goes on.  

These organized eccentric motions produce variations in the ring's surface density. If the particles at any given semi-major axis follow the same orbit (called a streamline), and if they are evenly distributed in  longitude, then the local surface density $\sigma$ is inversely proportional to the radial distance between adjacent streamlines:
\begin{equation}
\sigma=\frac{\sigma_o}{\partial r/ \partial a},
\label{streamline}
\end{equation}
where $\sigma_o$ is the unperturbed surface density and $r$ is the radial position of the streamline:
\begin{equation}
r=a-ae\cos(\theta-\varpi).
\end{equation}
Let us consider a region around a particular semi-major axis $a_o$ where the pericenter's angular location is given by the approximate expression:
\begin{equation}
\varpi(a)=\varpi(a_o)+\frac{\partial\dot{\varpi}}{\partial a}(t_f-t_i)(a-a_o).
\label{peria}
\end{equation}
At this location, the streamline equation becomes
\begin{equation}
r=a-A_r\cos[\theta-\varpi(a_o)+k_r(a-a_o)],
\end{equation}
where $A_r=ae$ is the magnitude of the radial excursions and 
\begin{equation}
k_r=\left|\frac{\partial\dot{\varpi}}{\partial a}\right|(t_f-t_i).
\label{eqkr}
\end{equation}
If we assume these $A_r$ and $k_r$ parameters are roughly constant on radial scales of order $k_r^{-1}$, then the derivative of $r$ with respect to $a$ is approximately:
\begin{equation}
\frac{\partial r}{\partial a}=1+A_rk_r\sin[\theta-\varpi(a_o)+k_r(a-a_o)].
\label{str}
\end{equation}
So long as $A_rk_r<<1$, we may combine equations~\ref{streamline} and~\ref{str},  and obtain the following approximate expression for the perturbations in the ring density
\begin{equation}
\sigma\simeq\sigma_o\left[1-A_rk_r\sin[\theta-\varpi(a_o)+k_r(a-a_o)]\right].
\end{equation}
Furthermore, we may now approximate the semi-major axes $a$ and $a_o$ with the observed radii $r$ and $r_o$. Hence the fractional density variations in
the ring can be written as:
\begin{equation}
\frac{\delta\sigma}{\sigma}\simeq-A_rk_r\sin[\theta-\varpi(r_o)+k_r(r-r_o)].
\end{equation}

The ring therefore exhibits a spiral pattern in its density analogous to the spiral corrugation, and at a given longitude $\theta$, the ring's density (or, equivalently, its normal optical depth $\tau_n$) will vary periodically with an amplitude $A_rk_r$ and wavenumber $k_r$. As with the corrugations, if we relaxed our assumptions that $A_r$ and $k_r$ are relatively constant or that $A_rk_r<<1$, we would still find a periodic signal with wavelength $k_r$, but with a slightly different amplitude. Since this spiral pattern arises from the ring-particles' eccentric motions, we will refer to this pattern as an ``eccentric spiral''  below.

The evolution of both  patterns' radial wavelengths are governed by radial gradients in orbit evolution rates, and previous studies of the corrugations demonstrated that most of this evolution can be ascribed to $J_2$ and higher-order components of Saturn's  gravitational field. More specifically, Saturn's large $J_2$ causes orbital  nodes and pericenters to evolve over time at the following rates (Murray and Dermott 1999 Equations 6.249 and 6.250):\nocite{MurrayDermott}
\begin{equation}
\dot{\varpi}=\frac{3}{2}J_2n\left(\frac{R_s}{r}\right)^2
\end{equation}
\begin{equation}
\dot{\Omega}=-\frac{3}{2}J_2n\left(\frac{R_s}{r}\right)^2,
\end{equation}
where $R_s=60,330$ km is the fiducial planetary radius used in the calculation of $J_2$ and $n=\sqrt{GM/r^3}$ is the zeroth-order estimate of the local mean motion ($G$ being the gravitational constant and $M$ being Saturn's mass).  If $J_2$ was the only source of precession, then to first order the wavelengths of the corrugation and the eccentric spiral would follow identical trends:
\begin{equation}
k_z=\frac{21}{4}J_2\sqrt{\frac{GM}{r^5}}\left(\frac{R_s}{r}\right)^2(t_f-t_i),
\end{equation}
\begin{equation}
k_r=\frac{21}{4}J_2\sqrt{\frac{GM}{r^5}}\left(\frac{R_s}{r}\right)^2(t_f-t_i).
\end{equation}
If we include higher-order perturbations to $\dot{\varpi}$ and $\dot\Omega$, then the above trends are slightly modified and the two winding rates differ by a few percent, which turns out to be consistent with the observations (see Section~\ref{spiral}). However, the above expressions provide a good first-order approximation of the trends  these patterns should exhibit. Hence our analytical routines must be able to cope with patterns whose  wavenumbers increase linearly with time and vary with radius approximately like $r^{-9/2}$. Also, since the same event that tilted the ring could have induced eccentric motions, these routines need to be able to untangle optical-depth variations and corrugations with nearly identical wavelengths.

\section{Data Reduction Procedures}
\label{methods}

The amplitudes and wavelengths of the periodic patterns in the ring's optical depth and vertical position are derived from individual Cassini images using a multi-step process. First,  the image data are converted  into maps of the ring's brightness as a function of radius and longitude. Second,  the azimuthal brightness trends in these maps are fit to a model that yields separate profiles of the ring's normal optical depth and vertical structure. Finally, Fourier transforms are used to determine the wavelengths and amplitudes of the relevant periodic patterns in the profiles. 

\subsection{From images to brightness maps}
\label{maps}

\begin{figure}
\centerline{\resizebox{5in}{!}{\includegraphics{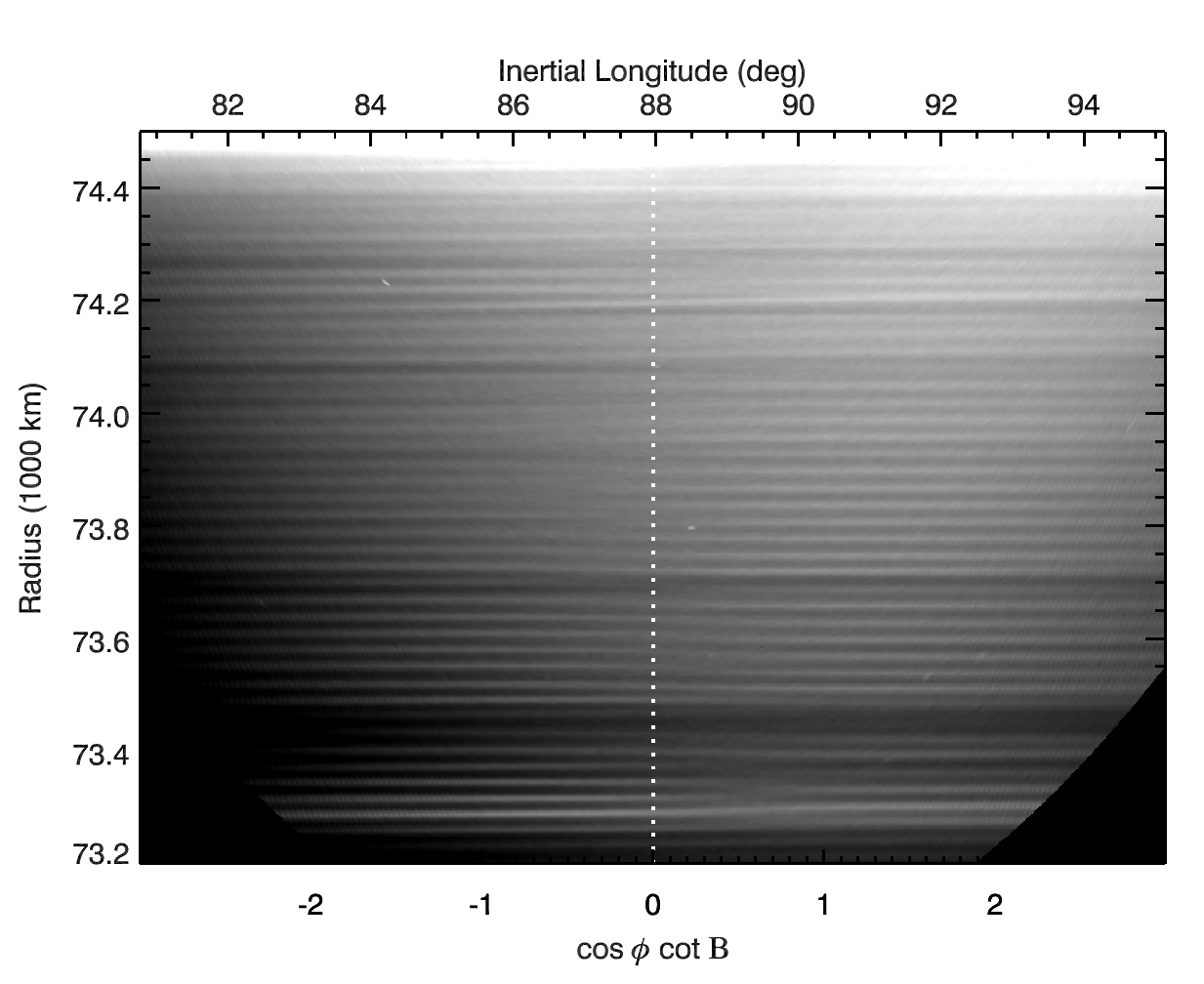}}}
\caption{A re-projected image of the brightness variations in the outer D ring image, derived from the Cassini Image N1571969357 shown in Figure~\ref{waveim}. This graphic shows more clearly how the ring's brightness varies with radius and longitude, and in particular, the dotted line corresponds to the exact ring ansa (i.e. $\cos\phi \cot B=0$, see Section~\ref{mprofs}), where the brightness variations from vertical structures should be minimal. That radial profile can therefore produce a measure of the true optical depth variations in the ring, while differences between profiles on either side of this line provide information about the magnitude of the vertical corrugation.}
\label{waverpj}
\end{figure}

This study focuses on images obtained by the Narrow Angle Camera (NAC) of the Imaging Science Subsystem onboard the Cassini spacecraft \citep{Porco04, West10}. Each image was calibrated using the standard CISSCAL routines to convert raw data numbers into $I/F$, a standard measure of brightness that is unity for a Lambertian surface illuminated at normal incidence (see {\tt http://pds-rings.seti.org/cassini/iss/calibration.html}). Each image was initially navigated using the appropriate SPICE kernels,  and this navigation was subsequently refined based on the positions of known stars in the field of view and verified against the position of the C-ring's inner edge. Once navigated, the radius and longitude viewed by each pixel in the image can be determined. Combined with the known spacecraft position, this information allows us to derive parameters like the phase and emission angles, and how they vary across the rings.

In prior examinations of these periodic structures in Saturn's rings \citep{Hedman07, Hedman11}, the image data were converted into profiles of the ring's brightness versus radius by averaging over a range of longitudes. This procedure has the benefit of improving signal-to-noise, but also removes any information about azimuthal trends in brightness. Such azimuthal trends provide the information needed to isolate vertical corrugations from opacity variations, so for this analysis we instead re-project the image data onto a regular grid of radii and longitudes. Figure~\ref{waverpj} shows an example of one such re-projected data set  (derived from the image shown in Figure~\ref{waveim}). The range of longitudes covered in these re-projected ``maps'' is rather limited (only $15^\circ$ in this case), so they do not provide clear evidence that the observed structures are  spiral patterns as opposed to purely radial patterns. However, these maps do contain sufficient information to generate radial profiles of both the ring's relative normal optical depth and its vertical structure.

\subsection{From brightness maps to radial profiles}
\label{mprofs}


While the re-projected brightness maps provide a clearer picture of how the ring's brightness varies with radius and azimuth, they are still cumbersome for the purposes of extracting precise wavelength and amplitude estimates. Hence separate profiles of the ring's opacity and vertical position as functions of radius are derived from each map by fitting the azimuthal brightness trends at each radius to a model of the ring's structure. This model includes various simplifying assumptions and approximations, but it should be sufficient to recover the desired pattern parameters.

First, let us assume that {\sl the ring has a sufficiently low optical depth that the observed brightness is proportional to the optical depth along the line of sight.} This is a reasonable assumption for the outer D-ring, whose normal optical depth is of order $10^{-3}$ \citep{Hedman07}, and it allows us to write the ring brightness signal $S_r$  as the following function of the ring optical depth and viewing geometry (which is valid regardless of whether the rings are viewed from the lit or unlit side):
\begin{equation}
S_r=\frac{1}{4}{\varpi_0P(\alpha)}\frac{\tau_n}{|\hat{n}\cdot\hat{o}|},
\label{tau1}
\end{equation} 
where $\varpi_0$ and $P(\alpha)$ are the effective mean single-scattering albedo and phase function of the D-ring particles, $\tau_n$ is the ring's optical depth when viewed at normal incidence, $\hat{n}$ is the unit vector normal to the ring surface and  $\hat{o}$ is the unit vector pointing from the surface to the observer. Note that the ``normal optical depth'' $\tau_n$ is the optical depth observed when the line of sight is perpendicular to the local surface, regardless of any warps or corrugations. Hence $\tau_n$ is proportional to the local surface density $\sigma$ discussed in Section~\ref{theory} above.

\begin{figure}
\centerline{\resizebox{5in}{!}{\includegraphics{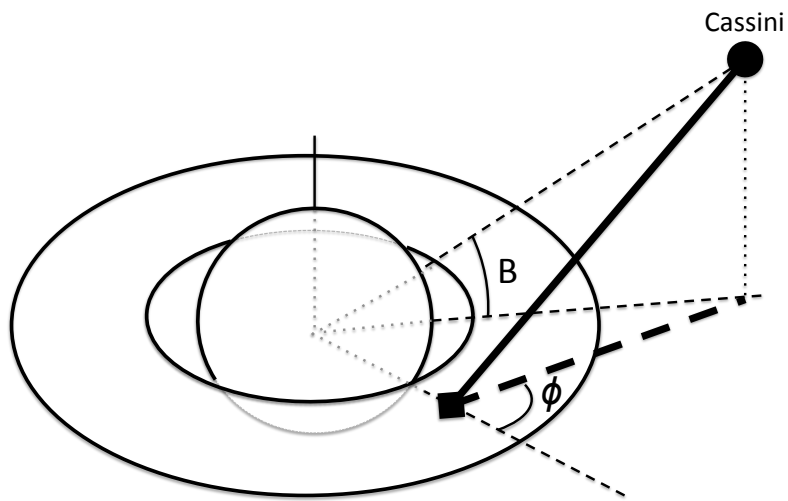}}}
\caption{Diagram showing the relevant observation geometry for this analysis. The spacecraft is located at the dot and views a location on the ring marked with the square. The angle $B$ is the spacecraft's elevation angle above the ring plane, while $\phi$ is the azimuthal angle between a radial direction and the observed line of sight.}
\label{ringdiag}
\end{figure}

Next, assume that {\sl within each image, the ring's vertical position, normal optical depth, and scattering properties are only functions of radius.} While the structures of interest here are almost certainly spiral patterns, in practice the spirals are wrapped so tightly that they can be well approximated as radial structures on the scale of individual images or ring maps. Assuming the vertical displacement $z$ is only a function of radius allows us to express the surface normal to the warped ring surface as: 
\begin{equation}
\hat{n}=\frac{\hat{z}-\hat{r}(dz/dr)}{\sqrt{1+(dz/dr)^2}},
\end{equation}
where $\hat{z}$ and $\hat{r}$ are unit vectors aligned perpendicular to Saturn's equatorial plane and with the local radial direction, respectively. The dot product of $\hat{n}$ with the unit vector pointing from the rings to the observer is:
\begin{equation}
|\hat{n}\cdot\hat{o}|
=|\sin B|\left|\frac{1-\cot B \cos \phi(dz/dr)}{\sqrt{1+(dz/dr)^2}}\right|,
\label{mu0}
\end{equation}
where $B$ is the spacecraft's elevation angle above the rings and $\phi$ is the azimuthal angle between the line of sight and the local radial direction (see Figure~\ref{ringdiag}). The vertical displacements therefore produce signals that depend upon the observed azimuth $\phi$ in a predictable way. Furthermore, if we also assume $\tau_n$ and $\varpi_0P(\alpha)$ are only functions of radius, then the vertical structure is the sole source of azimuthal brightness variations at a given radius.

We then  assume, for the sake of simplicity, that {\sl the vertical slopes within the pattern are sufficiently small that we may consider only their first-order perturbations to the observed brightness data.} This can be justified based on the {\sl post facto} derived slope profiles, which indicate $dz/dr<0.1$. Furthermore, in all of the maps used in this analysis, we only fit data 
where $|\cot B\cos\phi|<2$, so $\cot B\cos\phi(dz/dr)$ is also a small number. 
This means that we do not have to worry about situations where the line of sight crosses through the warped ring multiple times \citep{Gresh86}. The signal from the ring can therefore be obtained by simply inserting the above expression for  $|\hat{n}\cdot\hat{o}|$ into Equation~\ref{tau1}, which gives:
\begin{equation}
S_r=\frac{\varpi_0P(\alpha)\tau_n}{4|\sin B|}
\left|\frac{\sqrt{1+(dz/dr )^2}}{1-\cot B \cos \phi (dz/dr)}\right|.
\label{taufull}
\end{equation}
Furthermore, since both $dz/dr$ and $\cot B \cos \phi (dz/dr)$ are small quantities, we may approximate the above expression as:
\begin{equation}
S_r=\frac{\varpi_0P(\alpha)\tau_n}{4|\sin B|}\left[{1+\cot B \cos \phi (dz/dr)}\right].
\label{tauap}
\end{equation}

Equation~\ref{tauap} only provides the signal from the ring itself.  In practice, there are also instrument backgrounds, which can vary with both radius and azimuth in complex ways \citep{West10}. However, we deliberately exclude images or parts of images where such complex stray light patterns are prominent, so any residual background signals should be smooth compared to the patterns considered here. If we call this background level $S_b$, then the total signal in a given image should be:
\begin{equation}
S=\left(S_b+\frac{\varpi_0P(\alpha)\tau_n}{4|\sin B|}\right)+\frac{\varpi_0P(\alpha)\tau_n}{4|\sin B|} \left(\frac{dz}{dr}\right)\cot B \cos \phi .
\label{tauap1}
\end{equation}
We may also define the quantity $s(r)=S(r)|\sin B|$ (called ``normal I/F'' below) which would be the brightness of the ring viewed at normal incidence if $z$ and $S_b$ were identically zero. In this scenario, $s$ is a useful quantity  because its dependence on viewing angle is slightly simpler than $S$:
\begin{equation}
s=\left(S_b|\sin B|+\frac{1}{4}\varpi_0P(\alpha)\tau_n\right)+\frac{1}{4}\varpi_0P(\alpha)\tau_n \left(\frac{dz}{dr}\right)\cot B \cos \phi .
\label{tauap2}
\end{equation}

In any individual image, the azimuth angle $\phi$  between observer's projected line of sight and the local radial direction is a monotonic function of the observed ring longitude. Hence we can translate the observed longitudes for a given map into values of the parameter $\cot B \cos \phi$, (Indeed, the horizontal coordinate of the map in Figure~\ref{waverpj} is given in terms of both longitude and  $\cot B \cos \phi$). Since each row in the map gives the apparent brightness of the rings as a function of $\cot B \cos \phi$ at a given radius, 
we may fit these data to the following function:
\begin{equation}
s(r)=p_0(r)+p_2(r)\cot B \cos \phi.
\label{model}
\end{equation}
and obtain estimates of the parameters $p_0$ and $p_2$ at each radius. From Equation~\ref{tauap2}, we can identify these fit parameters with the following quantities:
\begin{equation}
p_0=S_b|\sin B|+\frac{1}{4}\varpi_0P(\alpha)\tau_n
\end{equation}
\begin{equation}
p_2=\frac{1}{4}\varpi_0P(\alpha)\tau_n\left(\frac{dz}{dr}\right).
\end{equation}
Note that if the fractional variations in ${\varpi_0P(\alpha)\tau_n}$ are small, then
the variations in $p_2$ will be dominated by variations in the vertical slopes, while the variations in $p_0$ should only be generated by variations in the product ${\varpi_0P(\alpha)\tau_n}$ (assuming $S_b$ does not vary much on the relevant scales). 

If we could be certain that $S_b=0$, then it would be easy to transform $p_0$ and $p_2$ into estimates of the fractional optical depth variations and vertical slopes. However, in practice we cannot assume this to be the case, and so we must estimate $S_b$ and subtract it from $p_0$. In a few images, a portion of the imaged ring lies in Saturn's shadow, and since $S_r=0$ in Saturn's shadow, we can use the residual signal in these regions to estimate $S_b$. Unfortunately, in most images no shadowed ring regions are visible, and so we need another method of estimating the background. Based on previous work, we know that the periodic patterns occupy a portion of the D ring that is close to much fainter regions \citep{Hedman07}. At low phase angles, the ring's brightness drops sharply interior to 73,200 km, while at high phase angles there are relatively faint regions exterior to 74,000 km. While neither of these regions is completely empty, each is sufficiently dark to provide a rough estimate of $S_b|\sin B|$. Hence we may define the quantity
\begin{equation}
p_1=p_0-\min(p_0),
\end{equation}
where $\min(p_0)$ is the minimum value of the $p_0$ profile between  73,000 and 74,500 km. While $p_1$ may be a simple and crude estimator of $\varpi_0P(\alpha)\tau_n/4$, more complex background models (such as a slope) are not practical because suitable dark regions usually cannot be found on both sides of the periodic patterns.
Fortunately, as demonstrated below, the background model does not greatly affect the pattern wavelength estimates.

\begin{figure}
\centerline{\resizebox{5in}{!}{\includegraphics{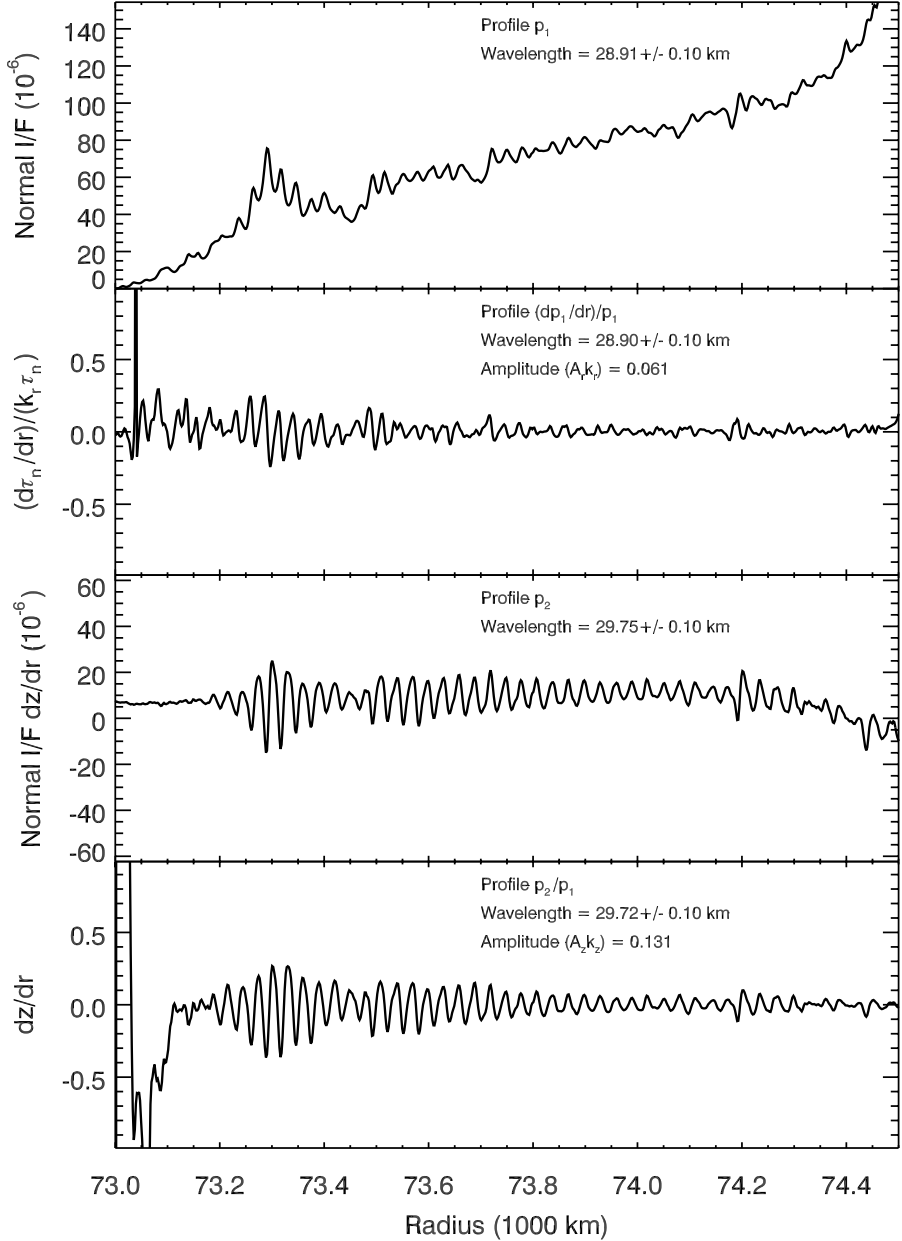}}}
\caption{Profiles of optical-depth variations and vertical slopes derived from image N1571969357 (shown in Figure~\ref{waveim}) by fitting data where $|\cos\phi\cot B|<0.5$. The top panel shows the profile $p_1$, with a constant background subtracted. Periodic variations in the optical depth are evident here. The second panel shows the profile of $(p_1'/p_1)\lambda_r/2\pi=(d\tau_n/dr)/k_r\tau_n$, which isolates the short-frequency periodic signals in this profile. The third panel shows the profile $p_2$, which is proportional to the product $\tau_n dz/dr$, while the bottom panel shows the profile $p_2/p_1$, which should ideally equal $dz/dr$. Each panel has the wavelength of the periodic signature provided (note that for the top panel the wavelength is derived from the radial derivative of the profile $p_1'$), while amplitudes of the $\delta \tau_n/\tau_n$ and $dz/dr$ variations are given in the second and fourth panels.}
\label{profex}
\end{figure}

\begin{figure}
\centerline{\resizebox{5in}{!}{\includegraphics{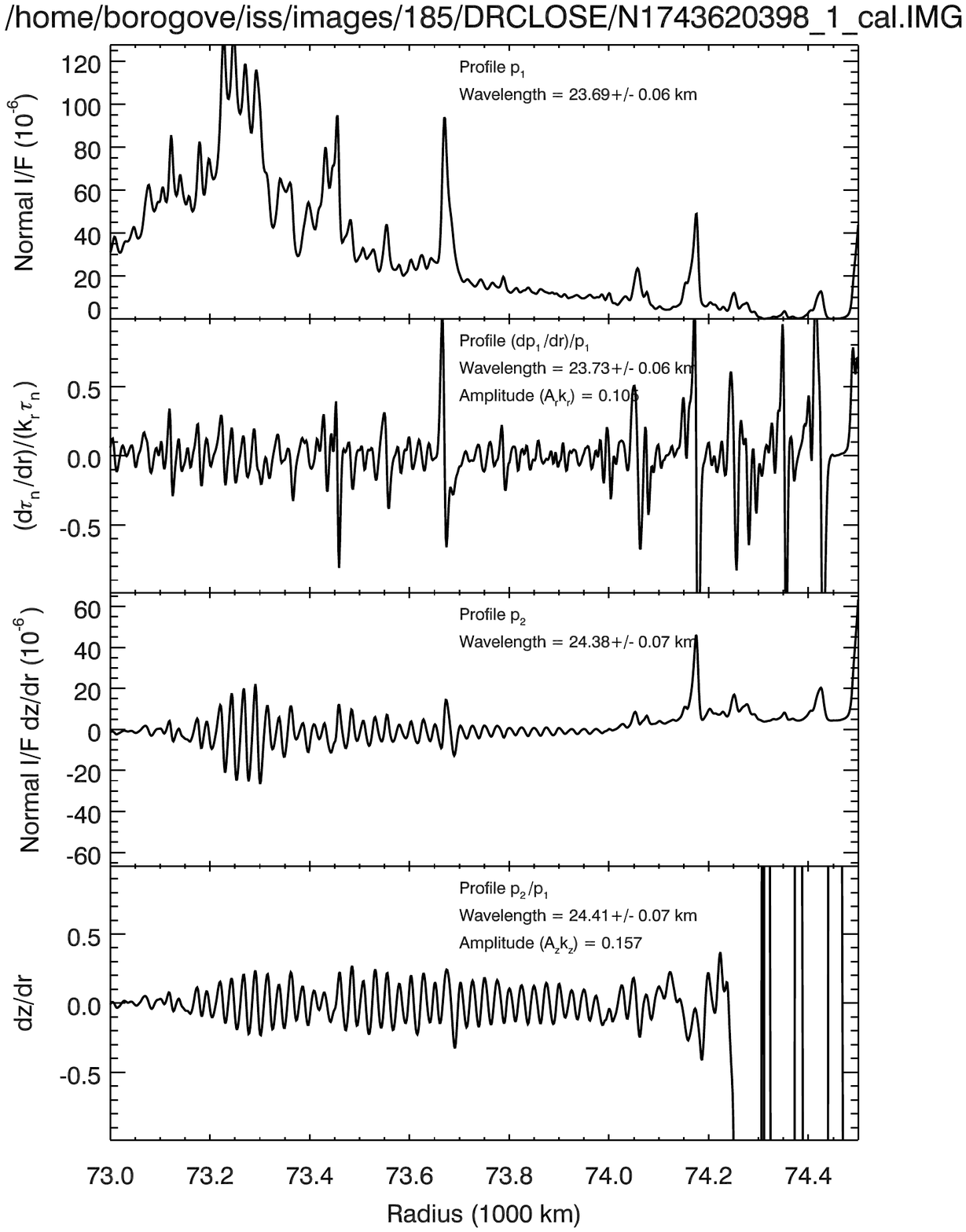}}}
\caption{Profiles of optical depth variations and vertical slopes derived from image N1743620398 by fitting data where $|\cos\phi \cot B|<1.5$. The top panel shows the profile $p_1$, with a constant background subtracted. Periodic variations in the optical depth are evident here. The second panel shows the profile of $(p_1'/p_1)\lambda_r/2\pi=(d\tau_n/dr)/k_r\tau_n$, which isolates the short-frequency periodic signals in this profile. The third panel shows the profile $p_2$, which is proportional to the product $\tau_n dz/dr$, while the bottom panel shows the profile $p_2/p_1$, which should ideally equal $dz/dr$. Each panel has the wavelength of the periodic signature provided (note that for the top panel the wavelength is derived from the radial derivative of the profile $p_1'$), while amplitudes of $\delta \tau_n/\tau_n$ and $dz/dr$ are given in the second and fourth panels.}
\label{profex2}
\end{figure}

Figures ~\ref{profex} and~\ref{profex2} illustrate examples of the profiles $p_1$ and $p_2$ derived using the above procedures. Both these profiles exhibit clear periodic signals. However, the periodic pattern in $p_1$ is superposed on top of other ring structures. These background features complicate efforts to measure the wavelength of the periodic pattern. Fortunately, a cleaner profile of the periodic signal can be obtained by simply taking the radial derivative of the profile $p_1'=dp_1/dr$. As Figures~\ref{profex} and~\ref{profex2} demonstrate,  the derivative operation suppresses the various long-wavelength trends and thus isolates the periodic pattern. Furthermore, this operation should not change the wavelength of  the pattern, and the pattern's amplitude in the $p'_1$ profile is just the pattern's amplitude in the $p_1$ profile multiplied by $2\pi/\lambda_r$, where $\lambda_r$ is the pattern's wavelength. 

The theoretical calculations in Section~\ref{theory} yield predictions for the parameters $dz/dr$ and $\delta \tau_n/\tau_n=\delta \sigma/\sigma$. Profiles of these quantities can be derived from the profiles $p_1, p'_1$ and $p_2$. The estimator of  vertical slope profile is the ratio of the $p_2$ and $p_1$ profiles:
\begin{equation}
\widehat{\frac{dz}{dr}}=\frac{p_2}{p_1}
\end{equation}
The estimator of the fractional optical-depth variations is a bit more complicated. The profile $p_1'$ provides the cleanest measure of the periodic patterns in the ring's brightness, and if we assume that neither the particle albedo nor the particle phase function varies on the scale of 30 km,  we may take $p_1'$ as an estimator of $\varpi_oP(\alpha)\tau_n'/4$. To obtain the fractional brightness variations, we must divide this profile by a measure of the average brightness. We estimate the average brightness using a smoothed version of the $p_1$ profile. Thus we may define the estimator
\begin{equation}
\widehat{\frac{\tau_n'}{\tau_n}}=\frac{p_1'}{\bar{p}_1}
\end{equation}
where $\bar{p}_1$ is a version of the $p_1$ profile that has been smoothed over  120 km. Note that unlike the vertical slopes and the fractional brightness variations computed in Section~\ref{theory} (which are both unitless),  $p_1'/\bar{p}_1$ has units of  km$^{-1}$. Thus to facilitate comparisons, we will consider the quantity:
\begin{equation}
\frac{\lambda_r}{2\pi}\widehat{\frac{\tau_n'}{\tau_n}}=\frac{p_1'}{k_r\bar{p}_1},
\end{equation}
where $\lambda_r$ is the wavelength  and $k_r$ is the wavenumber of the optical-depth variations derived from a Fourier analysis of the profile. The above quantity is unitless and should have the same amplitude and wavelength as $\delta \sigma/\sigma$. 

However, even though $p_1'/\bar{p}_1$ and $p_2/p_1$ are sensible estimators of the rings' vertical slopes and optical depth variations, they both involve ratios of signals, and both these ratios are sensitive to the assumed background level. By contrast, the profiles $p_1'$ and $p_2$ have the advantage that they are independent observable quantities derived from the images (note that $p_1'=p_0'$ so long as the background level varies slowly). Wavelength estimates derived from $p_1'$ and $p_2$ therefore might be more robust than those derived from $p_1'/\bar{p_1}$ and $p_2/p_1$. In order to control for this possibility and ensure we obtain the most reliable wavelength estimates possible, we will consider both sets of profiles and compare the results. Therefore, our analysis yields five profiles from each image:
\begin{itemize}
\item
$p_1$, which is an estimate of the ring's brightness when viewed from normal incidence (top panel in Figures~\ref{profex}-\ref{profex2}).
\item
$p_1'$, the radial derivative of $p_1$ which provides a cleaner measure of the relevant variations in the ring's opacity.
\item
$p_1'/\bar{p}_1$, which provides a quantitative estimate of the fractional density variations (second panel in Figures~\ref{profex}-\ref{profex2}).
\item
$p_2$, which provides a clean periodic signal that is predominantly due to the vertical corrugation (third panel in Figures~\ref{profex}-\ref{profex2}).
\item
$p_2/p_1$, which provides a quantitative estimate of the ring's vertical slopes (bottom panel in Figures~\ref{profex}-\ref{profex2}).
\end{itemize}
The Fourier techniques described in the following section are applied to the last four profiles.

\subsection{From profiles to wavelength and amplitude estimates}

The amplitude and wavelength of the periodic patterns in the various profiles are obtained using Fourier techniques very similar to those previously used by \citet{Hedman11}. As in that work, an important consideration for this step in the analysis is that patterns generated by differential nodal regression or differential apsidal precession have wavelengths that vary systematically and continuously with distance from the planet  (see Section~\ref{theory} above). In particular, we expect these wavelengths to scale roughly as $r^{9/2}$ (indeed this was true for the corrugations in the C ring, see Hedman {\it et al.} 2011). As can be seen in Figures~\ref{profex} and~\ref{profex2}, periodic patterns are evident over an 800-km-wide region between 73,200 km and 74,000 km. If these patterns' wavelengths vary like $r^{9/2}$, then the pattern's wavelength will vary by 5\% across this region. Shifts of this magnitude are just barely detectable in individual profiles, but cannot be ignored if we wish to obtain precise wavelength estimates. To deal with this phenomenon, we transform the observed ring radii $r$ into the re-scaled distance parameter $d$:
\begin{equation}
d=r\frac{2}{7}\left(\frac{r_o}{r}\right)^{9/2}, 
\end{equation}
where $r_o=$73,600 km. In this coordinate system, the ``rescaled wavelength'' of a periodic structure becomes:
\begin{equation}
{\Lambda}={\lambda}\left(\frac{r_o}{r}\right)^{9/2},
\end{equation}
where $\lambda=2\pi/k$ is the true radial wavelength (note that the factor of $2/7$ in the definition of $d$ ensures that $\Lambda=\lambda$ when $r=r_o$).  For patterns generated by differential nodal regression or apsidal precession, $\lambda$ should scale approximately as $r^{9/2}$ , so $\Lambda$ should be approximately constant.\footnote{This rescaled wavelength will not be exactly constant because the higher-order harmonics in the gravitational field perturb the relevant precession rates. However, the variations in $\Lambda$ due to these perturbations are less than 0.3\% across the region of interest, and neglecting these corrections changes the final estimates of the pattern wavelengths by less than 0.1\%. Hence we chose not to complicate our analysis to remove these small residual trends.} This transformation therefore allows us to take the Fourier spectrum of the entire region between 73,200 km and 74,000 km and obtain precise wavelength estimates. It  also makes these wavelength estimates insensitive to variations in the amplitude of the pattern across the region, which can influence what part of the wave contributes most to the peak in the Fourier transform. Since $\Lambda=\lambda$ when $r=r_o=73,600$ km, the wavelengths obtained by this procedure can be regarded as estimates of the pattern's wavelength at 73,600 km, which falls near  the middle of the region where the pattern in evident.

For each profile, we compute an over-resolved Fourier spectrum of the rescaled data by evaluating the Fourier transform  for a tightly spaced array of $\Lambda$ values ($\delta \Lambda/\Lambda =0.001$). These spectra contain a strong peak at the wavelength of the desired periodic signal. Fitting this peak to a Gaussian yields estimates of the pattern's wavelength and amplitude. Only data where the amplitude of the Fourier transform is at least half the peak value and within 10\% of the peak wavelength are included in the fit. This fit yields the following parameters:
\begin{itemize}
\item The location of the peak in the Fourier spectrum, which is an estimate of  $\Lambda$.
\item The amplitude $\mathcal{A}$ of the pattern, derived from the peak amplitude of the Fourier spectrum. For the profiles $p'_1/\bar{p}_1$ and $p_2/p_1$, $\mathcal{A}$ provides an estimate of the product $A_rk_r$ and $A_zk_z$, respectively (see Section 2).
\item The Gaussian width of the peak in the Fourier spectrum $W_\Lambda$, 
		which is a useful tool for determining the quality of the wavelength data.
\end{itemize}
Note that the width parameter $W_\Lambda$ is not necessarily an estimate of the uncertainty in the $\Lambda$ because the peak location can be determined to a small fraction of its width if the signal-to-noise is high, which is often the case for the profiles considered here. Indeed, for many of the profiles the resolution of the profile is the ultimate factor limiting our ability to determine $\Lambda$. The uncertainty in $\Lambda$ due to the finite resolution of the profiles can be estimated with the parameter $E_\Lambda=\sqrt{2}\Lambda\delta r/$(800 km), where $\delta r$ is the radial resolution of the relevant profile. Note we do not attempt to estimate uncertainties on the amplitudes of the patterns because these are likely to be dominated by systematic phenomena that are difficult to quantify {\sl a priori}.

\section{Observations and Results}
\label{results}

\begin{table}
\caption{Summary of wavelength and amplitude estimates}
\label{obssumtab}
\resizebox{6in}{!}{ \begin{tabular}{|c|c|c|c|c|c|c|c|c|c|}\hline
UTC time & Phase & $B$ & Radial & $A_r$ & $A_z$ & $\Lambda_r$ from $dp_1/dr$ (km) & $\Lambda_r$ from $dp_1/dr(\bar{p}_1)^{-1}$ (km) & $\Lambda_z$ from $p_2$ (km) & $\Lambda_z$ from $p_2/p_1$ (km) \\
 & Angle &  & Res. & (km) & (km) & Value$\pm$Error (Width) & Value$\pm$Error (Width) & Value$\pm$Error (Width) & Value$\pm$Error (Width) \\
\hline
1995-325T21:04:19 & -- & -2.6$^\circ$ & 4.7 km & -- & -- & -- & -- & 58.40$\pm$0.61 (1.60)$^a$ & -- \\ 
1995-325T21:04:19 & -- & -2.6$^\circ$ & 4.7 km & -- & -- & -- & -- & 59.92$\pm$0.64 (1.67)$^b$ & -- \\ 
2005-120T13:14:45 &  38.5$^\circ$ & -19.5$^\circ$ & 4.5 km &  0.28 &  0.60 & 32.11$\pm$0.25 (0.53) & 32.11$\pm$0.25 (0.55) & 33.09$\pm$0.26 (0.56) & 33.08$\pm$0.26 (0.57) \\ 
2005-140T17:15:52 &   1.1$^\circ$ & -20.7$^\circ$ & 1.6 km &  0.30 &  0.77 & 32.25$\pm$0.09 (0.54) & 32.23$\pm$0.09 (0.56) & 32.93$\pm$0.09 (0.53) & 32.93$\pm$0.09 (0.52) \\ 
2005-159T03:14:15 &  18.1$^\circ$ & -16.8$^\circ$ & 1.1 km &  0.86 &  2.31 & 32.04$\pm$0.06 (0.49) & 32.03$\pm$0.06 (0.54) & 32.98$\pm$0.06 (0.58) & 32.79$\pm$0.06 (0.65) \\ 
2005-248T03:24:27 &  12.5$^\circ$ & -15.8$^\circ$ & 1.2 km &  0.47 &  1.04 & 31.69$\pm$0.07 (0.49) & 31.69$\pm$0.07 (0.52) & 32.67$\pm$0.07 (0.58) & 32.60$\pm$0.13 (0.60) \\ 
2006-363T07:38:29 & 131.9$^\circ$ & -16.7$^\circ$ & 3.1 km &  0.73 &  1.33 & 29.80$\pm$0.16 (0.51) & 29.84$\pm$0.16 (0.41) & 30.72$\pm$0.19 (0.56) & 30.79$\pm$0.40 (0.46) \\ 
2007-036T21:09:23 &  67.4$^\circ$ &  30.5$^\circ$ & 4.3 km &  0.21 &  0.44 & 29.72$\pm$0.23 (0.46) & 29.73$\pm$0.23 (0.45) & 30.78$\pm$0.23 (0.47) & 31.06$\pm$0.24 (0.45) \\ 
2007-045T17:28:39 & 161.9$^\circ$ &  25.9$^\circ$ & 4.4 km &  0.23 &  0.84 & 29.68$\pm$0.23 (0.61) & 29.67$\pm$0.23 (0.39) & 30.20$\pm$0.24 (0.63) & 30.19$\pm$0.24 (0.48) \\ 
2007-064T15:41:40 & 160.2$^\circ$ &   6.8$^\circ$ & 2.9 km &  0.36 &  0.63 & 29.54$\pm$0.15 (0.58) & 29.57$\pm$0.15 (0.41) & 30.57$\pm$0.16 (0.60) & 30.50$\pm$0.16 (0.48) \\ 
2007-298T01:33:48 &  26.7$^\circ$ &  -2.4$^\circ$ & 2.0 km &  0.28 &  0.62 & 28.91$\pm$0.10 (0.42) & 28.90$\pm$0.10 (0.45) & 29.75$\pm$0.10 (0.45) & 29.72$\pm$0.10 (0.48) \\ 
2009-206T09:31:16 & 160.3$^\circ$ &  -7.0$^\circ$ & 2.8 km &  0.35 &  0.60 & 26.92$\pm$0.14 (0.43) & 26.93$\pm$0.14 (0.38) & 27.73$\pm$0.14 (0.53) & 27.75$\pm$0.14 (0.40) \\ 
2009-239T18:48:50 &  11.7$^\circ$ &   7.1$^\circ$ & 1.2 km &  0.41 &  0.74 & 26.97$\pm$0.06 (0.41) & 26.97$\pm$0.06 (0.42) & 27.74$\pm$0.06 (0.40) & 27.74$\pm$0.06 (0.40) \\ 
2009-263T20:03:51 &   7.9$^\circ$ &   8.5$^\circ$ & 1.4 km &  0.51 &  1.40 & 26.96$\pm$0.06 (0.40) & 26.96$\pm$0.06 (0.41) & 27.52$\pm$0.07 (0.43) & 27.50$\pm$0.07 (0.45) \\ 
2010-010T17:05:09 & 157.8$^\circ$ & -21.3$^\circ$ & 1.4 km &  0.42 &  1.16 & 26.61$\pm$0.06 (0.46) & 26.70$\pm$0.06 (0.39) & 27.27$\pm$0.07 (0.47) & 27.45$\pm$0.07 (0.38) \\ 
2010-185T07:23:24 & 143.7$^\circ$ & -18.7$^\circ$ & 2.3 km &  0.37 &  1.12 & 26.08$\pm$0.11 (0.44) & 26.11$\pm$0.11 (0.33) & 26.86$\pm$0.11 (0.41) & 26.85$\pm$0.11 (0.36) \\ 
2012-180T05:10:49 & 150.1$^\circ$ & -17.6$^\circ$ & 2.0 km &  0.47 &  1.06 & 24.27$\pm$0.10 (0.38) & 24.34$\pm$0.15 (0.30) & 25.00$\pm$0.09 (0.38) & 25.00$\pm$0.09 (0.34) \\ 
2012-182T11:10:09 &  24.2$^\circ$ &   3.3$^\circ$ & 3.0 km &  0.53 &  1.34 & 24.33$\pm$0.13 (0.30) & 24.34$\pm$0.13 (0.32) & 25.04$\pm$0.14 (0.32) & 25.06$\pm$0.14 (0.33) \\ 
2012-249T20:06:16 &  36.8$^\circ$ &  -3.5$^\circ$ & 4.2 km &  0.12 &  0.37 & 24.11$\pm$0.18 (0.30) & 24.10$\pm$0.18 (0.30) & 24.86$\pm$0.19 (0.31) & 24.83$\pm$0.19 (0.31) \\ 
2012-292T02:21:57 & 130.6$^\circ$ &  18.0$^\circ$ & 1.5 km &  0.63 &  1.75 & 23.96$\pm$0.06 (0.33) & 23.96$\pm$0.06 (0.26) & 24.54$\pm$0.07 (0.47) & 24.41$\pm$0.08 (0.32) \\ 
2012-315T23:28:37 & 136.4$^\circ$ &  14.5$^\circ$ & 1.5 km &  0.42 &  1.12 & 23.79$\pm$0.07 (0.34) & 23.89$\pm$0.07 (0.25) & 24.44$\pm$0.07 (0.39) & 24.50$\pm$0.07 (0.33) \\ 
2013-092T18:05:09 & 144.3$^\circ$ &   5.5$^\circ$ & 1.5 km &  0.40 &  0.61 & 23.69$\pm$0.06 (0.33) & 23.73$\pm$0.06 (0.30) & 24.38$\pm$0.07 (0.34) & 24.41$\pm$0.07 (0.30) \\ 
2013-187T20:18:39 & 136.7$^\circ$ &  14.7$^\circ$ & 1.9 km &  0.48 &  1.20 & 23.44$\pm$0.08 (0.51) & 23.48$\pm$0.08 (0.25) & 24.09$\pm$0.08 (0.38) & 24.17$\pm$0.09 (0.29) \\ 
\hline\end{tabular}}
$^a$ Using original geometry for HST occultation, from \citet{Hedman07}

$^b$ Using new geometry for HST occultation provided by R. G. French (private communication 2013).
\end{table}

The above data reduction procedures require images that both have high enough resolution to detect patterns with wavelengths around 30 km and sufficient longitude coverage to isolate the signals due to optical depth and vertical slope variations. We therefore performed a comprehensive search for clear-filter images of the appropriate region with resolutions better than 10 km/pixel and sufficient signal-to-noise to detect the patterns. We excluded most of these images either because the pattern was not well resolved or because only a narrow range of $\cos\phi\cot B$ were visible, so we could not reliably isolate the vertical structures from optical-depth patterns. Images were also excluded if  any of the profiles yielded a peak in the Fourier transform with a Gaussian width larger than 1 km, as this indicated that the periodic signal was not cleanly observed in the profile. While some of these excluded images could potentially provide viable amplitude or wavelength measurements with more elaborate analytical techniques, we felt that such complications were not worthwhile at this point. 

In the end, 199 images were deemed suitable for this particular analysis.  Tables~\ref{obstab}-\ref{obstab3} in Appendix A list these images, along with the relevant observation geometry parameters and the amplitude and wavelength information extracted using the above algorithms. The vast majority of these images  are part of a few observation sequences where the relevant part of the D-ring was imaged repeatedly over a relatively short period of time.  These sequences provide replicate measurements of the patterns at a particular time and thus allow us to evaluate the robustness of our data reduction procedures and verify our estimates of the wavelength uncertainties. For each sequence of images, the tables provide the mean of the wavelength and amplitude estimates, the scatter in those measurements, and the predicted scatter based on the resolution of the relevant sequence (corresponding to the $E_\Lambda$ parameter discussed above). For the observing sequences on UTC Days 2007-045, 2007-064, 2010-185, 2012-180, 2012-292, 2012-315 and 2013-187,  the observed scatter is comparable to or lower than the predicted scatter, indicating that our estimates of the uncertainty based on the image resolution are reasonable. For the short sequences on UTC Days 2005-248 and 2005-363, each of which consists of two images of different parts of the D ring, we find the difference between the two wavelength estimates derived from the $p_2/p_1$ profile is about twice the expected scatter. However, even in these cases the other three profiles yield wavelength estimates that are consistent to within their error bars. This suggests issues with the background levels may be contaminating the $p_2/p_1$ profile for these observations. Even so, overall it appears that the estimates of the wavelength errors are sensible. 

These long observation sequences have the potential to skew analyses and fits to these data because the formal statistical errors derived from many repeated measurements are very low, but do not account for  any common systematic errors such as those due to navigation uncertainties or  background levels. Hence for the remainder of this analysis we will not consider the 199 measurements individually, but instead reduce the data from each of the observation sequences in Tables~\ref{obstab}-~\ref{obstab3} to a single estimate of the relevant pattern wavelengths and amplitudes. The estimate for each parameter is the (unweighted) average value of the estimates from all the relevant images, and the error is the larger of the scatter in the measurements and the average $E_\Lambda$ parameter for the observations. In the end, for each parameter we obtain nine data points from various observing sequences and another twelve data points derived from individual images (listed at the start of Table~\ref{obstab}). Table~\ref{obssumtab} gives the relevant pattern parameters and observation times for these 21 observation epochs. This includes wavelength estimates derived from all four profiles ($p_1, p_2, p'_1/\bar{p}_1$ and $p_2/p_1$), but only amplitude estimates from the $p'_1/\bar{p}_1$ and $p_2/p_1$ profiles, since the latter are the only ones that can be directly related to the vertical and radial displacements $A_r$ and $A_z$.

Table~\ref{obssumtab} also includes a measurement of the corrugation wavelength derived from a 1995 Hubble Space Telescope occultation described previously in \citet{Hedman07}. Normally, an occultation only provides a radial cut through the ring, and thus cannot provide separate estimates of the corrugation and optical depth structure. However, it turns out that this occultation occurred far from the ring ansa, where $\cos\phi\cot B \simeq 5.5$. Thus the periodic signature from the corrugation is far larger the optical depth variation, and so we can assume this profile provides a fairly clean estimate of $\Lambda_z$. However, because of the low ring opening angle, the wavelength measurement depends sensitively  on the assumed occultation geometry. Recently, R. French has performed a comprehensive reconstruction of all the available occultation data \citep{French10}, which had the result of increasing the measured wavelength of the pattern from 58.4$\pm$0.6 km to 59.9$\pm$0.7 km. We will consider both values here and assume the difference between the two wavelength estimates provides a conservative measure of the potential systematic errors in this early observation.

\begin{figure}
\centerline{\resizebox{5in}{!}{\includegraphics{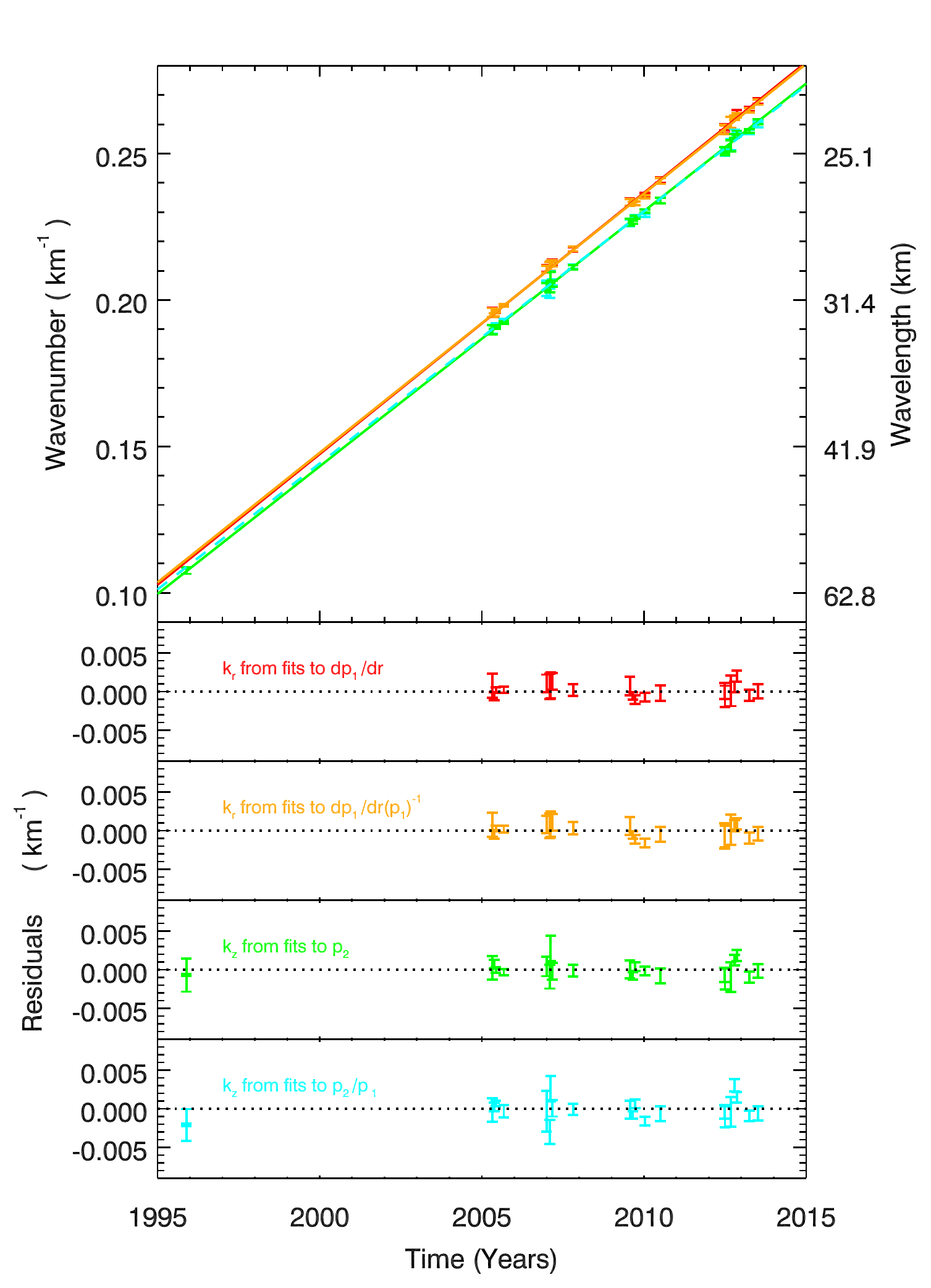}}}
\caption{Plot showing the measured pattern wavenumbers versus time and the residuals from the best-fitting linear trends. The top panel shows the wavenumber of the patterns versus time. Two different estimates of the opacity variations' wavenumber $K_r$ are shown in red and orange, while two estimates of the corrugation wavenumbers $K_z$ are show in green and cyan. (The different estimates of the same parameters come from analyses of different profiles, as indicated in the lower panels.) All four data sets follow clear linear trends (indicated by lines of corresponding colors). The bottom four panels show the residuals from the best-fitting trends. Note that in the bottom panels there are two values for the 1995 data point. These correspond to the two different estimates of the pattern wavelength, with the value from the older geometrical solution being the upper value.}
\label{waveshift}
\end{figure}

\begin{table}
\caption{Temporal trends in the wavelength patterns}
\label{trendtab}
\resizebox{6.5in}{!}{\begin{tabular}{|l|cccc|cccc|c|}\hline Analysis & $\lambda_r$ at & $dk_r/dt$ & Disturbance & $\chi^2/DOF$ & $\lambda_z$ at & $dk_z/dt$ & Disturbance & $\chi^2/DOF$ & $\underline{dk_z/dt}$  \\
 & Equinox (km) & ($10^{-5}$ km$^{-1}$/day) & Epoch (JD) &  & Equinox (km) & ($10^{-5}$ km$^{-1}$/day) & Epoch (JD) & & $dk_r/dt$ \\ \hline
All $dp_1/dr$ and $p_2$ profiles & 26.95$\pm$ 0.02 & 2.444$\pm$0.015 & 2445513.4$\pm$55.0 & 20.72/19 & 27.69$\pm$ 0.02 & 2.389$\pm$0.014 & 2445555.2$\pm$54.3 & 23.74/19 & 0.9777$\pm$0.0082 \\
\hspace{.25in} (including HST occultation, old geometry) & &&& & 27.69$\pm$ 0.02 & 2.387$\pm$0.012 & 2445547.0$\pm$46.8 & 23.83/20 & 0.9768$\pm$0.0077 \\
\hspace{.25in} (including HST occultation, new geometry) & &&& & 27.69$\pm$ 0.02 & 2.402$\pm$0.012 & 2445604.4$\pm$46.2 & 26.92/20 & 0.9829$\pm$0.0077 \\ \hline
All $dp_1/dr(\bar{p}_1)^{-1}$ and $p_2/p_1$ profiles & 26.98$\pm$ 0.02 & 2.424$\pm$0.015 & 2445447.1$\pm$56.4 & 19.16/19 & 27.69$\pm$ 0.02 & 2.348$\pm$0.015 & 2445389.1$\pm$61.8 & 43.43/19 & 0.9685$\pm$0.0087 \\
\hspace{.25in} (including HST occultation, old geometry) & &&& & 27.69$\pm$ 0.02 & 2.358$\pm$0.013 & 2445431.2$\pm$51.1 & 44.97/20 & 0.9728$\pm$0.0080 \\ 
\hspace{.25in} (including HST occultation, new geometry) & &&& & 27.69$\pm$ 0.02 & 2.375$\pm$0.013 & 2445498.5$\pm$50.5 & 54.03/20 & 0.9797$\pm$0.0080 \\ \hline
\end{tabular}}\end{table}

\begin{table}
\caption{Wavelength and amplitude ratios of the periodic patterns}
\label{rattab}
{\begin{tabular}{|l|cc|c|}\hline Analysis & $k_z/k_r$ & $\chi^2/DOF$ & $A_z/A_r$ \\ \hline
All $dp_1/dr$ and $p_2$ profiles & 0.9733$\pm$0.0009 & 12.63/20 & 2.21$\pm$0.46 \\
All $dp_1/dr(\bar{p}_1)^{-1}$ and $p_2/p_1$ profiles & 0.9747$\pm$0.0010 & 15.34/20 & 2.41$\pm$0.54 \\ 
\hline\end{tabular}}\end{table}

Figure~\ref{waveshift} plots the estimates of the pattern wavenumbers $K_z=2\pi/\Lambda_z$ and $K_r=2\pi/\Lambda_r$ versus time (note 
that $K_z$ and $K_r$ are estimates of the real wavenumbers $k_z$ and $k_r$ at 73,600 km). Both $K_z$ and $K_r$ are clearly increasing with time at nearly constant rates, as predicted in Section~\ref{theory}. Furthermore, the $rms$ residuals of these data from the appropriate linear trends are less than $0.003$ km$^{-1}$, which is much smaller than the scatter found in our previous analysis of these patterns (see Figure 23 of Hedman {\it et al.} 2007). This indicates that the excess scatter in our earlier investigation of the corrugations was indeed due to interference between the patterns, and that the new calculations are yielding consistent wavelength estimates.

A closer look at the fit parameters, listed in Table~\ref{trendtab}, reveals some interesting differences in the data derived from the various profiles. This table provides estimates of  the winding rates of the patterns $dK_r/dt$ and $dK_z/dt$, the ``disturbance epoch'' $t_i$ when $K_z$ or $K_r$ would be zero, and the estimated wavelength value at equinox (JD 2455054), which is not only near the mid-point of the Cassini observations, but also the epoch of the C-ring measurements \citep{Hedman11}. It also includes the $\chi^2$ for each linear fit.  For the corrugation patterns, these include fits for the Cassini data alone, as well as fits including both estimates of the pattern wavelength in the 1995 HST occultation data. The two estimates of the eccentric spiral patterns (derived from profiles $p'_1$ and $p'_1/\bar{p}_1$, respectively) yield consistent estimates for the winding rate, disturbance epoch and wavelength at equinox, and have comparable $\chi^2$ to a linear fit. By contrast, the two estimates of the vertical corrugation's wavelength (from $p_2$ and $p_2/p_1$) yield estimates of the winding rate and disturbance epoch that differ by more than their individual error bars would predict. The linear fit to the wavelength estimates derived from the $p_2/p_1$ profiles shows a much worse $\chi^2$  than the same fit to the wavelength estimates derived from $p_2$, and the fit parameters change more depending on whether or not the 1995 occultation data are included. These findings suggests that the wavelength estimates derived from  $p_2/p_1$ are not as good as the ones derived from $p_2$ alone. This is a bit surprising, since the $p_2$ profile is proportional to the product of $\tau_n$ and $dz/dr$, so one might have expected there to be systematic biases in the $p_2$ wavelength estimates due to contamination from the eccentric spiral pattern. Instead, it appears that uncertainties involved in the background subtraction are producing larger shifts in the wavelength estimates. However, it is worth noting that both methods yield very consistent estimates of the corrugation wavelength at equinox.

Compared with the wavelengths, the amplitudes of the patterns show a much larger dispersion, which is not surprising, since the apparent amplitude of the pattern is much more sensitive to small errors in the background subtraction procedures. At present, we have not identified any sensible trends in the amplitudes with time or any physically interesting quantity. Hence we regard the numbers in Table~\ref{obssumtab} as only very rough indicators of the ring-particles' epicyclic motions, and will not attempt to conduct a precise analysis of these values.

Many systematic errors that could influence our estimates of the patterns' amplitudes and wavelengths, like incomplete background subtraction, should affect both patterns roughly equally. We therefore might expect the ratios $k_z/k_r$ and $A_z/A_r$ to be more robustly determined than the individual parameters. Table~\ref{rattab}  provides the average ratios of the wavenumbers and amplitudes of the two patterns. For the wavenumber ratio $k_z/k_r=K_z/K_r$ we computed weighted average of the 21 Cassini measurements of $k_z/k_r$ that can be derived from Table~\ref{obssumtab}, as well as the $\chi^2$ parameter for a model where $k_z/k_r$  is constant. The $\chi^2$ values are well below the degrees of freedom, which indicates that the wavenumber ratio is consistent among the various observations. For the amplitude ratio $A_z/A_r$, we do not have robust uncertainty estimated on the individual estimates, so we instead compute the simple average and the $rms$ scatter in the 21 Cassini estimates. Note that while $A_z$ and $A_r$ can only be separately estimated from the $p'_1/\bar{p}_1$ and $p_2/p_1$ profiles, the ratio $A_z/A_r$ can be derived from the ratio of amplitudes of the patterns in the $p_2$ and $p_1$ profiles. Both methods yield essentially the same result, with a mean amplitude ratio around 2.3, with a scatter of around 0.5, which is far less than the scatter in the individual estimates of $A_z$ or $A_r$ in Table~\ref{obssumtab}. This ratio therefore does indeed appear to be better determined than the individual pattern amplitudes.

\section{Verifying the nature of the periodic patterns}
\label{spiral}

The winding rates and wavelength ratios in Tables~\ref{trendtab} and~\ref{rattab} provide strong evidence that these patterns are indeed corrugations and eccentric spiral patterns winding up under the influence of the planet's gravity field,  as laid out in Section~\ref{theory} above. At a very basic level, the observed data clearly show that the wavenumbers of both patterns are increasing linearly with time (see Figure~\ref{waveshift}), as predicted. Furthermore, quantitative comparisons of the observed winding rates and wavelength ratios to theoretical predictions provide very stringent tests on our model for these patterns.

First of all, we may note that the two patterns follow similar, but not identical trends. This basic observation is consistent with the expected evolution of corrugations and eccentric spirals. The first-order calculation outlined in Section~\ref{theory} indicates that the winding rate for an eccentric spiral should be approximately equal to the winding rate of the vertical corrugation, as observed. If we refine these calculations to include higher-order corrections to the relevant precession rates, we can even explain the observed differences between the two rates. Typically, the axisymmetric part of a planet's gravitational potential is expressed in terms of the following series:
\begin{equation}
V=-\frac{GM}{r}\left[1-\sum_{i=2}^\infty J_i\left(\frac{R_s}{r}\right)^iP_i(\sin\Theta)\right],
\end{equation}
where $\Theta$ is latitude, $R=$ 60,300 km is the assumed planetary radius, $P_i$ are Legendre polynomials of degree $i$ and $J_i$ are a series of numerical coefficients \citep{MurrayDermott}. Note that for a fluid planet only even $i$ should have non-negligible coefficients, so typically the planet's gravity can be described by the coefficients  $J_2, J_4, J_6...$. If we keep all terms out to fourth order in $R_s/r$, then the apsidal precession and nodal regression rates are (Murray and Dermott 1999, Equations 6.249, 6.250):
\begin{equation}
\dot{\varpi}=n\left[\frac{3}{2}J_2\left(\frac{R_s}{r}\right)^2
-\frac{15}{4}J_4\left(\frac{R_s}{r}\right)^4\right],
\end{equation}
\begin{equation}
|\dot{\Omega}|=n\left[\frac{3}{2}J_2\left(\frac{R_s}{r}\right)^2
-\frac{9}{4}J_2^2\left(\frac{R_s}{r}\right)^4
-\frac{15}{4}J_4\left(\frac{R_s}{r}\right)^4\right],
\end{equation}
where $n=\sqrt{GM/r^3}$ is the zeroth-order estimate of the ring particle's mean motion. Using these more complex expressions in Equations~\ref{eqkz} and~\ref{eqkr}, we obtain the winding rates:
\begin{equation}
\frac{dk_r}{dt}=\left|\frac{\partial\dot{\varpi}}{\partial r}\right|=\frac{n}{r}\left[\frac{21}{4}J_2\left(\frac{R_s}{r}\right)^2
-\frac{165}{8}J_4\left(\frac{R_s}{r}\right)^4\right],
\end{equation}
\begin{equation}
\frac{dk_z}{dt}=\left|\frac{\partial\dot{\Omega}}{\partial r}\right|=\frac{n}{r}\left[\frac{21}{4}J_2\left(\frac{R_s}{r}\right)^2
-\frac{99}{8}J_2^2\left(\frac{R_s}{r}\right)^4
-\frac{165}{8}J_4\left(\frac{R_s}{r}\right)^4\right].
\end{equation}
Thus, at this level of approximation, the normalized difference in the two winding rates should be:
\begin{equation}
\frac{\dot{k}_r-\dot{k}_z}{\dot{k}_r+\dot{k}_z}=\frac{33}{28}J_2\left(\frac{R_s}{r}\right)^2.
\end{equation}
According to \citet{Jacobson06}, $J_2=0.01629071$  so we can estimate that at  $r=$ 73,600 km, this normalized difference  will be 0.0129. Including even higher-order corrections (using $J_4=-0.00092583,$ $J_6=0.00008614$, and $J_8=-10^{-6}$, consistent with Jacobson {\it et al.} 2006),  yields a more accurate estimate of this difference of 0.0142, such that the ratio of the  two winding rates is $\dot{k}_z/\dot{k}_r=0.9720$. All the rate ratios listed in Table~\ref{trendtab} are consistent with this value, strongly supporting our hypothesis that the optical depth variations are due to an eccentric spiral. 

We can also compare the individual winding rate estimates with theoretical predictions, but this requires more careful treatment of the higher-order components of Saturn's gravitational field. For D-ring features the ratio $R_s/r$ is not much different from one, which means observable parameters  like winding rates are sensitive to a linear combination of many $J_i$.  This issue is discussed in detail in \citet{Hedman14}, which examined the precession of an eccentric ringlet in the inner D ring known as D68. Just as in that work, we will first examine the observed winding rates using the standard language of gravitational harmonics, but then use a simplified model to clarify whether the observed values are consistent with other measurements of Saturn's gravity field.

\begin{figure}
\centerline{\resizebox{5in}{!}{\includegraphics{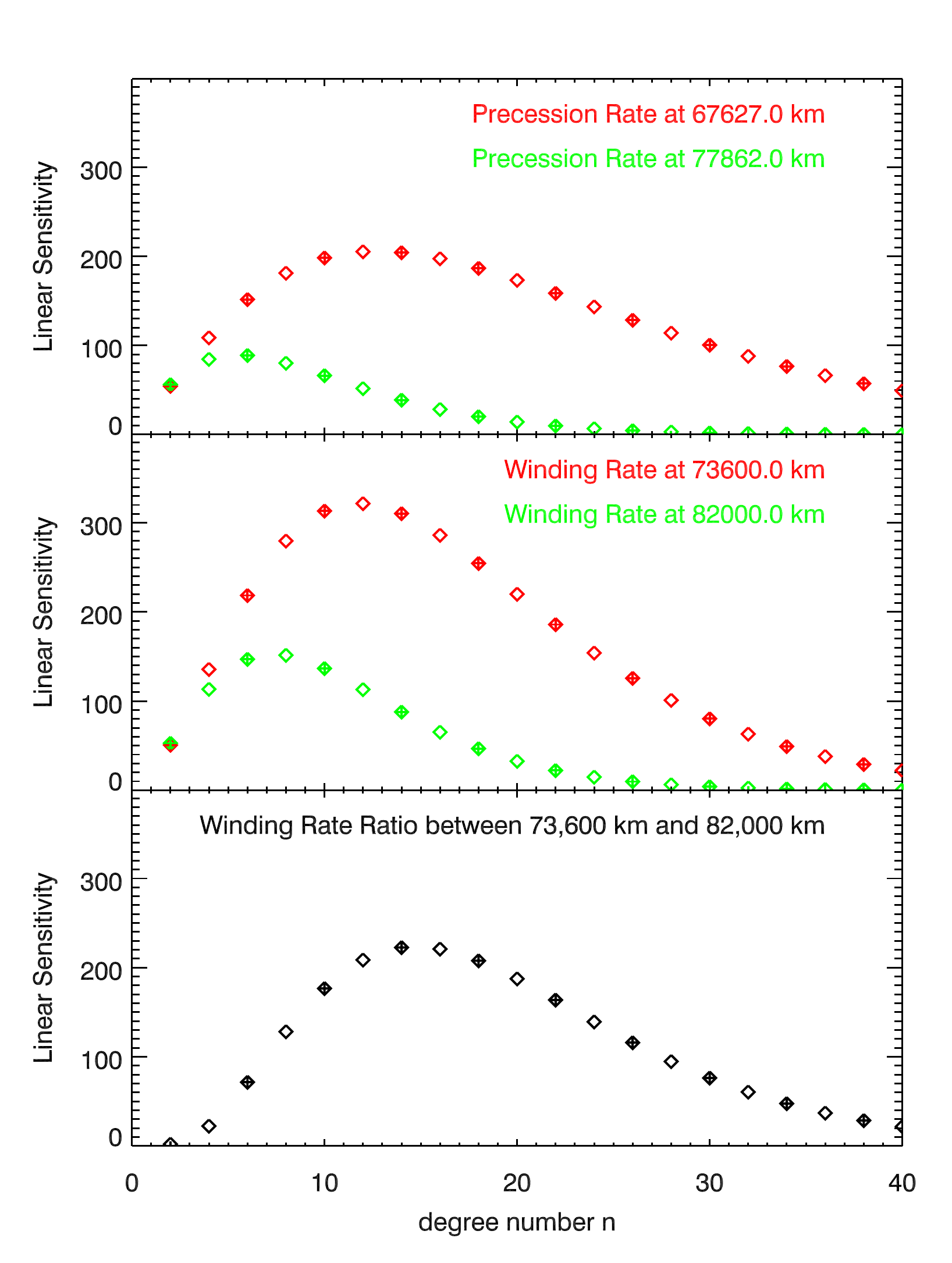}}}
\caption{Plots showing the fractional linear sensitivity of various orbital evolution parameters in the C and D rings to Saturn's gravitational harmonics as a function of the degree number. In all plots filled symbols correspond to positive $\mathcal{S}_i$ and empty symbols are negative $\mathcal{S}_i$. The top panel shows the sensitivity coefficients for a particle's orbital precession rate at two different distances from Saturn's center, corresponding to the D68 and Titan ringlets. The middle panel shows the sensitivity of the corrugation winding rate at two different locations (the sensitivity of the eccentric spiral winding rate follows nearly the same trends). The bottom panel shows the sensitivity of the winding rate ratio between the two locations illustrated in the middle panel.}
\label{sensplot}
\end{figure}

In general, given a suitable observable parameter $\mathcal{P}$, one can compute the linear (fractional) sensitivity of this parameter to a small change in any of the gravity harmonics $J_i$:
\begin{equation}
\mathcal{S}_i=\frac{1}{\mathcal{P}}\frac{\partial\mathcal{P}}{\partial J_i}.
\end{equation}
Figure~\ref{sensplot} plots the sensitivity coefficients $\mathcal{S}_i$ for the winding rates of a corrugation situated at 73,600 km in  the D ring and 82,000 km in the C ring (The winding rates of the corresponding eccentric spiral patterns have nearly the same sensitivity coefficients, and so are not included here). For the sake of comparison, we also plot  the sensitivity curves for the precession rates of two eccentric ringlets previously calculated by \citet{Hedman14}: the Titan ringlet in the inner C ring (77,862 km) and D68 in the inner D ring (67,627 km). Note that both the D-ring corrugation winding rate and D68's precession rate are sensitive to a broad range of $J_i$, with $\mathcal{S}_i$ reaching a maximum around $J_{12}$. This coincidence is a bit surprising since D68 is closer to the planet than the corrugation, so $R_s/r$ is closer to unity for D68 than it is for the corrugation. However, the  winding rate depends upon the gradient of the precession rate, which makes it more sensitive to higher-order harmonics and  partially compensates for the difference in the two features' locations.

These sensitivity coefficients allow observed ring parameters to be transformed into constraints on the gravitational harmonics. In particular, any one of the observed corrugation winding rates at 73,600 km in Table~\ref{trendtab} (which we will here designate as $\dot{k}_{z,obs}$) yields the following constraint on $J_i$:
\begin{equation}
\frac{\dot{k}_{z,obs}-\dot{k}_{z,mod}}{\dot{k}_{z,mod}}=\sum(J_i-J_{i,mod})\mathcal{S}^z_i,
\label{sol1}
\end{equation}
where $\dot{k}_{z,mod}=2.3840\times10^{-5}$ km$^{-1}$/day, while $J_{i, mod}$ and $\mathcal{S}^z_i$ are listed in Table~\ref{params}.  Similarly, any of the eccentric spiral pattern winding rates at 73,600 km  $\dot{k}_{r,obs}$ yields the constraint
\begin{equation}
\frac{\dot{k}_{r,obs}-\dot{k}_{r,mod}}{\dot{k}_{r,mod}}=\sum(J_i-J_{i,mod})\mathcal{S}^r_i,
\end{equation}
\label{sol2}
where $\dot{k}_{r,mod}=2.4528\times10^{-5}$ km$^{-1}$/day and the coefficients $\mathcal{S}^r_i$ are listed in Table~\ref{params}.

\begin{table}
\caption{The parameters used in Equation~\ref{sol1}-~\ref{sol2}}
\label{params}
\centerline{\begin{tabular}{|c|c|c|c||c|c|c|c|} \hline
i & $\mathcal{S}^z_i$ &  $\mathcal{S}^r_i$ & $J_{i,mod}$ & i & $\mathcal{S}^z_i$ & $\mathcal{S}^r_i$ & $J_{i,mod}$ \\ \hline
2 & +50 & +52 & 16290.71$\times10^{-6}$ & 22 & +185 & +187 & 0 \\
4 & -135 & -138 &  -935.85$\times10^{-6}$  & 24 & -154 & -155 & 0 \\
6 & +218 & +222 &  86.14$\times10^{-6}$ &26 & +126 & +127 & 0 \\
8 & -280 & -284 & -10$\times10^{-6}$ & 28 & -101 & -102 & 0 \\
10 & +313 & +317 & 0 & 30 & +80 & +81 & 0 \\
12 & -322 & -326 & 0& 32 &  -63 & -64 & 0 \\
14 & +310 & +314 & 0 & 34 &  +49 & +50 & 0 \\
16 & -286 & -289 & 0 & 36 & -38 & -38 & 0 \\
18 & +254& +257 & 0 & 38 &+29 & + 29  & 0 \\
20 & -220 & -222 & 0 & 40 & -22 & -22  & 0 \\
 \hline
\end{tabular}}
\end{table}

While these constraints can be incorporated into various fitting routines, it is not immediately obvious whether these constraints are consistent with each other  or with other observations. To address these shortcomings, we may use the highly simplified model of Saturn's gravity field  developed in \citet{Hedman14}. For this model $J_2$ and $J_4$ are assumed to be known \citep{Jacobson06}, and the higher-order harmonics are assumed to be produced by two phenomena: the planet's rotation-induced oblateness and its equatorial jet. The former is modeled using the harmonics of a MacLaurin spheroid \citep{Hubbard12, Hubbard13}, while the latter is approximated as a massive wire wrapped around the planet's equator. This model of the planet has effectively two free parameters, which are here taken to be the planet's $J_6$ and the mass of the wire. 

\begin{figure}
\centerline{\resizebox{5in}{!}{\includegraphics{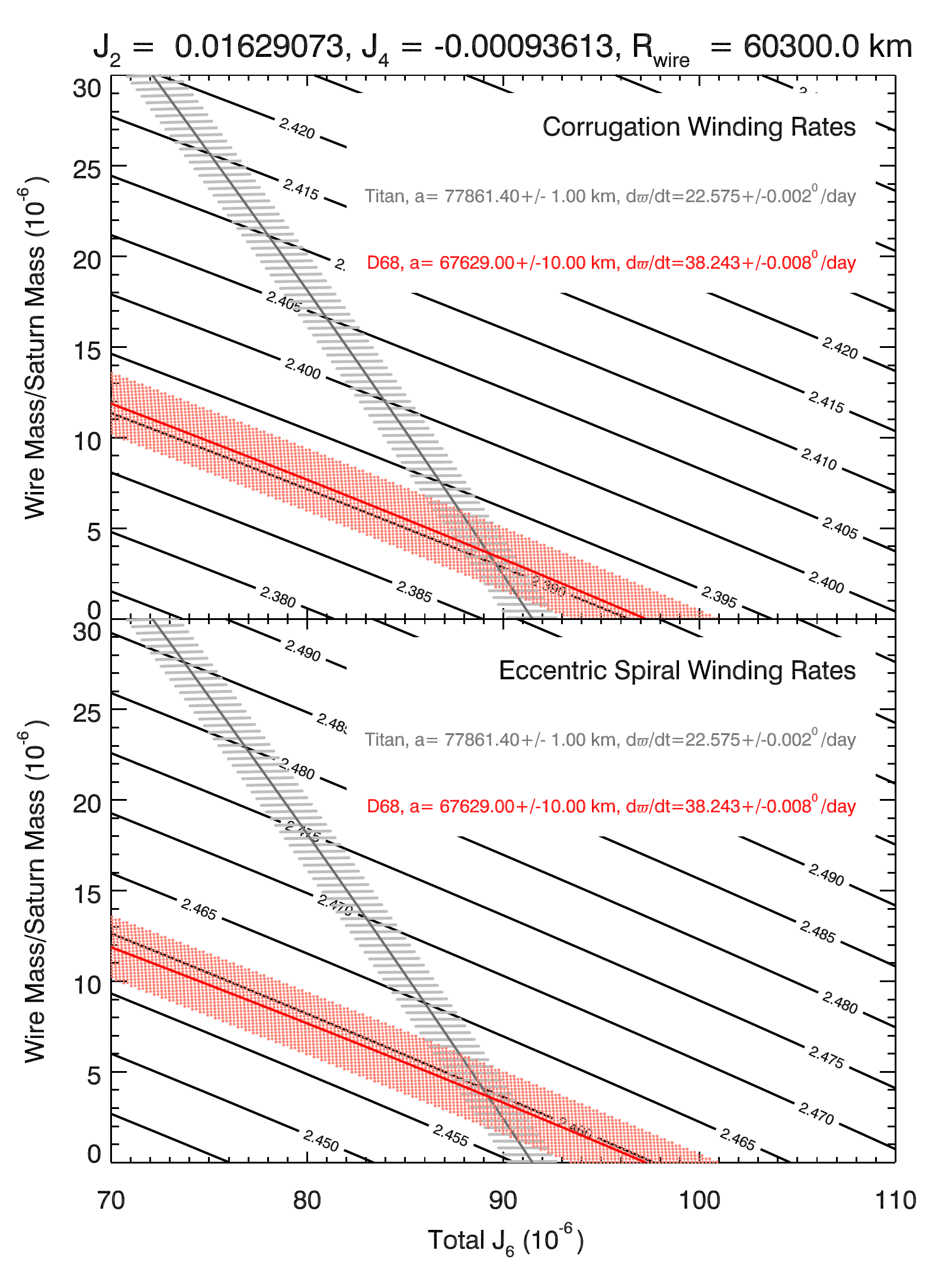}}}
\caption{Expected spiral pattern winding rates in a simplified model of Saturn's gravity field. The two panels show the winding rates  (in units of $10^{-5}$ km$^{-1}$/day) of the corrugation (top)  and the eccentric spiral (bottom) as functions of the assumed $J_6$ and wire mass. Overplotted on these contours are the constraints on these parameters served from the forced eccentricity of the  Titan ringlet and the precession rate of D68 reported in \citet{Nicholson14} and \citet{Hedman14}, respectively. Note the contours of constant winding rates are nearly parallel to the constraint from D68.}
\label{j6wirewarp}
\end{figure}

Figure~\ref{j6wirewarp} shows the expected winding rates for the corrugation and the eccentric spiral as a function of the parameters in this simplified gravity field model, along with the constraints derived from the forced eccentricity of the Titan ringlet and the precession rate of D68 \citep{Nicholson14, Hedman14}. Note in particular that contours of constant winding rate are nearly parallel to the line corresponding to the observed precession rate of D68. This is consistent with both these  phenomena having similarly shaped sensitivity curves in Figure~\ref{sensplot}.

Given the observed precession rate of D68, we would expect a corrugation winding rate of $(2.390\pm0.003)\times10^{-5}$ km$^{-1}$/day and an eccentric spiral winding rate of   $(2.450\pm0.003)\times10^{-5}$ km$^{-1}$/day. These numbers are perfectly consistent with the observed winding rates for the analysis that used the profiles $p'_1$ and $p_2$ (see Table~\ref{trendtab}). As mentioned above, these profiles will probably provide the most robust wavelength estimates, since they are less sensitive to background levels. The winding rates derived from the $p'_1/\bar{p}_1$ and $p_2/p_1$ profiles, by contrast, typically fall below this prediction, which suggests that these ratio profiles provide a less accurate estimate of the corrugation wavelengths, consistent with our previous suppositions based on the $\chi^2$ of the relevant fits (see Section~\ref{results}). Thus we may conclude that both of the winding rates derived here are consistent with any reasonable theoretical predictions that can reproduce the precession rate of D68. This result gives us even more confidence that these periodic brightness variations do indeed represent corrugations and eccentric spirals evolving under the influence of Saturn's gravitational field.

\section{Investigating the ring-disturbing event}
\label{event}

The above analysis of the wavelength trends provides strong evidence that the periodic patterns observed in the D ring consist of a vertical corrugation and an eccentric spiral winding up due to differential nodal regression and apsidal precession, respectively. Now we may turn our attention to the origins of these patterns. Both the eccentric spirals and the vertical corrugations described in Section~\ref{theory} arise from some event that produces organized epicyclic motions of ring particles. More specifically, \citet{Hedman07, Hedman11} suggested that a swarm of meteoritic debris struck the rings, perturbing the orbits of the ring particles. The wavelength and amplitude estimates derived above allow us to explore this scenario in more detail than previously possible. In particular, we have three pieces of information that can provide useful information about how the rings were disturbed:
\begin{itemize}
\item The relative wavelengths of the corrugation and eccentric spiral patterns, which indicate that the two patterns formed at similar times, and thus likely had a common origin.
\item The relative amplitudes of the two patterns, which depend upon the approach angle of the impacting debris.
\item The relative wavelengths of the corrugation in the D and C rings, which contain additional information
about the pre-impact trajectory of the debris. 
\end{itemize}
We will consider each of these topics in turn below, followed by a brief discussion of how the above constraints could potentially help us ascertain the origin of the ring-disturbing material.

\subsection{The eccentric spiral and corrugation have a common origin}

Table~\ref{trendtab} provides estimates of the ``disturbance epoch" for the corrugation and eccentric spiral. This epoch corresponds to the time when the wavelength of the pattern is infinite, and so all the particles in the rings have aligned pericenters or nodes. The derived disturbance epochs for the two patterns have overlapping error bars, so the data are consistent with both patterns arising from the same event. 

We can obtain this same result by considering the wavenumber ratios $k_z/k_r$ derived from individual observations. 
If both patterns arose from the same event, then not only would the winding rates of the two patterns  be a predictable ratio (see Section~\ref{spiral}), but the wavenumbers of the two patterns should also have the same ratio in any individual image.  Table~\ref{rattab} shows that, regardless of the profiles considered, the observations yield an average $k_z/k_r$ of between 0.973 and 0.975, very close to the predicted value of 0.972. The scatter among the various estimates of $k_z/k_r$ is also consistent with their estimates' error-bars (note the $\chi^2/DOF$ values in Table~\ref{rattab}). Again, this suggests that the eccentric motions of the ring particles were generated by the same event that tilted the ring.

Simultaneous excitation of  ring particles' eccentricities and inclinations is easy enough to explain if  these excitations were caused by a collision with interplanetary debris. So long as the debris struck the rings at an angle, the impacts would produce both vertical and radial epicyclic motions for the ring particles. By contrast, it is not obvious that other potential ring-tilting events, such as a shift in Saturn's gravity field, would necessarily produce simultaneous radial and vertical perturbations.

\subsection{Amplitude ratios and impacting debris angles}

Table~\ref{obssumtab} lists the values of the pattern amplitudes $A_z$ and $A_r$ derived from the various observations. The scatter of these measurements is quite large, which is not unreasonable given that these parameters are very sensitive to the assumed background level. Still, we find that $A_r$ is generally between 200 and 500 meters, while $A_z$ is between 600 and 1200 meters. Any background-dependent factors should cancel out in the ratio $A_z/A_r$, and indeed we find that this ratio has a much better-defined value of 2.3$\pm$0.5 (see Table~\ref{rattab}), regardless of which profiles we consider. This implies that the typical inclinations of the ring particles are 2-3 times larger than their present eccentricities. The epicyclic motions of the particles in these patterns are about an order of magnitude less than the patterns' wavelengths, so collisions among the ring particles should not be able to efficiently dissipate these organized eccentricities and inclinations. Indeed, fitting a linear trend to the  $A_z/A_r$ measurements requires any steady temporal variation in this ratio to be less than 0.1/year. Thus we may reasonably assume that the initial disturbance induced an average inclination in the ring particles that was 2-3 times larger than the induced eccentricities.

If we assume that the initial eccentricities and inclinations were produced by collisions with meteoritic debris, we can translate the $A_z/A_r$ ratio into a constraint on the trajectory of the incoming debris.  Consider a ring particle initially on a circular orbit with a semi-major axis $a$, and corresponding mean motion $n$. This particle is struck by a meteoroid that  imparts some of its momentum to the ring particles,  and so there is a small change in the ring particle's velocity $\delta {\bf v}$. Let the radial, azimuthal and vertical components of $\delta {\bf v}$ be $\delta v_r$, $\delta v_\lambda$ and $\delta v_z$, respectively.  If $\delta v_z$ is nonzero, the ring particle has a finite orbital inclination. Specifically, the inclination induced by the collision is:\begin{equation}
I=\frac{|\delta v_z|}{an}.
\end{equation}
If $\delta v_r$ is nonzero, then the ring particle moves radially, and thus has a finite eccentricity. Also if $\delta v_\lambda$ is nonzero, then the ring particle is not moving at the proper speed around the planet to maintain a circular orbit, which also requires the particle be on an eccentric orbit. In general, the resulting eccentricity is:
\begin{equation}
e=\frac{1}{an}\sqrt{\delta v_r^2+4\delta v_\lambda^2}.
\end{equation}
Thus the ratio of the induced inclination to the induced eccentricity is:
\begin{equation}
\frac{I}{e}=\frac{|\delta v_z|}{\sqrt{\delta v_r^2+4 \delta v_\lambda^2}}.
\label{ieratio}
\end{equation}

Of course, the orientation of the $\delta {\bf v}$ vector for any particular collision will  depend upon parameters like the impact parameter between the ring particle and the meteoroid. However, if we consider many collisions, the ensemble average $\delta {\bf v}$ should be directly proportional to  the average debris velocity ${\bf v}_d$ (the constant of proportionality depends on the mass distribution of the debris). Hence, so long as collisions among the ring particles efficiently dissipate the velocity dispersion among the ring particles, but do not alter the ring-particles' mean orbital elements at each $a$, the amplitude ratio $A_z/A_r$ should approximately equal the ratio of the average $I/e$ of the ring particles' orbits, which in turn can be expressed as the following function of the debris' impact velocity:
  \begin{equation}
\frac{A_z}{A_r}=\frac{\langle I\rangle}{\langle e \rangle}=\frac{|v_{d,z}|}{\sqrt{v_{d,r}^2+4 v_{d,\lambda}^2}},
\label{ampratio}
\end{equation}
where $v_{d,r}, v_{d,\lambda}$ and $v_{d,z}$ are the radial, azimuthal and vertical components of the incident debris velocity where they intersect the ring plane (note that $v_{d,\lambda}$ is the azimuthal velocity of the debris relative the average ring particle). Our estimate of the amplitude ratio $A_z/A_r=2.3 \pm 0.5$ therefore implies that the debris struck the rings at a fairly steep angle. Indeed, depending on how the in-plane motion was partitioned between $v_{d,r}$ and $v_{d,\lambda}$, the debris would need to hit the rings at an elevation angle between 60$^\circ$ and 80$^\circ$ to produce the observed $A_z/A_r$.

\subsection{Wavelength trends and the impacting debris trajectory}


\begin{figure}
\centerline{\resizebox{4.5in}{!}{\includegraphics{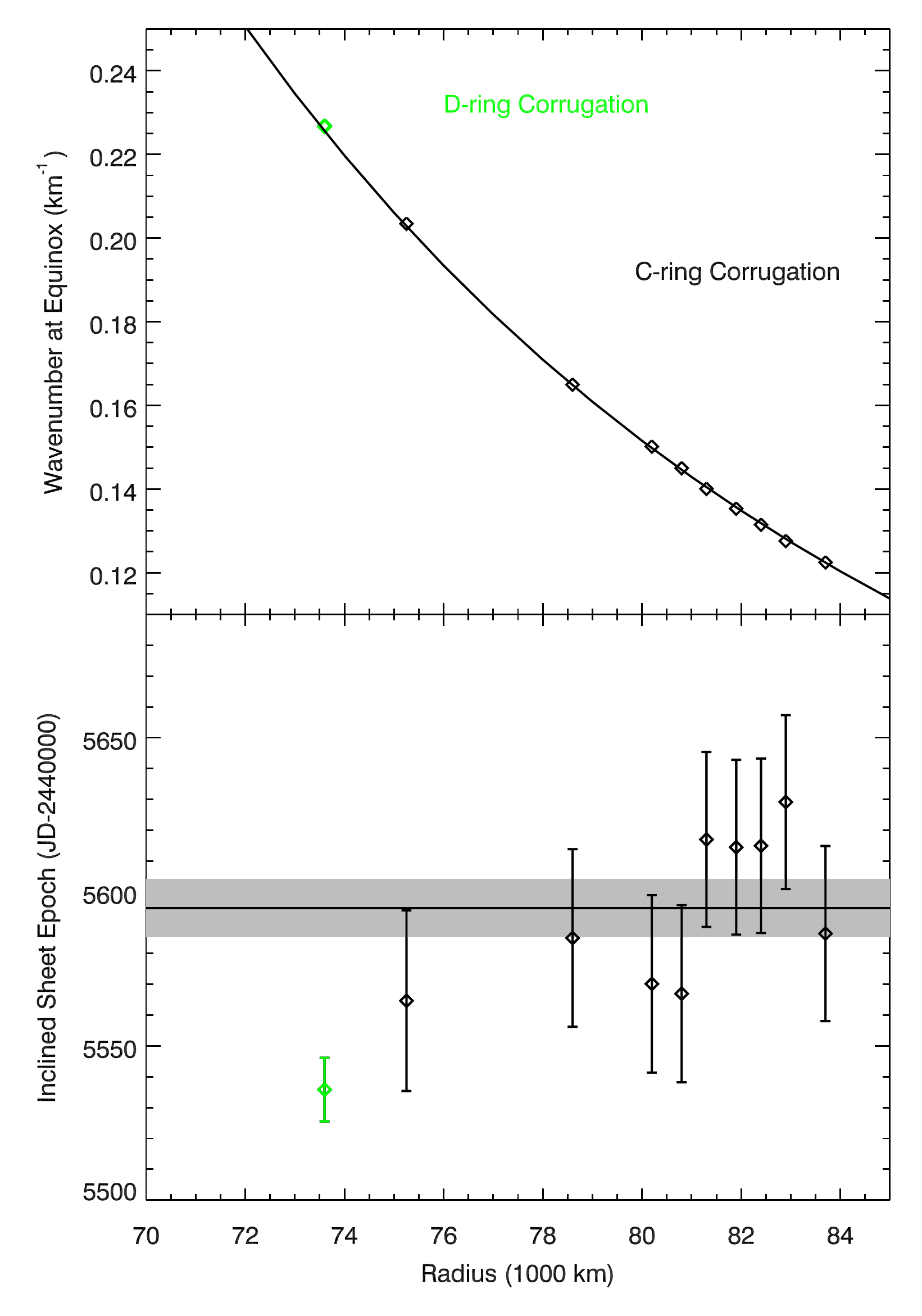}}}
\caption{A plot showing the corrugation's wavelength estimates at equinox (JD 2455054) across Saturn's  C and D rings. The top panel shows the wavenumber estimates from \citet{Hedman11} versus radius, along with the D-ring wavenumber  derived in this paper (in green). The curve shows the predicted trend assuming that all these rings tilted at the same time and that the planet's gravity field is perfectly described by the \citet{Jacobson06} model. Note the D-ring data point falls slightly above this trend. The bottom panel shows the observed wavenumbers divided by the predicted winding rate assuming the \citet{Jacobson06} gravity model, which corresponds to the predicted epoch of ring inclination. The shaded region depicts the estimate of the inclined sheet epoch based on the C-ring observations (with only statistical uncertainties). 
Here the D-ring estimate falls well below the C-ring predictions, suggesting that either the gravity model is incorrect, or the two parts of the rings were tilted at different times.}
\label{cdcomp}
\end{figure}

Additional information about the impact event can be obtained from the trends in the corrugations' wavelength with radius across the C and D rings. Table~\ref{trendtab} includes estimates of the corrugation wavelength on JD 2455054 at 73,600 km derived from our simple linear fits. Regardless of whether we use the $p_2$ or the $p_2/p_1$ profile, we find that this wavelength is $27.69\pm0.03$ km. This number can be compared with the precise estimates of the corrugation wavelengths in the C ring at the same epoch \citep{Hedman11}. The top panel of Figure~\ref{cdcomp} shows the relevant wavenumbers, along with the predicted trend assuming these corrugations formed at the same time and evolved  in a manner consistent with the \citet{Jacobson06} model of Saturn's gravitational field. The D-ring data point falls slightly  above this expected trend, and indeed the measured wavelength is 0.7\% shorter than this model would predict. This difference, while small, is statistically significant. The dispersion of the C ring data about this trend yields  a $\chi^2$ of 6 for eight degrees of freedom, so the C-ring wavenumbers are consistent  the \citet{Jacobson06} gravity model being correct and the entire C-ring being tilted at the same time. However, if we include the D-ring observation the $\chi^2/DOF$ increases to 23.6/9, which has a probability to exceed of less than 0.5\%. This discrepancy between the D-ring and the C-ring data could be explained in one of three ways: 
\begin{itemize}
\item There is a  systematic error in some wavelength estimates.
\item The \citet{Jacobson06} model does not perfectly describe Saturn's gravitational field.
\item Different parts of the ring became tilted at different times.
\end{itemize}
We will explore each of these possibilities below, and argue that the third option appears to be most likely. Furthermore, we will show that if different parts of the rings tilted at different times, then that information provides further information about the trajectory of the impacting debris.

\subsubsection{Systematic errors}

Thus far, we have been unable to find a systematic error that could increase the estimates of the D-ring corrugation wavelengths by the  0.7\% required to make the data fully consistent with the \citet{Jacobson06} gravity field model. Increasing the D-ring wavelength by this amount would require a D-ring corrugation wavelength at equinox (day 2009-239) of 27.9 km, which is inconsistent with the direct measurements of the corrugation wavelengths obtained around the time of equinox on days 206 and 237 of 2009 (see Table~\ref{obssumtab}). Furthermore, if our D-ring wavelength estimates are systematically low by 0.7\%, then the corrugation winding rate would be underestimated by 0.7\%, so the actual winding rate would be at most 2.384$\times10^{-5}$km$^{-1}$/day, which would be inconsistent with the D68 precession rate estimate (see Figure~\ref{j6wirewarp}). Thus there seems to be no way that we can shift the D-ring corrugation wavelength estimate that would be fully consistent with current models of Saturn's gravity field. 

We are also unaware of any suitable systematic error in the \citet{Hedman11} estimates of the C-ring corrugation wavelengths. \citet{Hedman11} demonstrated that the ring's local surface gravity did influence the corrugations' winding rates, but for the wavelength estimates used here this is unlikely to alter the wavelengths by more than 0.1\%. Thus, unless the C-ring's surface mass density has been grossly underestimated, it seems unlikely that the unmodeled gravitational perturbations from nearby ring material could significantly influence the relevant winding rates. Also, the C-ring estimates were computed using the same basic algorithms and yield a data set that is consistent with the standard gravity field model, so it is difficult to imagine how any computational error could shift the C-ring data 0.7\% relative to the D-ring measurement.

\subsubsection{Anomalies in Saturn's gravity field}

If the discrepancy does not reflect a systematic error, then the next logical option is that the standard gravity field does not accurately predict how much the winding rate varies with radius between the D and C rings. In order to explore this possibility quantitatively, let us reduce the trend shown in Figure~\ref{cdcomp} to a single parameter: the corrugation wavenumber ratio between 73,600 km and 82,000 km at equinox. While the wavenumber at 73,600 km can be easily computed from the  D-ring corrugation wavelengths of $27.69\pm0.03$ km given in Table~\ref{trendtab},  to obtain a correspondingly precise wavenumber estimate at 82,000 km we must combine all the wavenumber estimates for the region between 80,000 and 85,000 km provided by \citet{Hedman11}. To reduce these data to a single effective wavenumber, we first transform the individual wavenumber estimates to scaled wavenumbers by multiplying  by $(r/82,000 km)^{-4.5}$, and fitting these numbers to a line. This calculation yields a precise estimate of the corrugation wavelength at 82,000 km: $46.63\pm0.05$ km, which means the wavenumber ratio  $({k}^{73.6}_{z}/{k}^{82}_{z})_{obs}=1.684\pm0.003$. If we assume that both sets of corrugations formed at the same time, this estimate of $({k}^{73.6}_{z}/{k}^{82}_{z})_{obs}$ is also a measure of  $(\dot{k}^{73.6}_{z}/\dot{k}^{82}_{z})_{obs}$, which is 1.673 for the \citet{Jacobson06} gravity model. The predicted and observed numbers differ by about 0.7\%, or about three sigma, so this number properly reproduces the deviations seen in Figure~\ref{cdcomp}.

\begin{figure}[tbp]
\centerline{\resizebox{4in}{!}{\includegraphics{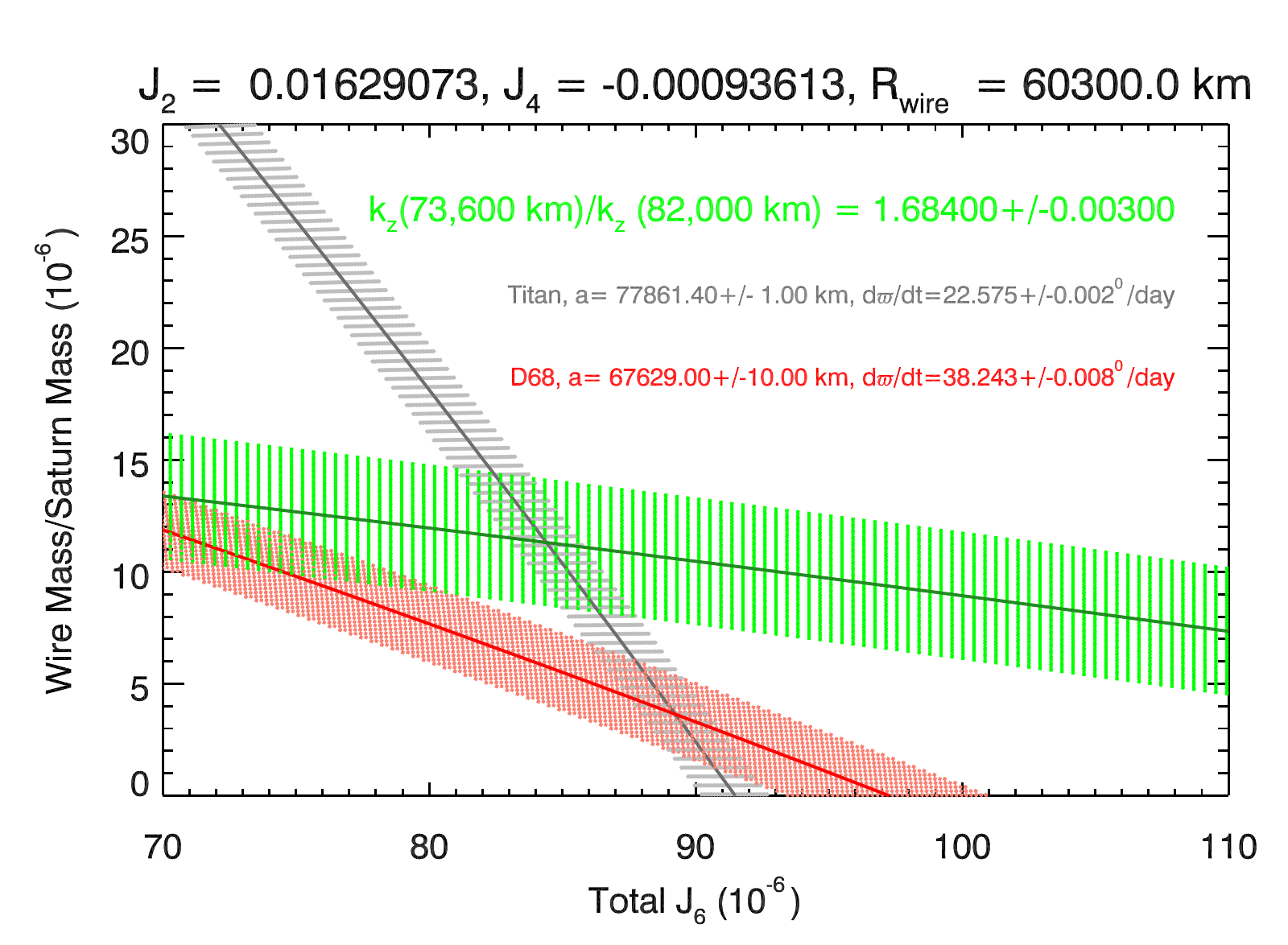}}}
\caption{Plot showing the implied constraints on Saturn's gravitational field provided by the ratio of the corrugation wavenumbers at 73,600 km and 82,000 km, 
compared with the constraints obtained from the forced eccentricity of the Titan ringlet and the precession rate of D68 reported in \citet{Nicholson14} and \citet{Hedman14}, respectively.
Note that there is no region that falls within the one-sigma acceptance bands of all three observations, and the winding rate ratio requires the wire mass to be around $10^{-5}$ Saturn's mass.}
\label{j6wirewr}
\end{figure}

Like the individual corrugation winding rates, the ratio of corrugation winding rates is sensitive to a broad range of gravitational harmonics. The bottom panel of Figure~\ref{sensplot} shows the nominal sensitivity of the ratio $(\dot{k}^{73.6}_{z}/\dot{k}^{82}_{z})_{obs}$ to harmonics of various degrees. Compared with D68's precession rate and the corrugation winding rate, this ratio is even more sensitive to the higher-order components of Saturn's gravitational field. Again, the large number of harmonics that could influence the observable parameters makes the problem under-determined and complicates efforts to evaluate whether the various constraints on the planet's gravitational field are consistent. By contrast, the simplified model of the gravity field discussed above, which has only two free parameters, can more clearly represent the relationships between the various measurements.

Figure~\ref{j6wirewr} illustrates the region of parameter space consistent with these estimates of the corrugation wavenumber ratios for the simplified gravity field model. Note that the strip consistent with the  observed wavenumber ratio is nearly horizontal in this parameter space, which is reasonable given that this ratio is even more sensitive to higher-order harmonics than  D68's precession rate (see Figure~\ref{sensplot}). Note that the constraint from the Titan ringlet is based on the new analysis by \citet{Nicholson14}, which is tighter than the \citet{NP88} constraint used in \citet{Hedman14}.

No set of gravitational field parameters is consistent with all three constraints at the one-sigma level, implying that there is some tension among these different observations. Worse, the winding-rate-ratio solutions all require  a rather large wire mass of $\sim10^{-5}$ Saturn's mass. Such a wire mass corresponds to $|J_{2i}| \sim 10^{-6}$  for all $2i>12$, which is an order of magnitude higher than physically realistic internal models that include gravity-field contributions from Saturn's winds \citep{Kaspi13b}. Combined, the tension with the other gravity field constraints and the implausibility of the implied solutions suggest that the discrepancy between the D and C ring corrugation wavelength cannot be attributed to Saturn's gravitational field. 

We have explored other, more complex gravity field models in order to ascertain if these could resolve the discrepancies between the various constraints shown in Figure~\ref{j6wirewr}. For example, we considered the possibility that
the gravity of the massive B ring could be influencing the corrugation winding rates in the C ring. This ring's gravitational field would indeed reduce the winding rate in the C ring relative to the D ring, but again the mass required seems unreasonably high. To shift the wavelength around 82,000 km by 0.7\%, the mass of the B ring would need to be $\sim10^{-5}$ the mass of the planet or $\sim10^{21}$ kg, or over an order of magnitude more massive than  Mimas. While recent work has suggested that the B-ring could be denser and more massive than previously thought \citep{Robbins10, Hedman13spec}, no one has yet suggested such an extreme mass for the ring.  We also considered the possibility that non-gravitational forces acting on the small particles in the D ring could cause the D-ring pattern to wind faster than the \citet{Jacobson06} model would predict. However, this explanation for the wavelength ratio discrepancy is problematic because the measured value of the winding rate is actually consistent with the theoretical predictions. Indeed, just as it would be difficult to attribute the high wavelength ratio to a systematic error in the D-ring wavelength estimate, it seems unlikely that the wavelength ratio can be due to an accelerated D-ring winding rate. We therefore tentatively conclude that the discrepancy between the D and C ring corrugation wavelengths is unlikely to be due to any phenomenon that influences the winding rates of the two patterns.

\subsubsection{Tilting the C and D rings at different times}

Finally, we consider the possibility that the wavelength ratio is different from the predicted value because the two patterns formed at different times. The bottom panel of Figure~\ref{cdcomp} shows when each part of the ring would need to have been tilted in order for the corrugation to have its measured wavelength at epoch, assuming the \citet{Jacobson06} model of Saturn's gravity field is correct. This plot reveals that the discrepancy in the wavelengths measured in the two rings could be explained if the D-ring became tilted approximately 60 days before the middle C-ring.\footnote{This timing difference is comparable to the uncertainty in the disturbance epoch given in Table~\ref{trendtab}, but in this scenario differences in the timing of the ring-tilting events across the ring can be measured more accurately than the absolute age of the disturbance epoch because the former is based on a comparison of wavelengths directly observed by Cassini and thus does not require extrapolating beyond the available observations.}  If the ring was tilted by collisions with meteoritic debris, this would mean that debris fell on different parts of the ring over the course of several months. This is not as unreasonable as it might seem because previous calculations of particle trajectories for the debris released during the break up of Shoemaker-Levy 9 in 1992 indicate that fine material could have rained down on the planet for around two months in 1994 \citep{Sekanina94}. 

\begin{figure}
\resizebox{6in}{!}{\includegraphics{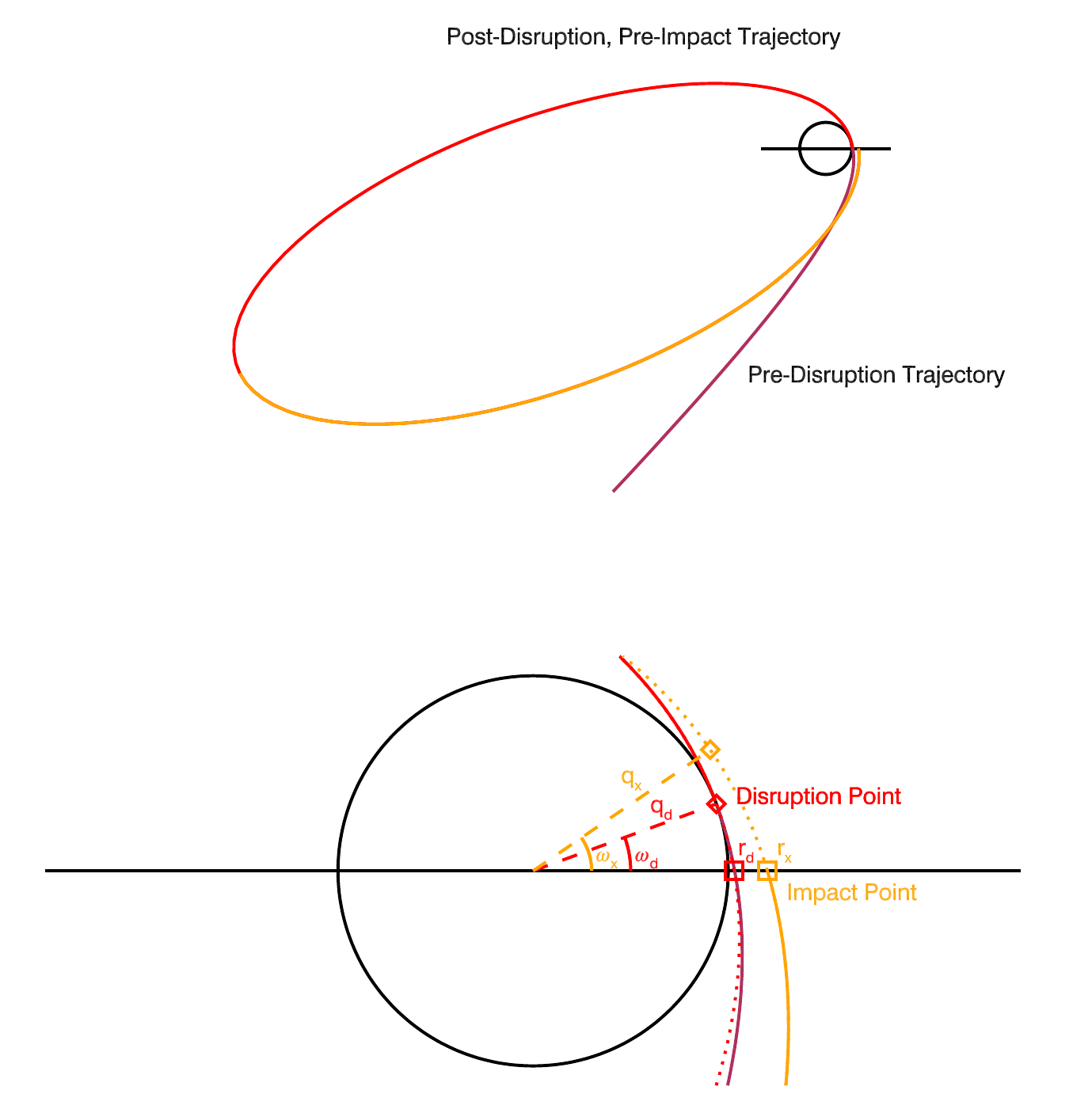}}
\caption{Diagram summarizing the geometry of the impact scenarios considered here. An object is assumed to approach Saturn on an unbound orbit (maroon line), and is disrupted when it passes through Saturn's upper atmosphere. The captured debris is now on a highly elliptical orbit (red/orange line) and impacts the rings one orbit period later. The lower panel shows a close up of the region around Saturn, which illustrates the relevant orbital parameters. $q_d$ and $\omega_d$ are the pericenter distance and longitude of the particle's orbit after its disruption. If the particle's orbit did not change between it disruption and impact with the rings, it would return along the red dotted line and impact at the ring radius $r_d$. However, if other forces (e.g. solar tides) perturb the particles' orbit, then when it returns it will be following the orbit trajectory defined by $q_x$ and $\omega_x$, and impact the ring at a different distance $r_x$.}
\label{trajdiag}
\end{figure}

Inspired by the Shoemaker-Levy 9 example, we will consider scenarios where an object approached Saturn on an initially unbound orbit. This object broke up as it passed close by the planet and some of the debris from this disruption event then became trapped on highly elliptical orbits. This debris will then impact the rings when it comes back through the inner Saturn system one orbit period later (see Figure~\ref{trajdiag}). While this sequence of events is consistent with our favored model for how the corrugations were formed \citep{Hedman11}, such a scenario does not automatically generate the required correlation between the impact locations and impact times. 

The major challenge with such scenarios is that the processes that yield a broad range of impact times are not the same as those that generate substantial differences in the impact radii. A disruption event can produce debris with a range of trajectories and orbital elements. This debris can therefore naturally acquire a range of semi-major axes and orbit periods that will cause the debris to strike the rings over a range of times. However, since all the debris arises from the same event, if no other forces act on the particles they  will all have orbits that return to the same point near the planet, severely limiting the range of ring radii that the debris can reach. Thus we require something else to act on the captured debris between their formation and their collision with the rings that will allow a broad swath of ring radii to be struck. The most likely candidates for this perturbing force are solar tides or solar radiation pressure. \citet{Showalter11} showed that these forces could have a significant effect on the trajectories of debris from Shoemaker-Levy 9, allowing some of the smaller particles to strike Jupiter's rings instead of the planet itself. However, it is important to note that  perturbation forces from both solar radiation pressure and solar gravity are nearly fixed in inertial space. Such conservative forces cannot efficiently change a debris-particle's semi-major axes and thus cannot change when the debris hits the rings. Therefore, we will need to consider both the dispersion of orbital elements generated by the object's disruption and the subsequent evolution of their orbital elements under the influence of solar perturbations if we want to produce the observed spreads in both the impact times and positions.

A full numerical analysis of possible incoming trajectories for the debris, capture scenarios, and orbital evolution processes is beyond the scope of this paper. Instead we will perform some simple analytical calculations that will not only illustrate the necessity of both processes mentioned above, but also demonstrate that together these forces can potentially furnish the observed trends in the impact times and radii. These calculations  should not be regarded as an attempt to prove one particular sequence of events {\em must} have delivered material into the rings. Instead, they simply show that there is at least one {\em possible} sequence of events that could be consistent with the observations.

Let us first consider the object's fragmentation, which could in principle occur due to tidal disruption, a collision with ring material, or excessive ram pressure in Saturn's atmosphere. For these calculations we will assume that the object breaks up in Saturn's upper atmosphere. This particular scenario not only has fewer free parameters than the others (e.g. the disruption event needs to occur roughly 60,000 km from the planet's center), it  also could be consistent with the measurements (see below). The break-up event produces a cloud of debris with an average semi-major axis $a$.  By Kepler's third law, the orbit period of the debris particles $T$ satisfies:
\begin{equation}
T^2=\frac{4\pi^2}{GM}a^3,
\end{equation}
where $G$ is the universal gravitational constant, $M$ is Saturn's mass and $GM = $37931208 km$^3$/s$^2$ \citep{Jacobson06}.  If two debris particles have semi-major axes that differ by $\delta a << a$, then they will impact the rings at times that differ by:
\begin{equation}
\delta T=3\pi(GM)^{-1/2}a^{1/2}\delta a.
\label{dteq}
\end{equation}
(We assume that the orbit period is sufficiently long that the difference in true anomalies of the impacting debris makes a negligible contribution to the impact times). The observed $\delta T$ between the D and C rings is $\sim$60 days, so to be self-consistent the debris' orbital periods must be significantly longer than 60 days, which means $a$ must be larger than 3,000,000 km or 50 Saturn radii.

If we assume the debris formed in a single disruption, then the orbit of every bit of debris must pass through the point in space where the break-up occurred. If we assume the object was disrupted by ram pressure in Saturn's upper atmosphere, then that point will be one Saturn radius ($\sim$ 60,000 km) from Saturn's center. Furthermore, since the atmospheric column density increases rapidly with depth, any debris that survives the encounter with the planet  must have only skimmed by the planet's atmosphere. Hence we may infer that the disruption point corresponds to the orbital pericenter for all the particles. Let us define the debris' average orbital pericenter immediately after the disruption event as  $q_d=a(1-e_d)$, where $e_d$ is the debris' average orbital eccentricity after the disruption event (see Figure~\ref{trajdiag}). 

The strongest forces applied  the various debris particles as they pass through the atmosphere act parallel to the particles'  direction of motion, so we may reasonably assume that all the debris orbits in roughly the same plane. This means that if no forces subsequently perturb the orbital trajectories of the debris, then it will encounter the rings at the same true anomaly $f_d$.  In other words, if we use the ring plane as the debris orbits' reference plane,  all the debris will have the same argument of pericenter $\omega_d$ (see Figure~\ref{trajdiag}).

The radial location where the debris would intercept the ring if it remained on the same orbit after its disruption is given by the standard orbit equation
\begin{equation}
r_d=\frac{a(1-e_d^2)}{1+e_d\cos f_d}=\frac{q_d(1+e_d)}{1+e_d\cos \omega_d}.
\label{rdeq}
\end{equation}
While all the debris particles have the same $q_d$ and $\omega_d$, their eccentricities may differ. If  two debris particles have initial orbital eccentricities that differ by $\delta e_d$, then the impact radii (again assuming no other perturbations to the orbit) will be separated by:
\begin{equation}
\delta r_d=\frac{q_d\delta e_d}{1+e_d\cos \omega_d}\left(1-\frac{(1+e_d)\cos \omega_d}{1+e_d\cos \omega_d}\right)
\end{equation}
or, using Equation~\ref{rdeq} to eliminate $\omega_d$:
\begin{equation}
\delta r_d=\frac{r_d}{e_d(1+e_d)}\left(\frac{r_d-q_d}{q_d}\right)\delta e_d
\end{equation}
and if we assume $e\simeq 1$, then this can be approximated as:
\begin{equation}
\delta r_d=\frac{1}{2}r_d\left(\frac{r_d-q_d}{q_d}\right) \delta e_d
\end{equation}
Since all the particles must retain the same pericenter distance $q_d=a(1-e_d)$, two particles whose orbital semi-major axis differ by $\delta a$ must have eccentiricities that are initially separated by:
\begin{equation}
\delta e_d=\frac{(1-e_d)}{a}{\delta a}=\frac{q_d\delta a}{a^2}
\label{edeq}
\end{equation}
so we may re-cast this difference in collision locations in terms of $\delta a$:
\begin{equation}
\delta r_d=\frac{r_d(r_d-q_d)}{2 a^2}\delta a.
\label{drdeq}
\end{equation}

Combining Equations~\ref{drdeq} and~\ref{dteq}, we can eliminate $\delta a$ and obtain  $\delta T/\delta r_d$ as a function of $a$:
\begin{equation}
\frac{\delta T}{\delta r_d}=\frac{6\pi}{\sqrt{GM}}\frac{a^{5/2}}{r_d(r_d-q_d)}.
\label{dtdreq}
\end{equation}
If we assume the spread of impact radii is entirely due to the spread of orbital elements associated with the disruption event, then  we can estimate the semi-major axis the debris needed to have to produce the observed corrugation wavelength trends. Solving Equation~\ref{dtdreq} for $a$, we find
\begin{equation}
a=\left[\frac{\sqrt{GM}}{6\pi}r_d(r_d-q_d)\frac{\delta T}{\delta r_d}\right]^{2/5}
\label{adeq}
\end{equation}
In this case, we assume $r_d$ is the actual impact location $r_x$, so  $r_d \simeq 80,000$ km, while $q_d \simeq 60,000$ km. Furthermore, if we assume that the observed spread in impact locations $\delta r_x$ is the same as the spread of impact locations produced by the disruption event alone $\delta r_d$, then we can say $\delta T/\delta r_d=\delta T/\delta r_x$. Fitting a linear trend to the observed data in the bottom panel of Figure~\ref{cdcomp}, we find that $\delta T/\delta r_x = 670\pm150$ s/km. Inserting all these numbers into Equation~\ref{adeq}, we obtain $a=415,000\pm38,000$ km or $6.9\pm0.6 R_s$, which corresponds to an orbit period of around 3 days. 

The short orbit period required to produce the observed $\delta T/\delta r_x=\delta T/\delta r_d$ in this model is clearly  inconsistent with the observed $\delta T \simeq 60$ days, and nicely illustrates why the disruption event is unlikely to be the only process involved in dispersing debris across the rings. The fundamental problem is that $\delta r_d \propto r_d^2\delta a/a^2$, so a small semi-major axis is required to produce a big enough separation in the impact radii. This is insensitive to the exact nature and location of the breakup because the ring impact must occur near pericenter. Indeed, we have explored variants of this basic scenario where the inclination and ascending node of the debris' orbits are allowed to vary, but we were unable to substantially alter this basic result.


Thus we are forced to consider the orbital evolution of the debris particles as they orbit the planet on their eccentric orbits between disruption and ring impact. As mentioned above, the Sun is the most likely source of orbital perturbations, and for the sake of simplicity we will assume the particles feel a force from the Sun that is fixed in inertial space. Such a force will not change the debris particles' semi-major axes, but can alter their eccentricities and arguments of pericenter. Since particles on larger semi-major axes (i.e. with longer orbital periods) are exposed to these forces for more time, these changes in orbital parameters induced by the solar forces  will naturally be correlated with the spread in semi-major axes $\delta a$ generated by the disruption event.

We may compute the changes in eccentricity and longitude of ascending node using the standard perturbation equations \citep{Burns76, Burns76err}. For the orbital eccentricity, the relevant perturbation equation is:
\begin{equation}
\frac{de}{dt}=n\sqrt{1-e^2}\left[\frac{F_r}{F_G}\sin f+\frac{F_\lambda}{F_G}\left(\cos f+\frac{e+\cos f}{1+e\cos f}\right)\right],
\end{equation}
where $F_r$ and $F_\lambda$ are the radial and azimuthal components of the perturbing force, $n=\sqrt{GM/a^3}$, $F_G=GMm/a^2$ is the average force of Saturn's gravity exerted on the particle, and $f$ is the particle's orbital true anomaly. The perturbation equation for the argument of pericenter $\omega$ is:
\begin{equation}
\frac{d\omega}{dt}+\cos I \frac{d\Omega}{dt}={n}\frac{\sqrt{1-e^2}}{e}\left[-\frac{F_r}{F_G}\cos f+\frac{F_\lambda}{F_G}\sin f\frac{2+e\cos f}{1+e\cos f}\right],
\end{equation}
where $I$  is the orbital inclination and  $\Omega$ is the longitude of ascending node. For the sake of simplicity, let us assume that the inertial force does not cause $\Omega$ to evolve over time, so this equation can be re-written as:
\begin{equation}
\frac{d\omega}{dt}={n}\frac{\sqrt{1-e^2}}{e}\left[-\frac{F_r}{F_G}\cos f+\frac{F_\lambda}{F_G}\sin f\frac{2+e\cos f}{1+e\cos f}\right].
\end{equation}
If this force is in a fixed inertial direction, then we may specify: $F_r=F_I\cos(f-f_0)$ and $F_\lambda=-F_I\sin(f-f_0)$, and so we may re-write these expressions as:
\begin{equation}
\frac{de}{dt}=n\sqrt{1-e^2}\frac{F_I}{F_G}\left[\sin(f_0)-\sin(f-f_0)\left(\frac{e+\cos f}{1+e\cos f}\right)\right],
\end{equation}
\begin{equation}
\frac{d\omega}{dt}={-n}\frac{\sqrt{1-e^2}}{e}\frac{F_I}{F_G}\left[\cos(f-f_0)\cos f+\sin(f-f_0)\sin f\frac{2+e\cos f}{1+e\cos f}\right].
\end{equation}
Next, since the rings are so close to the planet, we may reasonably assume that the debris will impact the rings almost exactly one orbit period after the disruption event. Hence we may integrate the above expressions over one complete orbit period to estimate the  difference between the orbital elements at impact ($e_x,\varpi_x$) and disruption ($e_d, \varpi_d$) assuming that any changes in $e$ are small:
\begin{equation}
e_x-e_d=\int_0^T\frac{de}{dt}{dt}=\int_0^{2\pi}\frac{de}{dt}\frac{df}{\dot{f}},
\end{equation}
\begin{equation}
\omega_x-\omega_d=\int_0^T\frac{d\omega}{dt}{dt}=\int_0^{2\pi}\frac{d\omega}{dt}\frac{df}{\dot{f}}.
\end{equation}
Next we note that $r^2\dot{f}=na^2\sqrt{1-e^2}$ and $r=a(1-e^2)/(1+e\cos f)$, so we can re-write these  as:
\begin{equation}
e_x-e_d=\int_0^{2\pi}\frac{de}{dt}\frac{(1-e^2)^2}{n\sqrt{1-e^2}}\frac{df}{(1+e\cos f)^2}
\end{equation}
\begin{equation}
\omega_x-\omega_d=\int_0^{2\pi}\frac{d\omega}{dt}\frac{(1-e^2)^2}{n\sqrt{1-e^2}}\frac{df}{(1+e\cos f)^2}
\end{equation}
Inserting the above expressions for $de/dt$ and $d\omega/dt$ then yields:
\begin{equation}
e_x-e_d=(1-e^2)^2\frac{F_I}{F_G}\int^{2\pi}_0\left[\sin(f_0)-\sin(f-f_0)\left(\frac{e+\cos f}{1+e\cos f}\right)\right] \frac{df}{(1+e\cos f)^2},
\end{equation}
\begin{equation}
\small
\omega_x-\omega_d=-\frac{(1-e^2)^2}{e}\frac{F_I}{F_G}\int_0^{2\pi}\left[\cos(f-f_0)\cos f+\sin(f-f_0)\sin f\frac{2+e\cos f}{1+e\cos f}\right]\frac{df}{(1+e\cos f)^2}.
\end{equation}
We may now note symmetry requires any term that varies like $\sin f$ to vanish, so
\begin{equation}
e_x-e_d=(1-e^2)^2\frac{F_I}{F_G}\sin f_0\int^{2\pi}_0\left[1+\left(\frac{\cos f(e+\cos f)}{1+e\cos f}\right)\right] \frac{df}{(1+e\cos f)^2},
\end{equation}
\begin{equation}
\small
\omega_x-\omega_d=-\frac{(1-e^2)^2}{e}\frac{F_I}{F_G}\cos f_0\int_0^{2\pi}\left[\frac{1+\sin^2 f+ e\cos f}{(1+e\cos f)^3}\right]{df}.
\end{equation}
Evaluating the relevant integrals yields:
\begin{equation}
e_x-e_d=\sqrt{1-e^2}\frac{F_I}{F_G}3\pi\sin f_0,
\end{equation}
\begin{equation}
\omega_x-\omega_d=-\frac{\sqrt{1-e^2}}{e}\frac{F_I}{F_G}3\pi\cos f_0.
\end{equation}

Thus far, we have not been explicit about whether the eccentricity on the right hand side of the above expressions is $e_x$ or $e_d$. So long as the changes in $e$ are small, we can approximate $e$ as $e_d$, in which case the orbital eccentricities and arguments of pericenter at ring impact can be written as:
\begin{equation}
e_x=e_d+\sqrt{1-e_d^2}\frac{F_I}{F_G}3\pi\sin f_0,
\end{equation} 
\begin{equation}
\omega_x=\omega_d-\frac{\sqrt{1-e_d^2}}{e_d}\frac{F_I}{F_G}3\pi\cos f_0.
\end{equation}
Furthermore, if we consider two particles which emerge from the disruption event with orbital eccentricities that differ by $\delta e_d$, we expect the final eccentricities and pericenters to differ by:
\begin{equation}
\delta e_x=\left(1-\frac{e_d}{\sqrt{1-e_d^2}}\frac{F_I}{F_G}3\pi\sin f_0\right)\delta e_d,
\end{equation} 
\begin{equation}
\delta\omega_x=\frac{1}{e_d^2\sqrt{1-e_d^2}}\frac{F_I}{F_G}3\pi\cos f_0 \delta e_d.
\end{equation} 
Note that for the sake of simplicity,  we have assumed that the inertial perturbation force $F_I/F_G$ is the same for all
the debris particles. This is equivalent to assuming that solar gravity (rather than solar radiation pressure)
is the dominant perturbation. Such an approximation is likely to be valid for debris particles more than a few microns wide.

For the sake of clarity, we can now cast these expressions in terms of $q_d$, $a$ and $\delta a$,
using the identities $\delta e_d=q_d \delta a/a^2$ and $(1-e_d)=q_d/a$. We can also assume that
$q_d/a<<1$, so that we can expand each term above and keep only those that are lowest-order in
$q_d/a$.
\begin{equation}
e_x=1-\frac{q_d}{a}+\sqrt{\frac{2q_d}{a}}\frac{F_I}{F_G}3\pi\sin f_0,
\label{ex1}
\end{equation} 
\begin{equation}
\omega_x=\omega_d-\sqrt{\frac{2q_d}{a}}\frac{F_I}{F_G}3\pi\cos f_0,
\label{om1}
\end{equation}
\begin{equation}
\delta e_x=\left(1-\frac{1}{\sqrt{2q_d/a}}\frac{F_I}{F_G}3\pi\sin f_0\right)\frac{q_d}{a}\frac{\delta a}{a},
\label{dex1}
\end{equation} 
\begin{equation}
\delta\omega_x=\frac{1}{2}\sqrt{\frac{2q_d}{a}}\frac{F_I}{F_G}3\pi\cos f_0 \frac{\delta a}{a}.
\label{dom1}
\end{equation} 

With these expressions, we can now derive expressions for the impact location $r_x$ and the separation in the impact locations $\delta r_x$ for two objects with semi-major axes that differ by $\delta a$. From the standard orbit equation, the mean impact radius is:
\begin{equation}
r_x=\frac{a(1-e_x^2)}{1+e_x\cos\omega_x}
\end{equation}
However, in practice it is easier to express this observable in terms of the final average pericenter distance of the debris (see Figure~\ref{trajdiag})
\begin{equation}
q_x=a(1-e_x)=q_d-a\sqrt{\frac{2q_d}{a}}\frac{F_I}{F_G}3\pi\sin f_0.
\label{qx1}
\end{equation}
Accordingly,  two particles with initial semi-major axes that differ by $\delta a$ will have final pericenter distances that
are separated by:
\begin{equation}
\delta q_x =-\frac{1}{2}\sqrt{\frac{2q_d}{a}}\frac{F_I}{F_G}3\pi\sin f_0\delta a.
\end{equation}
Combining these two equations, we find:
\begin{equation}
\delta q_x =\frac{1}{2}(q_x-q_d)\frac{\delta a}{a}.
\end{equation}
Similarly, we can use the above expression for $q_x$ to re-write $\delta e_x$ as:
\begin{equation}
\delta e_x =\frac{(q_x+q_d)}{2a}\frac{\delta a}{a}.
\end{equation}
Also, we can combine Equations~\ref{om1} and~\ref{dom1} to get the following expression for $\delta\omega_x$: 
\begin{equation}
\delta \omega_x =-\frac{(\omega_x-\omega_d)}{2}\frac{\delta a}{a}.
\end{equation}

In terms of $q_x$, $e_x$ and $\omega_x$, the mean impact radius is
\begin{equation}
r_x=\frac{q_x(1+e_x)}{1+e_x\cos\omega_x}
\label{rxeq}
\end{equation}
and the separation in the impact locations $\delta r_x$ for two objects with semi-major axes that differ by $\delta a$ is:
\begin{equation}
\delta r_x=\frac{r_x}{q_x}\delta q_x+\left(\frac{r_x}{1+e_x}-\frac{r_x^2\cos\omega_x}{q_x(1+e_x)}\right)\delta e_x+\frac{r_x^2\sin\omega_x}{q_x(1+e_x)}\delta \omega_x.
\end{equation}
Assuming $e_x \simeq 1$ and using the above expressions for $\delta q_x$, $\delta e_x$ and $\delta \omega_x$, we get:
\begin{equation}
\delta r_x=r_x\left[\frac{q_x-q_d}{2q_x}-\left(\frac{1}{2}-\frac{r_x\cos\omega_x}{2q_x}\right)\frac{(q_x+q_d)}{2a}+\frac{r_x\sin\omega_x}{2q_x}(\omega_d-\omega_x)\right]\frac{\delta a}{a}.
\end{equation}
We can now use equation~\ref{rxeq} to eliminate $\omega_x$ from the above expression:
\begin{equation}
\small
\delta r_x=r_x\left[\frac{q_x-q_d}{2q_x}-\left(\frac{1}{2}-\frac{2q_x-r_x}{2q_x}\right)\frac{(q_x+q_d)}{2a}+\frac{1}{2}\sqrt{\frac{r_x-q_x}{q_x}}\left(\omega_d\pm\sin^{-1}\sqrt{\frac{4q_x(r_x-q_x)}{r_x^2}}\right)\right]\frac{\delta a}{a}.
\label{drxx}
\end{equation}
where the choice of sign on the last term depends on whether the impact occurs before or after pericenter. The second term in the above expression contains  $(q_x+q_d)/a$, while the other terms instead involve $(q_x-q_d)/q_x$ and $(r_x-q_x)/q_x$. Since $q_x$ lies somewhere between $r_x$ and $q_d$, and since $a>>r_x$ and $a>>q_d$, we can neglect this term, and so we may approximate Equation~\ref{drxx} as:
\begin{equation}
\delta r_x=r_x\left[\frac{q_x-q_d}{2q_x}+\frac{1}{2}\sqrt{\frac{r_x-q_x}{q_x}}\left(\omega_d\pm\sin^{-1}\sqrt{\frac{4q_x(r_x-q_x)}{r_x^2}}\right)\right]\frac{\delta a}{a}.
\end{equation}
Finally, we note that neither of the remaining terms in the square brackets is of order unity, since $q_x-q_d$ and
$r_x-q_d$ are both less than $q_x$. Hence, we re-express $\delta r_x$ as:
\begin{equation}
\delta r_x=\left[\frac{r_x(q_x-q_d)}{2q_x(r_x-q_d)}+\frac{1}{2}\sqrt{\frac{r_x^2(r_x-q_x)}{q_x(r_x-q_d)^2}}\left(\omega_d\pm\sin^{-1}\sqrt{\frac{4q_x(r_x-q_x)}{r_x^2}}\right)\right]{(r_x-q_d)}\frac{\delta a}{a}.
\label{drxeq}
\end{equation}
The quantity in the square brackets, which we will here denote $\eta(r_x,q_x,q_d,\omega_d)$, can easily be of order unity. Indeed, if we assume $\omega_d=0$ and that the impact occurs prior to pericenter passage, then $\eta$ runs between 1.2 and 0.6 as $q_x$ runs between $q_d=60,000$ km and $r_x=$80,000 km. Of course, this range will shift depending on the assumed value of $\omega_d$ and the sign of $\omega_x$. However, $\eta$ only approaches zero when $\omega_d-\omega_x \simeq -20^\circ$, so there is a reasonable range of parameter space where $\eta$ will be of order unity and  $\delta r_x$ will be of order $a/r_d$ larger than $\delta r_d$. 

Combining Equation~\ref{drxeq} with Equation~\ref{dteq}, and eliminating $\delta a$, the impact time versus impact radius becomes:
\begin{equation}
\frac{\delta T}{\delta r_x}=\frac{3\pi}{\sqrt{GM}}\frac{a^{3/2}}{\eta (r_x-q_d)}.
\end{equation}
Solving this for $a$ yields:
\begin{equation}
a=\left(\frac{\sqrt{GM}}{3\pi}\eta (r_x-q_d)\frac{\delta T}{\delta r_x}\right)^{2/3}.
\end{equation}
Assuming $r_x-q_d \simeq 20,000$ km and $\delta T/\delta r_x \simeq 670$ s/km, we find that the average  semi-major axis of the debris is  $70\eta^{2/3} R_s$. The average orbit period of the debris is therefore 100$\eta$ days. Recall that the time between the D and C ring impacts is of order 60 days. Hence, so long as $\eta$ exceeds one, we have a marginally consistent model. Of course, if $\eta \simeq 1$, then our assumption that $\delta T/T$ and $\delta a/a$ are small quantities is not really valid. While the above calculations could in principle be refined to account for larger orbit variations, such complex calculations are beyond the scope of this paper. We also do not expect the relevant corrections to qualitatively change our results.

To further evaluate whether this model is reasonable, we can examine whether the required values of $\delta a$ and $F_I/F_G$ are sensible. First consider the force ratio. From Equations~\ref{qx1} and~\ref{om1} we have:
\begin{equation}
\frac{F_I}{F_G}\sin f_0=\frac{-1}{3\pi}\frac{(q_x-q_d)}{a}\sqrt{\frac{a}{2q_d}},
\end{equation}
\begin{equation}
\frac{F_I}{F_G}\cos f_0=\frac{-1}{3\pi}(\omega_x-\omega_d)\sqrt{\frac{2q_d}{a}}.
\end{equation}
Thus the force ratio is:
\begin{equation}
\frac{F_I}{F_G}=\frac{1}{3\pi}\left[\frac{(q_x-q_d)^2}{2aq_d}+\frac{2q_d}{a}(\omega_d-\omega_x)^2\right]^{1/2}.
\end{equation}
and again  using the  orbit equation to eliminate $\omega_x$:
\begin{equation}
\frac{F_I}{F_G}=\frac{1}{3\pi}\left[\frac{(q_x-q_d)^2}{2aq_d}+\frac{2q_d}{a}\left(\omega_d\pm\sin^{-1}\sqrt{\frac{4q_x(r_x-q_x)}{r_x^2}}\right)^2\right]^{1/2}.
\end{equation}
Finally, we pull out a factor of $\sqrt{(r_x-q_d)/a}$ to get:
\begin{equation}
\frac{F_I}{F_G}=\frac{1}{3\pi}\left[\frac{(q_x-q_d)^2}{2q_d(r_x-q_d)}+\frac{2q_d}{(r_x-q_d)}\left(\omega_d\pm\sin^{-1}\sqrt{\frac{4q_x(r_x-q_x)}{r_x^2}}\right)^2\right]^{1/2}\sqrt{\frac{r_x-q_d}{a}}
\end{equation}
The term in square brackets is another quantity of order unity that we will designate $\zeta^2$. Again, assuming $\omega_d=0$, we find $\zeta$ ranges between 2.6 and 0.6 as $q_x$ runs from $q_d=$60,000 km to $r_x=80,000$ km. Assuming $(r_x-q_d)=$ 20,000 km and $a=70\eta^{2/3} R_s$, we find $F_I/F_G=0.007\zeta\eta^{-1/3}$. Note that the prefactors partially cancel each other out, so  $F_I/F_G$ runs between .017 and 0.005 if $\omega_d=0$.

Now, if $F_I$ is due to solar gravity, we would expect $F_I/F_G=(M_\sun/M_{S})(a/a_{S})^2$, where $M_\sun/M \simeq 3500$ is the ratio of the Sun's mass to Saturn's mass, and $a_S=1.5\times 10^9$ km is Saturn's semi-major axis. Assuming $a=70\eta^{2/3} R_s$, we estimate $F_I/F_G \simeq 0.03 \eta^{4/3}$, which is of the same order as the required perturbing force. Hence it appears that the Sun's gravity can perturb the debris particles' orbits by the required amount.

Finally, we must consider the spread of semi-major axes $\delta a$ produced in the original disruption event. This variance in semi-major axes is directly related to the spread in velocities of the debris emerging from the disruption event. If we stipulate that the object arrives on a nearly parabolic orbit, then it will be  moving relative to the planet at a speed $v_0\simeq\sqrt{2GM/q_d}\sim35$ km/s when it is disrupted. Since the speed of the debris is given by the $vis-viva$ equation:
\begin{equation}
v^2=\frac{2GM}{r}-\frac{GM}{a},
\end{equation}
we may relate the spread in the speed of the debris particles $\delta v$ to the spread in their orbital semi-major axes $\delta a$:
\begin{equation}
\delta v=\frac{GM}{v_0 a^2}\delta a\simeq\sqrt{\frac{GM}{2}} \frac{q_d^{1/2}}{a^2} \delta a,
\label{veq}
\end{equation}
which we  re-write in terms of the spread in impact times using Equation~\ref{dteq}:
\begin{equation}
\delta v= \frac{1}{\sqrt{2}3\pi}GM \frac{q_d^{1/2}}{a^{5/2}}{\delta T}.
\label{vteq}
\end{equation}
Assuming $q_d=1 R_s$ and $a=70\eta^{2/3} R_s$ and $\delta T\simeq60$ days,  we find $\delta v\simeq100\eta^{-5/3}$ m/s.

 
Such a large velocity dispersion is extremely unlikely to arise from tidal disruption. However, it could occur if the object broke up while passing through Saturn's upper atmosphere. Debris passing through Saturn's upper atmosphere is decelerated by ram pressure \citep{PBT79, HG93, Hedman11}. So long as the debris enters and leaves Saturn's atmosphere at close to the escape speed, the changes in debris velocity within the atmosphere are relatively small perturbations, and the fractional change in velocity of a given piece of debris as it passes through the atmosphere can be approximated as:
\begin{equation}
\Delta v/v=\frac{3m_aN_a}{4\rho_d s_d},
\end{equation}
where $m_a\simeq 2$AMU  is the average molecular mass, $N_a$ is the total column density along the particle's path, $\rho_d\simeq 1000$kg/m$^3$ is the mass density of the particle and $s_d$ is its size (radius). In this limit, the difference in the final speeds between two debris particles $\delta v$ will be given by:
\begin{equation}
\delta v/v=\frac{3m_a}{4\rho_d}\delta(N_a/s_d),
\end{equation}
A difference in orbit speeds of 100 m/s between two bits of debris (which corresponds to a $\delta v/v \simeq$ 0.25\%) therefore requires either the size or the total column density encountered by the two particles to differ by only a factor of 0.25\%. Given the debris will consist of particles with a wide range of sizes spread over a finite region of space, this should not be a difficult condition to meet. Note also that the velocity dispersion in this scenario is mostly along the direction of motion, consistent with the above calculations.

It is also possible that sufficiently large velocity dispersions could have been created if  the object collided with a ring particle. Certainly, such an impact could provide enough kinetic energy to provide the required velocity dispersion. Unfortunately, the fragmentation and collision dynamics of such extreme hypervelocity collisions between ice-rich objects are still not well characterized at the relevant velocities, so it is not as easy to evaluate the viability of this scenario.

In conclusion, the required correlation between the impact times and impact locations can be achieved with plausible initial velocity dispersions and solar perturbations. The observational data therefore could be consistent with a scenario where an object broke up in Saturn's atmosphere, produced a cloud of debris on a highly eccentric orbit around the planet. This debris was then perturbed by solar tides so that it crashed into a broad swath of the rings one orbit period later. While alternative scenarios do exist and should be explored, this analysis indicates that there is at least one scenario that can accommodate all the observations. The simple calculations done above also suggest that the time between disruption and impact could be just a few months, but we caution that more sophisticated numerical integrations are needed to robustly determine the range of possible orbit periods.

\subsection{Speculations on the source of the ring-disturbing material}

Our previous examination of the C-ring patterns \citep{Hedman11} indicated that between $10^{11}$ and $10^{13}$ kg of material needed to rain down on the rings in 1983 to produce the observed corrugation amplitudes. This mass is comparable to that of a few-kilometer-wide comet, and since such small bodies are very difficult to detect in the outer solar system, it is extremely unlikely that this object would have been identified prior to its collision with the Saturn system. While the above analysis does not change our estimate of the impacting debris mass, it does provide new insights into the trajectory of the debris before it struck the rings. Specifically, the amplitude measurements indicate that the ring was struck by debris  that approached the rings from close to face-on, while the wavelength data suggest that the material striking the ring may have encountered Saturn some months before the rings were disturbed in the summer (D ring) or fall (C ring) of 1983. We may therefore ask whether these requirements are likely to be met by debris derived from known small-body populations in the outer solar system

In lieu of a detailed numerical simulation, we will investigate whether the ring-disturbing material could be part of a debris stream following the orbit of a larger object, analogous to the meteor streams found in the inner solar system. Using orbital elements from the Minor Planet Catalog ({\tt www.minorplanetcenter.net/iau/MPCORB/MPCORB.dat}), we searched for objects with orbits that came within 0.5 AU of Saturn in 1982 or 1983, and where debris following that orbit would approach Saturn's rings at an elevation angle greater than 50$^\circ$ (neglecting gravitational focusing effects). Two objects satisfied all these criteria: Centaur  32532 Thereus (2001 PT13) and Centaur 328884 (2010 LJ109). If we wanted to tie the ring-disturbing event to one of these objects, Thereus would be the better option because Saturn passed closer to Thereus' orbit than it did to 201 LJ109's orbit.  (0.13 AU versus 0.21 AU). Furthermore, Saturn seems to have passed closest to 2010 LJ109's orbit in the fall of 1983, while it passed Thereus' orbit in late spring.  Debris near Thereus' orbit would be more likely to reach Saturn's rings at the appropriate time than would debris following the other object's orbit. Nevertheless, the existence of both these objects demonstrates that the impacting debris could potentially be derived from objects on Centaur-like orbits.

Of course  Thereus and perhaps 2010 LJ109 are the only {\it known} objects whose orbits have the right shape and orientation to deliver material into the rings at the appropriate time and at the correct angle. Future surveys and numerical simulations could reveal other objects that could produce the required ring-tilting debris. Even so, these quick calculations do indicate that potential connections between Saturn and  Thereus  and/or 2010 LJ109 merit further scrutiny. For example, spectra of Thereus show variable water-ice band-strengths that have been interpreted as evidence for a recent impact \citep{Licandro05}, and it could be worth examining whether such an impact could produce the required ring-impacting debris.

\section{Summary}

The above analysis of the periodic patterns in the outer D ring provides a much more refined picture of what happened to Saturn's rings in 1983 and how the structures produced at that time evolved over the last 30+ years. This analysis has yielded the following results.

\begin{itemize}
\item The outer D ring possesses a vertical corrugation and an eccentric spiral, and the wavelengths of both these patterns are decreasing with time at rates that are consistent with current models of Saturn's gravity field.
\item The amplitude of the vertical corrugation is 2.3$\pm$0.5 times the amplitude of the eccentric spiral, which implies that the ring-disturbing event perturbed the ring particles' vertical motions more than their radial motions.
\item The D-ring was disturbed roughly 60 days before the middle C ring,  probably in the middle of July 1983.
\end{itemize}

Nothing in these new data contradicts the idea that these patterns were generated by a diffuse debris cloud striking
the rings. Indeed, within this context, the above observations provide new information about the event:

\begin{itemize}
\item The amplitude ratio of the vertical corrugation and the eccentric spiral indicates that the rings were struck at a high angle (at least 60$^\circ$ from the ringplane).
\item The differences in the corrugation wavelengths between the C and D rings could be consistent with debris formed by the disruption of an object in Saturn's atmosphere (or, perhaps, the rings) some months prior to the ring-disturbing event. 
\item The Centaurs Thereus and 2010 LJ104 have orbits that passed close to Saturn in 1983 and were inclined such that
debris moving along their orbits would strike the rings at a suitably high angle. 
\end{itemize}

\section*{Acknowledgements}
We wish to thank the Imaging Team, the Cassini Project and NASA for providing the data used for this analysis. This work was funded by the Cassini Data Analysis Program Grant NNX12AC29G. We also want to thank P.D. Nicholson, and M.S. Tiscareno for useful conversations. He would also like to thank John Smith for first bringing the potential connection between the 1983 event and Thereus to our attention, and R. G. French for providing updated geometry information on the 1995 occultation data. Finally we would like to thank two anonymous reviewers for providing helpful comments on an earlier version of this manuscript.


\section*{Appendix A: Detailed list of images used in this analysis}

\begin{table}
\caption{Detailed list of images used in this analysis}
\label{obstab}
\resizebox{6in}{!}{\begin{tabular}{|c|c|c|c|c|c|c|c|c|c|c|c|}\hline
Image & UTC time & Phase & $B$ & $\cos\phi/\tan B$ & Radial & $A_r$ & $A_z$ & $\Lambda_r$ from $dp_1/dr$ (km) & $\Lambda_r$ from $dp_1/dr(\bar{p}_1)^{-1}$ (km) & $\Lambda_z$ from $p_2$ (km) & $\Lambda_z$ from $p_2/p_1$ (km) \\
 Name &  & Angle &  & values fit & Res. & (km) & (km) & Value$\pm$Error (Width) & Value$\pm$Error (Width) & Value$\pm$Error (Width) & Value$\pm$Error (Width) \\
\hline
N1493559711 & 2005-120T13:14:45 &  38.5$^\circ$ & -19.5$^\circ$ & [-0.40, 0.40] & 4.5 km &  0.28 &  0.60 & 32.11$\pm$0.25 (0.53) & 32.11$\pm$0.25 (0.55) & 33.09$\pm$0.26 (0.56) & 33.08$\pm$0.26 (0.57) \\ 
N1495302191 & 2005-140T17:15:52 &   1.1$^\circ$ & -20.7$^\circ$ & [ 0.16, 0.36] & 1.6 km &  0.30 &  0.77 & 32.25$\pm$0.09 (0.54) & 32.23$\pm$0.09 (0.56) & 32.93$\pm$0.09 (0.53) & 32.93$\pm$0.09 (0.52) \\ 
N1496893302 & 2005-159T03:14:15 &  18.1$^\circ$ & -16.8$^\circ$ & [-0.15, 0.15] & 1.1 km &  0.86 &  2.31 & 32.04$\pm$0.06 (0.49) & 32.03$\pm$0.06 (0.54) & 32.98$\pm$0.06 (0.58) & 32.79$\pm$0.06 (0.65) \\ 
N1549402947 & 2007-036T21:09:23 &  67.4$^\circ$ &  30.5$^\circ$ & [-0.10, 0.20] & 4.3 km &  0.21 &  0.44 & 29.72$\pm$0.23 (0.46) & 29.73$\pm$0.23 (0.45) & 30.78$\pm$0.23 (0.47) & 31.06$\pm$0.24 (0.45) \\ 
N1571969357 & 2007-298T01:33:48 &  26.7$^\circ$ &  -2.4$^\circ$ & [-0.50, 0.50] & 2.0 km &  0.28 &  0.62 & 28.91$\pm$0.10 (0.42) & 28.90$\pm$0.10 (0.45) & 29.75$\pm$0.10 (0.45) & 29.72$\pm$0.10 (0.48) \\ 
N1627207994 & 2009-206T09:31:16 & 160.3$^\circ$ &  -7.0$^\circ$ & [-1.00, 1.00] & 2.8 km &  0.35 &  0.60 & 26.92$\pm$0.14 (0.43) & 26.93$\pm$0.14 (0.38) & 27.73$\pm$0.14 (0.53) & 27.75$\pm$0.14 (0.40) \\ 
N1630092669 & 2009-239T18:48:50 &  11.7$^\circ$ &   7.1$^\circ$ & [-1.00, 1.00] & 1.2 km &  0.41 &  0.74 & 26.97$\pm$0.06 (0.41) & 26.97$\pm$0.06 (0.42) & 27.74$\pm$0.06 (0.40) & 27.74$\pm$0.06 (0.40) \\ 
N1632170784 & 2009-263T20:03:51 &   7.9$^\circ$ &   8.5$^\circ$ & [-0.80, 0.50] & 1.4 km &  0.51 &  1.40 & 26.96$\pm$0.06 (0.40) & 26.96$\pm$0.06 (0.41) & 27.52$\pm$0.07 (0.43) & 27.50$\pm$0.07 (0.45) \\ 
N1641836932 & 2010-010T17:05:09 & 157.8$^\circ$ & -21.3$^\circ$ & [-0.10, 0.10] & 1.4 km &  0.42 &  1.16 & 26.61$\pm$0.06 (0.46) & 26.70$\pm$0.06 (0.39) & 27.27$\pm$0.07 (0.47) & 27.45$\pm$0.07 (0.38) \\ 
N1719748946 & 2012-182T11:10:09 &  24.2$^\circ$ &   3.3$^\circ$ & [-0.50, 0.50] & 3.0 km &  0.53 &  1.34 & 24.33$\pm$0.13 (0.30) & 24.34$\pm$0.13 (0.32) & 25.04$\pm$0.14 (0.32) & 25.06$\pm$0.14 (0.33) \\ 
N1725569950 & 2012-249T20:06:16 &  36.8$^\circ$ &  -3.5$^\circ$ & [-0.50, 0.50] & 4.2 km &  0.12 &  0.37 & 24.11$\pm$0.18 (0.30) & 24.10$\pm$0.18 (0.30) & 24.86$\pm$0.19 (0.31) & 24.83$\pm$0.19 (0.31) \\ 
N1743620398 & 2013-092T18:05:09 & 144.3$^\circ$ &   5.5$^\circ$ & [-1.50, 1.50] & 1.5 km &  0.40 &  0.61 & 23.69$\pm$0.06 (0.33) & 23.73$\pm$0.06 (0.30) & 24.38$\pm$0.07 (0.34) & 24.41$\pm$0.07 (0.30) \\ 
\hline
N1504584655 & 2005-248T03:42:39 &  11.8$^\circ$ & -15.8$^\circ$ & [-0.30,-0.00] & 1.2 km &  0.64 &  1.53 & 31.65$\pm$0.06 (0.49) & 31.66$\pm$0.06 (0.55) & 32.68$\pm$0.07 (0.58) & 32.50$\pm$0.07 (0.65) \\ 
N1504582470 & 2005-248T03:06:14 &  13.2$^\circ$ & -15.9$^\circ$ & [-0.30,-0.00] & 1.2 km &  0.29 &  0.55 & 31.73$\pm$0.07 (0.49) & 31.72$\pm$0.07 (0.50) & 32.66$\pm$0.07 (0.57) & 32.69$\pm$0.07 (0.56) \\ 
\hline
\multicolumn{6}{|c|}{Average Values} &  0.47 &  1.04 & 31.69 & 31.69 & 32.67 & 32.60 \\ 
\multicolumn{6}{|c|}{Observed Standard Deviations} &  0.24 &  0.69 & 0.06 & 0.04 & 0.01 & 0.13 \\ 
\multicolumn{6}{|c|}{Expected Standard Deviations}  &  &  & 0.07 & 0.07 & 0.07 & 0.07 \\ 
\hline
N1546070861 & 2006-363T07:34:55 & 131.9$^\circ$ & -16.7$^\circ$ & [-0.80,-0.00] & 3.0 km &  0.70 &  1.24 & 29.78$\pm$0.16 (0.51) & 29.81$\pm$0.16 (0.40) & 30.58$\pm$0.17 (0.54) & 30.51$\pm$0.17 (0.49) \\ 
N1546071289 & 2006-363T07:42:03 & 131.9$^\circ$ & -16.6$^\circ$ & [-0.00, 0.80] & 3.1 km &  0.76 &  1.42 & 29.82$\pm$0.16 (0.50) & 29.86$\pm$0.16 (0.42) & 30.85$\pm$0.17 (0.57) & 31.07$\pm$0.17 (0.42) \\ 
\hline
\multicolumn{6}{|c|}{Average Values} &  0.73 &  1.33 & 29.80 & 29.84 & 30.72 & 30.79 \\ 
\multicolumn{6}{|c|}{Observed Standard Deviations} &  0.04 &  0.13 & 0.03 & 0.04 & 0.19 & 0.40 \\ 
\multicolumn{6}{|c|}{Expected Standard Deviations}  &  &  & 0.16 & 0.16 & 0.17 & 0.17 \\ 
%
\hline
N1550157993 & 2007-045T14:53:17 & 161.7$^\circ$ &  27.3$^\circ$ & [-0.20, 0.20] & 4.4 km &  0.22 &  0.82 & 29.68$\pm$0.24 (0.60) & 29.68$\pm$0.24 (0.40) & 30.54$\pm$0.24 (0.59) & 30.34$\pm$0.24 (0.49) \\ 
N1550158309 & 2007-045T14:58:33 & 161.7$^\circ$ &  27.2$^\circ$ & [-0.20, 0.20] & 4.4 km &  0.24 &  1.01 & 29.67$\pm$0.24 (0.60) & 29.68$\pm$0.24 (0.40) & 30.18$\pm$0.24 (0.63) & 30.16$\pm$0.24 (0.47) \\ 
N1550158625 & 2007-045T15:03:49 & 161.7$^\circ$ &  27.2$^\circ$ & [-0.20, 0.20] & 4.4 km &  0.23 &  0.81 & 29.67$\pm$0.24 (0.60) & 29.68$\pm$0.24 (0.39) & 30.17$\pm$0.24 (0.64) & 30.06$\pm$0.24 (0.54) \\ 
N1550158941 & 2007-045T15:09:05 & 161.7$^\circ$ &  27.1$^\circ$ & [-0.20, 0.20] & 4.4 km &  0.24 &  0.82 & 29.67$\pm$0.23 (0.60) & 29.68$\pm$0.23 (0.39) & 30.26$\pm$0.24 (0.63) & 30.18$\pm$0.24 (0.51) \\ 
N1550159257 & 2007-045T15:14:21 & 161.8$^\circ$ &  27.1$^\circ$ & [-0.20, 0.20] & 4.4 km &  0.22 &  0.84 & 29.68$\pm$0.23 (0.60) & 29.68$\pm$0.23 (0.39) & 30.30$\pm$0.24 (0.65) & 30.26$\pm$0.24 (0.46) \\ 
N1550159573 & 2007-045T15:19:37 & 161.8$^\circ$ &  27.0$^\circ$ & [-0.20, 0.20] & 4.4 km &  0.22 &  0.96 & 29.68$\pm$0.23 (0.60) & 29.68$\pm$0.23 (0.39) & 30.28$\pm$0.24 (0.63) & 30.22$\pm$0.24 (0.45) \\ 
N1550159889 & 2007-045T15:24:53 & 161.8$^\circ$ &  27.0$^\circ$ & [-0.20, 0.20] & 4.4 km &  0.24 &  0.94 & 29.67$\pm$0.23 (0.60) & 29.67$\pm$0.23 (0.39) & 30.28$\pm$0.24 (0.66) & 30.24$\pm$0.24 (0.47) \\ 
N1550160205 & 2007-045T15:30:09 & 161.8$^\circ$ &  26.9$^\circ$ & [-0.20, 0.20] & 4.4 km &  0.22 &  0.75 & 29.67$\pm$0.23 (0.60) & 29.67$\pm$0.23 (0.39) & 30.29$\pm$0.24 (0.67) & 30.25$\pm$0.24 (0.51) \\ 
N1550160521 & 2007-045T15:35:25 & 161.8$^\circ$ &  26.9$^\circ$ & [-0.20, 0.20] & 4.4 km &  0.21 &  0.85 & 29.68$\pm$0.23 (0.60) & 29.68$\pm$0.23 (0.39) & 30.19$\pm$0.24 (0.64) & 30.17$\pm$0.24 (0.47) \\ 
N1550160837 & 2007-045T15:40:41 & 161.8$^\circ$ &  26.8$^\circ$ & [-0.20, 0.20] & 4.4 km &  0.25 &  1.30 & 29.67$\pm$0.23 (0.60) & 29.67$\pm$0.23 (0.39) & 30.18$\pm$0.24 (0.64) & 30.14$\pm$0.24 (0.45) \\ 
N1550161153 & 2007-045T15:45:57 & 161.8$^\circ$ &  26.8$^\circ$ & [-0.20, 0.20] & 4.4 km &  0.26 &  0.98 & 29.69$\pm$0.23 (0.60) & 29.69$\pm$0.23 (0.40) & 30.25$\pm$0.24 (0.63) & 30.20$\pm$0.24 (0.46) \\ 
N1550161469 & 2007-045T15:51:13 & 161.8$^\circ$ &  26.7$^\circ$ & [-0.20, 0.20] & 4.4 km &  0.24 &  0.74 & 29.70$\pm$0.23 (0.61) & 29.69$\pm$0.23 (0.40) & 30.28$\pm$0.24 (0.64) & 30.22$\pm$0.24 (0.61) \\ 
N1550161785 & 2007-045T15:56:29 & 161.8$^\circ$ &  26.7$^\circ$ & [-0.20, 0.20] & 4.4 km &  0.23 &  0.69 & 29.71$\pm$0.23 (0.60) & 29.68$\pm$0.23 (0.40) & 30.28$\pm$0.24 (0.66) & 30.30$\pm$0.24 (0.59) \\ 
N1550162101 & 2007-045T16:01:45 & 161.8$^\circ$ &  26.7$^\circ$ & [-0.20, 0.20] & 4.4 km &  0.22 &  0.85 & 29.71$\pm$0.23 (0.62) & 29.68$\pm$0.23 (0.41) & 30.26$\pm$0.24 (0.60) & 30.19$\pm$0.24 (0.44) \\ 
N1550162417 & 2007-045T16:07:01 & 161.8$^\circ$ &  26.6$^\circ$ & [-0.20, 0.20] & 4.4 km &  0.23 &  0.81 & 29.71$\pm$0.23 (0.61) & 29.69$\pm$0.23 (0.39) & 30.25$\pm$0.24 (0.65) & 30.20$\pm$0.24 (0.47) \\ 
N1550162733 & 2007-045T16:12:17 & 161.8$^\circ$ &  26.6$^\circ$ & [-0.20, 0.20] & 4.4 km &  0.21 &  0.78 & 29.71$\pm$0.23 (0.61) & 29.69$\pm$0.23 (0.39) & 30.15$\pm$0.24 (0.60) & 30.11$\pm$0.24 (0.49) \\ 
N1550163049 & 2007-045T16:17:33 & 161.9$^\circ$ &  26.5$^\circ$ & [-0.20, 0.20] & 4.4 km &  0.24 &  0.82 & 29.71$\pm$0.23 (0.62) & 29.69$\pm$0.23 (0.39) & 30.21$\pm$0.24 (0.63) & 30.14$\pm$0.24 (0.54) \\ 
N1550163365 & 2007-045T16:22:49 & 161.9$^\circ$ &  26.5$^\circ$ & [-0.20, 0.20] & 4.4 km &  0.24 &  0.85 & 29.72$\pm$0.23 (0.62) & 29.70$\pm$0.23 (0.40) & 30.25$\pm$0.24 (0.64) & 30.20$\pm$0.24 (0.50) \\ 
N1550163681 & 2007-045T16:28:05 & 161.9$^\circ$ &  26.4$^\circ$ & [-0.20, 0.20] & 4.4 km &  0.22 &  0.86 & 29.72$\pm$0.23 (0.62) & 29.70$\pm$0.23 (0.39) & 30.18$\pm$0.24 (0.63) & 30.16$\pm$0.24 (0.48) \\ 
N1550163997 & 2007-045T16:33:21 & 161.9$^\circ$ &  26.4$^\circ$ & [-0.20, 0.20] & 4.4 km &  0.16 &  0.79 & 29.71$\pm$0.23 (0.61) & 29.70$\pm$0.23 (0.39) & 29.97$\pm$0.24 (0.57) & 30.02$\pm$0.24 (0.47) \\ 
N1550164313 & 2007-045T16:38:37 & 161.9$^\circ$ &  26.3$^\circ$ & [-0.20, 0.20] & 4.4 km &  0.22 &  0.84 & 29.71$\pm$0.23 (0.63) & 29.68$\pm$0.23 (0.39) & 30.23$\pm$0.24 (0.63) & 30.22$\pm$0.24 (0.47) \\ 
N1550164629 & 2007-045T16:43:53 & 161.9$^\circ$ &  26.3$^\circ$ & [-0.20, 0.20] & 4.4 km &  0.23 &  0.89 & 29.70$\pm$0.23 (0.64) & 29.68$\pm$0.23 (0.39) & 30.21$\pm$0.24 (0.63) & 30.22$\pm$0.24 (0.48) \\ 
N1550164945 & 2007-045T16:49:09 & 161.9$^\circ$ &  26.2$^\circ$ & [-0.20, 0.20] & 4.4 km &  0.23 &  0.80 & 29.70$\pm$0.23 (0.64) & 29.68$\pm$0.23 (0.39) & 30.28$\pm$0.24 (0.67) & 30.25$\pm$0.24 (0.48) \\ 
N1550165261 & 2007-045T16:54:25 & 161.9$^\circ$ &  26.2$^\circ$ & [-0.20, 0.20] & 4.4 km &  0.24 &  0.87 & 29.69$\pm$0.23 (0.64) & 29.68$\pm$0.23 (0.39) & 30.22$\pm$0.24 (0.65) & 30.18$\pm$0.24 (0.48) \\ 
N1550165577 & 2007-045T16:59:41 & 161.9$^\circ$ &  26.1$^\circ$ & [-0.20, 0.20] & 4.4 km &  0.23 &  0.86 & 29.71$\pm$0.23 (0.63) & 29.68$\pm$0.23 (0.39) & 30.23$\pm$0.24 (0.66) & 30.18$\pm$0.24 (0.47) \\ 
N1550165893 & 2007-045T17:04:57 & 161.9$^\circ$ &  26.1$^\circ$ & [-0.20, 0.20] & 4.4 km &  0.24 &  0.92 & 29.71$\pm$0.23 (0.64) & 29.69$\pm$0.23 (0.39) & 30.20$\pm$0.24 (0.64) & 30.17$\pm$0.24 (0.47) \\ 
N1550166209 & 2007-045T17:10:13 & 161.9$^\circ$ &  26.1$^\circ$ & [-0.20, 0.20] & 4.4 km &  0.24 &  0.86 & 29.71$\pm$0.23 (0.64) & 29.68$\pm$0.23 (0.39) & 30.26$\pm$0.24 (0.65) & 30.24$\pm$0.24 (0.47) \\ 
N1550166525 & 2007-045T17:15:29 & 161.9$^\circ$ &  26.0$^\circ$ & [-0.20, 0.20] & 4.4 km &  0.24 &  0.93 & 29.71$\pm$0.23 (0.64) & 29.69$\pm$0.23 (0.39) & 30.23$\pm$0.24 (0.63) & 30.21$\pm$0.24 (0.46) \\ 
N1550166841 & 2007-045T17:20:45 & 161.9$^\circ$ &  26.0$^\circ$ & [-0.20, 0.20] & 4.4 km &  0.26 &  1.02 & 29.70$\pm$0.23 (0.64) & 29.68$\pm$0.23 (0.39) & 30.17$\pm$0.24 (0.62) & 30.16$\pm$0.24 (0.46) \\ 
N1550167157 & 2007-045T17:26:01 & 161.9$^\circ$ &  25.9$^\circ$ & [-0.20, 0.20] & 4.4 km &  0.24 &  0.97 & 29.70$\pm$0.23 (0.63) & 29.68$\pm$0.23 (0.39) & 30.15$\pm$0.24 (0.62) & 30.15$\pm$0.24 (0.46) \\ 
N1550167473 & 2007-045T17:31:17 & 161.9$^\circ$ &  25.9$^\circ$ & [-0.20, 0.20] & 4.4 km &  0.24 &  0.91 & 29.69$\pm$0.23 (0.63) & 29.68$\pm$0.23 (0.39) & 30.16$\pm$0.24 (0.63) & 30.19$\pm$0.24 (0.47) \\ 
N1550167789 & 2007-045T17:36:33 & 161.9$^\circ$ &  25.8$^\circ$ & [-0.20, 0.20] & 4.4 km &  0.20 &  0.78 & 29.69$\pm$0.23 (0.62) & 29.68$\pm$0.23 (0.39) & 30.08$\pm$0.24 (0.59) & 30.11$\pm$0.24 (0.47) \\ 
N1550168105 & 2007-045T17:41:49 & 161.9$^\circ$ &  25.8$^\circ$ & [-0.20, 0.20] & 4.4 km &  0.25 &  0.79 & 29.68$\pm$0.23 (0.63) & 29.68$\pm$0.23 (0.39) & 30.19$\pm$0.24 (0.63) & 30.22$\pm$0.24 (0.52) \\ 
N1550168421 & 2007-045T17:47:05 & 161.9$^\circ$ &  25.7$^\circ$ & [-0.20, 0.20] & 4.4 km &  0.25 &  0.87 & 29.67$\pm$0.23 (0.62) & 29.66$\pm$0.23 (0.39) & 30.12$\pm$0.24 (0.62) & 30.17$\pm$0.24 (0.50) \\ 
N1550168737 & 2007-045T17:52:21 & 162.0$^\circ$ &  25.7$^\circ$ & [-0.20, 0.20] & 4.4 km &  0.18 &  0.66 & 29.68$\pm$0.23 (0.62) & 29.66$\pm$0.23 (0.39) & 30.16$\pm$0.24 (0.62) & 30.18$\pm$0.24 (0.48) \\ 
N1550169053 & 2007-045T17:57:37 & 162.0$^\circ$ &  25.6$^\circ$ & [-0.20, 0.20] & 4.4 km &  0.24 &  0.87 & 29.67$\pm$0.23 (0.63) & 29.66$\pm$0.23 (0.39) & 30.22$\pm$0.24 (0.63) & 30.21$\pm$0.24 (0.47) \\ 
N1550169369 & 2007-045T18:02:53 & 162.0$^\circ$ &  25.6$^\circ$ & [-0.20, 0.20] & 4.4 km &  0.19 &  0.76 & 29.67$\pm$0.23 (0.62) & 29.66$\pm$0.23 (0.38) & 30.14$\pm$0.24 (0.61) & 30.17$\pm$0.24 (0.48) \\ 
N1550169685 & 2007-045T18:08:09 & 162.0$^\circ$ &  25.5$^\circ$ & [-0.20, 0.20] & 4.4 km &  0.25 &  0.93 & 29.66$\pm$0.23 (0.63) & 29.66$\pm$0.23 (0.39) & 30.12$\pm$0.24 (0.64) & 30.17$\pm$0.24 (0.49) \\ 
N1550170001 & 2007-045T18:13:25 & 162.0$^\circ$ &  25.5$^\circ$ & [-0.20, 0.20] & 4.4 km &  0.22 &  0.75 & 29.67$\pm$0.23 (0.63) & 29.65$\pm$0.23 (0.39) & 30.23$\pm$0.24 (0.69) & 30.28$\pm$0.24 (0.51) \\ 
N1550170317 & 2007-045T18:18:41 & 162.0$^\circ$ &  25.4$^\circ$ & [-0.20, 0.20] & 4.4 km &  0.18 &  0.63 & 29.67$\pm$0.23 (0.64) & 29.64$\pm$0.23 (0.39) & 30.14$\pm$0.24 (0.64) & 30.14$\pm$0.24 (0.51) \\ 
N1550170633 & 2007-045T18:23:57 & 162.0$^\circ$ &  25.4$^\circ$ & [-0.20, 0.20] & 4.4 km &  0.21 &  0.75 & 29.67$\pm$0.23 (0.64) & 29.65$\pm$0.23 (0.39) & 30.20$\pm$0.24 (0.62) & 30.22$\pm$0.24 (0.48) \\ 
N1550170949 & 2007-045T18:29:13 & 162.0$^\circ$ &  25.3$^\circ$ & [-0.20, 0.20] & 4.4 km &  0.24 &  0.88 & 29.66$\pm$0.23 (0.63) & 29.64$\pm$0.23 (0.39) & 30.18$\pm$0.24 (0.63) & 30.21$\pm$0.24 (0.48) \\ 
N1550171265 & 2007-045T18:34:29 & 162.0$^\circ$ &  25.3$^\circ$ & [-0.20, 0.20] & 4.4 km &  0.21 &  0.84 & 29.66$\pm$0.23 (0.63) & 29.65$\pm$0.23 (0.39) & 30.12$\pm$0.24 (0.61) & 30.14$\pm$0.24 (0.47) \\ 
N1550171581 & 2007-045T18:39:45 & 162.0$^\circ$ &  25.3$^\circ$ & [-0.20, 0.20] & 4.4 km &  0.23 &  0.86 & 29.65$\pm$0.23 (0.63) & 29.64$\pm$0.23 (0.39) & 30.17$\pm$0.24 (0.63) & 30.21$\pm$0.24 (0.47) \\ 
N1550171897 & 2007-045T18:45:01 & 162.0$^\circ$ &  25.2$^\circ$ & [-0.20, 0.20] & 4.4 km &  0.19 &  0.64 & 29.65$\pm$0.23 (0.62) & 29.62$\pm$0.23 (0.38) & 30.23$\pm$0.24 (0.66) & 30.22$\pm$0.24 (0.48) \\ 
N1550172213 & 2007-045T18:50:17 & 162.0$^\circ$ &  25.2$^\circ$ & [-0.20, 0.20] & 4.4 km &  0.23 &  0.83 & 29.64$\pm$0.23 (0.62) & 29.63$\pm$0.23 (0.39) & 30.12$\pm$0.24 (0.63) & 30.14$\pm$0.24 (0.48) \\ 
N1550172529 & 2007-045T18:55:33 & 162.0$^\circ$ &  25.1$^\circ$ & [-0.20, 0.20] & 4.4 km &  0.25 &  0.85 & 29.63$\pm$0.23 (0.61) & 29.63$\pm$0.23 (0.39) & 30.15$\pm$0.24 (0.63) & 30.20$\pm$0.24 (0.49) \\ 
N1550172845 & 2007-045T19:00:49 & 162.0$^\circ$ &  25.1$^\circ$ & [-0.20, 0.20] & 4.4 km &  0.16 &  0.77 & 29.64$\pm$0.23 (0.59) & 29.64$\pm$0.23 (0.38) & 29.81$\pm$0.23 (0.54) & 29.95$\pm$0.23 (0.47) \\ 
N1550173161 & 2007-045T19:06:05 & 162.0$^\circ$ &  25.0$^\circ$ & [-0.20, 0.20] & 4.4 km &  0.25 &  0.90 & 29.64$\pm$0.23 (0.60) & 29.65$\pm$0.23 (0.39) & 30.17$\pm$0.24 (0.62) & 30.22$\pm$0.24 (0.48) \\ 
N1550173477 & 2007-045T19:11:21 & 162.0$^\circ$ &  25.0$^\circ$ & [-0.20, 0.20] & 4.4 km &  0.24 &  0.86 & 29.65$\pm$0.23 (0.60) & 29.65$\pm$0.23 (0.39) & 30.20$\pm$0.24 (0.62) & 30.23$\pm$0.24 (0.49) \\ 
N1550173793 & 2007-045T19:16:37 & 162.0$^\circ$ &  24.9$^\circ$ & [-0.20, 0.20] & 4.4 km &  0.24 &  0.83 & 29.65$\pm$0.23 (0.60) & 29.64$\pm$0.23 (0.39) & 30.16$\pm$0.24 (0.63) & 30.20$\pm$0.24 (0.50) \\ 
N1550174109 & 2007-045T19:21:53 & 162.0$^\circ$ &  24.9$^\circ$ & [-0.20, 0.20] & 4.4 km &  0.24 &  0.76 & 29.65$\pm$0.23 (0.59) & 29.64$\pm$0.23 (0.39) & 30.15$\pm$0.24 (0.64) & 30.18$\pm$0.24 (0.50) \\ 
N1550174425 & 2007-045T19:27:09 & 162.0$^\circ$ &  24.8$^\circ$ & [-0.20, 0.20] & 4.4 km &  0.21 &  0.74 & 29.65$\pm$0.23 (0.59) & 29.64$\pm$0.23 (0.39) & 30.16$\pm$0.24 (0.62) & 30.18$\pm$0.24 (0.48) \\ 
N1550174741 & 2007-045T19:32:25 & 162.0$^\circ$ &  24.8$^\circ$ & [-0.20, 0.20] & 4.4 km &  0.21 &  0.77 & 29.65$\pm$0.23 (0.58) & 29.65$\pm$0.23 (0.38) & 30.16$\pm$0.23 (0.60) & 30.17$\pm$0.23 (0.47) \\ 
N1550175057 & 2007-045T19:37:41 & 162.0$^\circ$ &  24.7$^\circ$ & [-0.20, 0.20] & 4.4 km &  0.24 &  0.83 & 29.64$\pm$0.23 (0.58) & 29.64$\pm$0.23 (0.39) & 30.16$\pm$0.23 (0.62) & 30.17$\pm$0.23 (0.49) \\ 
N1550175373 & 2007-045T19:42:57 & 162.0$^\circ$ &  24.7$^\circ$ & [-0.20, 0.20] & 4.4 km &  0.24 &  0.91 & 29.66$\pm$0.23 (0.56) & 29.67$\pm$0.23 (0.39) & 30.25$\pm$0.24 (0.56) & 30.33$\pm$0.24 (0.47) \\ 
N1550175689 & 2007-045T19:48:13 & 162.0$^\circ$ &  24.6$^\circ$ & [-0.20, 0.20] & 4.4 km &  0.24 &  0.82 & 29.64$\pm$0.23 (0.58) & 29.65$\pm$0.23 (0.39) & 30.18$\pm$0.24 (0.63) & 30.21$\pm$0.24 (0.49) \\ 
N1550176005 & 2007-045T19:53:29 & 162.0$^\circ$ &  24.6$^\circ$ & [-0.20, 0.20] & 4.4 km &  0.23 &  0.82 & 29.64$\pm$0.23 (0.58) & 29.65$\pm$0.23 (0.39) & 30.23$\pm$0.24 (0.64) & 30.23$\pm$0.24 (0.47) \\ 
N1550176321 & 2007-045T19:58:45 & 162.0$^\circ$ &  24.5$^\circ$ & [-0.20, 0.20] & 4.4 km &  0.24 &  0.95 & 29.64$\pm$0.23 (0.58) & 29.65$\pm$0.23 (0.39) & 30.20$\pm$0.24 (0.64) & 30.22$\pm$0.24 (0.47) \\ 
N1550176627 & 2007-045T20:03:51 & 162.0$^\circ$ &  24.5$^\circ$ & [-0.20, 0.20] & 4.4 km &  0.16 &  0.55 & 29.66$\pm$0.23 (0.58) & 29.64$\pm$0.23 (0.40) & 30.57$\pm$0.24 (0.59) & 30.43$\pm$0.24 (0.47) \\ 
\hline
\multicolumn{6}{|c|}{Average Values} &  0.23 &  0.84 & 29.68 & 29.67 & 30.20 & 30.19 \\ 
\multicolumn{6}{|c|}{Observed Standard Deviations} &  0.02 &  0.11 & 0.03 & 0.02 & 0.10 & 0.07 \\ 
\multicolumn{6}{|c|}{Expected Standard Deviations}  &  &  & 0.23 & 0.23 & 0.24 & 0.24 \\ 
\hline\end{tabular}}
\end{table}

\begin{table}
\caption{Detailed list of images used in this analysis (continued)}
\resizebox{6in}{!}{\begin{tabular}{|c|c|c|c|c|c|c|c|c|c|c|c|}\hline
Image & UTC time & Phase & $B$ & $\cos\phi/\tan B$ & Radial & $A_r$ & $A_z$ & $\Lambda_r$ from $dp_1/dr$ (km) & $\Lambda_r$ from $dp_1/dr(\bar{p}_1)^{-1}$ (km) & $\Lambda_z$ from $p_2$ (km) & $\Lambda_z$ from $p_2/p_1$ (km) \\
 Name &  & Angle &  & values fit & Res. & (km) & (km) & Value$\pm$Error (Width) & Value$\pm$Error (Width) & Value$\pm$Error (Width) & Value$\pm$Error (Width) \\
\hline
N1551798977 & 2007-064T14:42:52 & 161.3$^\circ$ &   8.0$^\circ$ & [-1.00, 1.00] & 2.9 km &  0.40 &  0.74 & 29.54$\pm$0.15 (0.57) & 29.59$\pm$0.15 (0.38) & 30.55$\pm$0.16 (0.60) & 30.47$\pm$0.16 (0.46) \\ 
N1551799369 & 2007-064T14:49:24 & 161.2$^\circ$ &   7.9$^\circ$ & [-1.00, 1.00] & 2.9 km &  0.37 &  0.70 & 29.56$\pm$0.15 (0.58) & 29.59$\pm$0.15 (0.39) & 30.55$\pm$0.16 (0.60) & 30.48$\pm$0.16 (0.46) \\ 
N1551799761 & 2007-064T14:55:56 & 161.1$^\circ$ &   7.7$^\circ$ & [-1.00, 1.00] & 2.9 km &  0.39 &  0.72 & 29.56$\pm$0.15 (0.58) & 29.60$\pm$0.15 (0.39) & 30.57$\pm$0.16 (0.59) & 30.50$\pm$0.16 (0.47) \\ 
N1551800153 & 2007-064T15:02:28 & 161.0$^\circ$ &   7.6$^\circ$ & [-1.00, 1.00] & 2.9 km &  0.38 &  0.70 & 29.53$\pm$0.15 (0.58) & 29.58$\pm$0.15 (0.39) & 30.55$\pm$0.16 (0.60) & 30.48$\pm$0.16 (0.47) \\ 
N1551800545 & 2007-064T15:09:00 & 160.9$^\circ$ &   7.5$^\circ$ & [-1.00, 1.00] & 2.9 km &  0.36 &  0.66 & 29.54$\pm$0.15 (0.58) & 29.58$\pm$0.15 (0.40) & 30.58$\pm$0.16 (0.60) & 30.49$\pm$0.16 (0.47) \\ 
N1551800937 & 2007-064T15:15:32 & 160.7$^\circ$ &   7.4$^\circ$ & [-1.00, 1.00] & 2.9 km &  0.36 &  0.64 & 29.53$\pm$0.15 (0.58) & 29.57$\pm$0.15 (0.40) & 30.58$\pm$0.16 (0.60) & 30.50$\pm$0.16 (0.48) \\ 
N1551801329 & 2007-064T15:22:04 & 160.6$^\circ$ &   7.2$^\circ$ & [-1.00, 1.00] & 2.9 km &  0.38 &  0.66 & 29.54$\pm$0.15 (0.59) & 29.57$\pm$0.15 (0.40) & 30.58$\pm$0.16 (0.60) & 30.51$\pm$0.16 (0.48) \\ 
N1551801721 & 2007-064T15:28:36 & 160.5$^\circ$ &   7.1$^\circ$ & [-1.00, 1.00] & 2.9 km &  0.35 &  0.61 & 29.54$\pm$0.15 (0.59) & 29.57$\pm$0.15 (0.40) & 30.57$\pm$0.16 (0.60) & 30.50$\pm$0.16 (0.48) \\ 
N1551802113 & 2007-064T15:35:08 & 160.4$^\circ$ &   7.0$^\circ$ & [-1.00, 1.00] & 2.9 km &  0.35 &  0.61 & 29.51$\pm$0.15 (0.58) & 29.56$\pm$0.15 (0.40) & 30.55$\pm$0.16 (0.60) & 30.48$\pm$0.16 (0.49) \\ 
N1551802505 & 2007-064T15:41:40 & 160.3$^\circ$ &   6.8$^\circ$ & [-1.00, 1.00] & 2.9 km &  0.37 &  0.63 & 29.52$\pm$0.15 (0.58) & 29.57$\pm$0.15 (0.41) & 30.57$\pm$0.16 (0.60) & 30.49$\pm$0.16 (0.48) \\ 
N1551802897 & 2007-064T15:48:12 & 160.1$^\circ$ &   6.7$^\circ$ & [-1.00, 1.00] & 2.9 km &  0.39 &  0.63 & 29.55$\pm$0.15 (0.58) & 29.59$\pm$0.15 (0.42) & 30.61$\pm$0.16 (0.60) & 30.54$\pm$0.16 (0.48) \\ 
N1551803289 & 2007-064T15:54:44 & 160.0$^\circ$ &   6.6$^\circ$ & [-1.00, 1.00] & 2.9 km &  0.37 &  0.63 & 29.51$\pm$0.15 (0.58) & 29.55$\pm$0.15 (0.41) & 30.55$\pm$0.16 (0.60) & 30.47$\pm$0.16 (0.49) \\ 
N1551803681 & 2007-064T16:01:16 & 159.9$^\circ$ &   6.4$^\circ$ & [-1.00, 1.00] & 2.9 km &  0.36 &  0.62 & 29.53$\pm$0.15 (0.58) & 29.56$\pm$0.15 (0.41) & 30.55$\pm$0.16 (0.60) & 30.48$\pm$0.16 (0.49) \\ 
N1551804073 & 2007-064T16:07:48 & 159.8$^\circ$ &   6.3$^\circ$ & [-1.00, 1.00] & 2.9 km &  0.36 &  0.60 & 29.56$\pm$0.15 (0.58) & 29.58$\pm$0.15 (0.42) & 30.60$\pm$0.16 (0.60) & 30.52$\pm$0.16 (0.48) \\ 
N1551804465 & 2007-064T16:14:20 & 159.6$^\circ$ &   6.2$^\circ$ & [-1.00, 1.00] & 2.9 km &  0.36 &  0.60 & 29.55$\pm$0.15 (0.58) & 29.57$\pm$0.15 (0.42) & 30.58$\pm$0.16 (0.60) & 30.50$\pm$0.16 (0.49) \\ 
N1551804857 & 2007-064T16:20:52 & 159.5$^\circ$ &   6.0$^\circ$ & [-1.00, 1.00] & 2.9 km &  0.35 &  0.58 & 29.53$\pm$0.15 (0.58) & 29.55$\pm$0.15 (0.42) & 30.58$\pm$0.16 (0.60) & 30.50$\pm$0.16 (0.49) \\ 
N1551805249 & 2007-064T16:27:24 & 159.4$^\circ$ &   5.9$^\circ$ & [-1.00, 1.00] & 2.9 km &  0.35 &  0.59 & 29.54$\pm$0.15 (0.58) & 29.55$\pm$0.15 (0.42) & 30.58$\pm$0.16 (0.61) & 30.51$\pm$0.16 (0.49) \\ 
N1551805641 & 2007-064T16:33:56 & 159.3$^\circ$ &   5.8$^\circ$ & [-1.00, 1.00] & 2.9 km &  0.33 &  0.56 & 29.53$\pm$0.15 (0.57) & 29.55$\pm$0.15 (0.41) & 30.57$\pm$0.16 (0.60) & 30.49$\pm$0.16 (0.50) \\ 
N1551806025 & 2007-064T16:40:22 & 159.1$^\circ$ &   5.6$^\circ$ & [-1.00, 1.00] & 2.9 km &  0.31 &  0.51 & 29.56$\pm$0.15 (0.59) & 29.57$\pm$0.15 (0.43) & 30.60$\pm$0.16 (0.61) & 30.52$\pm$0.16 (0.50) \\ 
\hline
\multicolumn{6}{|c|}{Average Values} &  0.36 &  0.63 & 29.54 & 29.57 & 30.57 & 30.50 \\ 
\multicolumn{6}{|c|}{Observed Standard Deviations} &  0.02 &  0.06 & 0.01 & 0.01 & 0.02 & 0.02 \\ 
\multicolumn{6}{|c|}{Expected Standard Deviations}  &  &  & 0.15 & 0.15 & 0.16 & 0.16 \\ 
%
\hline
N1656912493 & 2010-185T04:42:31 & 140.4$^\circ$ & -18.4$^\circ$ & [-0.20, 0.20] & 2.5 km &  0.34 &  1.15 & 26.02$\pm$0.12 (0.41) & 26.08$\pm$0.12 (0.32) & 26.85$\pm$0.12 (0.40) & 26.86$\pm$0.12 (0.35) \\ 
N1656912887 & 2010-185T04:49:05 & 140.5$^\circ$ & -18.4$^\circ$ & [-0.20, 0.20] & 2.5 km &  0.30 &  0.98 & 26.02$\pm$0.12 (0.41) & 26.10$\pm$0.12 (0.32) & 26.85$\pm$0.12 (0.41) & 26.86$\pm$0.12 (0.35) \\ 
N1656913281 & 2010-185T04:55:39 & 140.6$^\circ$ & -18.4$^\circ$ & [-0.20, 0.20] & 2.5 km &  0.35 &  1.12 & 26.02$\pm$0.11 (0.41) & 26.08$\pm$0.12 (0.32) & 26.86$\pm$0.12 (0.41) & 26.85$\pm$0.12 (0.35) \\ 
N1656913675 & 2010-185T05:02:13 & 140.8$^\circ$ & -18.5$^\circ$ & [-0.20, 0.20] & 2.5 km &  0.33 &  1.10 & 26.02$\pm$0.11 (0.42) & 26.07$\pm$0.11 (0.32) & 26.91$\pm$0.12 (0.40) & 26.87$\pm$0.12 (0.35) \\ 
N1656914069 & 2010-185T05:08:47 & 140.9$^\circ$ & -18.5$^\circ$ & [-0.20, 0.20] & 2.5 km &  0.35 &  1.15 & 26.03$\pm$0.11 (0.42) & 26.07$\pm$0.11 (0.32) & 26.92$\pm$0.12 (0.40) & 26.84$\pm$0.12 (0.36) \\ 
N1656914463 & 2010-185T05:15:21 & 141.0$^\circ$ & -18.5$^\circ$ & [-0.20, 0.20] & 2.5 km &  0.33 &  1.15 & 26.03$\pm$0.11 (0.41) & 26.08$\pm$0.11 (0.32) & 26.84$\pm$0.12 (0.40) & 26.82$\pm$0.12 (0.35) \\ 
N1656914857 & 2010-185T05:21:55 & 141.2$^\circ$ & -18.5$^\circ$ & [-0.20, 0.20] & 2.5 km &  0.36 &  1.14 & 26.02$\pm$0.11 (0.41) & 26.08$\pm$0.11 (0.32) & 26.74$\pm$0.12 (0.43) & 26.84$\pm$0.12 (0.35) \\ 
N1656915251 & 2010-185T05:28:29 & 141.3$^\circ$ & -18.5$^\circ$ & [-0.20, 0.20] & 2.4 km &  0.37 &  1.15 & 26.04$\pm$0.11 (0.41) & 26.09$\pm$0.11 (0.33) & 26.85$\pm$0.12 (0.40) & 26.84$\pm$0.12 (0.35) \\ 
N1656915645 & 2010-185T05:35:03 & 141.4$^\circ$ & -18.5$^\circ$ & [-0.20, 0.20] & 2.4 km &  0.37 &  1.20 & 26.06$\pm$0.11 (0.41) & 26.09$\pm$0.11 (0.33) & 26.85$\pm$0.12 (0.40) & 26.83$\pm$0.12 (0.35) \\ 
N1656916039 & 2010-185T05:41:37 & 141.5$^\circ$ & -18.5$^\circ$ & [-0.20, 0.20] & 2.4 km &  0.33 &  1.07 & 26.07$\pm$0.11 (0.41) & 26.11$\pm$0.11 (0.33) & 26.86$\pm$0.12 (0.41) & 26.82$\pm$0.12 (0.36) \\ 
N1656916433 & 2010-185T05:48:11 & 141.7$^\circ$ & -18.5$^\circ$ & [-0.20, 0.20] & 2.4 km &  0.37 &  1.09 & 26.07$\pm$0.11 (0.42) & 26.09$\pm$0.11 (0.33) & 26.85$\pm$0.12 (0.41) & 26.83$\pm$0.12 (0.36) \\ 
N1656916827 & 2010-185T05:54:45 & 141.8$^\circ$ & -18.5$^\circ$ & [-0.20, 0.20] & 2.4 km &  0.39 &  1.21 & 26.08$\pm$0.11 (0.42) & 26.10$\pm$0.11 (0.32) & 26.84$\pm$0.12 (0.41) & 26.84$\pm$0.12 (0.35) \\ 
N1656917221 & 2010-185T06:01:19 & 141.9$^\circ$ & -18.5$^\circ$ & [-0.20, 0.20] & 2.4 km &  0.34 &  1.06 & 26.10$\pm$0.11 (0.42) & 26.12$\pm$0.11 (0.33) & 26.88$\pm$0.11 (0.40) & 26.86$\pm$0.11 (0.35) \\ 
N1656917615 & 2010-185T06:07:53 & 142.1$^\circ$ & -18.6$^\circ$ & [-0.20, 0.20] & 2.4 km &  0.39 &  1.20 & 26.12$\pm$0.11 (0.42) & 26.13$\pm$0.11 (0.33) & 26.86$\pm$0.11 (0.41) & 26.85$\pm$0.11 (0.36) \\ 
N1656918009 & 2010-185T06:14:27 & 142.2$^\circ$ & -18.6$^\circ$ & [-0.20, 0.20] & 2.4 km &  0.38 &  1.19 & 26.15$\pm$0.11 (0.43) & 26.15$\pm$0.11 (0.33) & 26.90$\pm$0.11 (0.41) & 26.81$\pm$0.11 (0.36) \\ 
N1656918403 & 2010-185T06:21:01 & 142.3$^\circ$ & -18.6$^\circ$ & [-0.20, 0.20] & 2.4 km &  0.38 &  1.19 & 26.13$\pm$0.11 (0.43) & 26.12$\pm$0.11 (0.33) & 26.84$\pm$0.11 (0.41) & 26.84$\pm$0.11 (0.35) \\ 
N1656918797 & 2010-185T06:27:35 & 142.5$^\circ$ & -18.6$^\circ$ & [-0.20, 0.20] & 2.4 km &  0.37 &  1.17 & 26.13$\pm$0.11 (0.43) & 26.13$\pm$0.11 (0.33) & 26.79$\pm$0.11 (0.42) & 26.86$\pm$0.11 (0.34) \\ 
N1656919191 & 2010-185T06:34:09 & 142.6$^\circ$ & -18.6$^\circ$ & [-0.20, 0.20] & 2.4 km &  0.38 &  1.17 & 26.15$\pm$0.11 (0.43) & 26.15$\pm$0.11 (0.33) & 26.84$\pm$0.11 (0.42) & 26.85$\pm$0.11 (0.36) \\ 
N1656919585 & 2010-185T06:40:43 & 142.7$^\circ$ & -18.6$^\circ$ & [-0.20, 0.20] & 2.4 km &  0.38 &  1.22 & 26.16$\pm$0.11 (0.44) & 26.16$\pm$0.11 (0.33) & 26.85$\pm$0.11 (0.41) & 26.86$\pm$0.11 (0.36) \\ 
N1656919979 & 2010-185T06:47:17 & 142.9$^\circ$ & -18.6$^\circ$ & [-0.20, 0.20] & 2.4 km &  0.37 &  1.12 & 26.17$\pm$0.11 (0.45) & 26.17$\pm$0.11 (0.35) & 26.87$\pm$0.11 (0.42) & 26.86$\pm$0.11 (0.35) \\ 
N1656920373 & 2010-185T06:53:51 & 143.0$^\circ$ & -18.6$^\circ$ & [-0.20, 0.20] & 2.3 km &  0.36 &  1.06 & 26.18$\pm$0.11 (0.46) & 26.18$\pm$0.11 (0.34) & 26.90$\pm$0.11 (0.41) & 26.83$\pm$0.11 (0.37) \\ 
N1656920767 & 2010-185T07:00:25 & 143.1$^\circ$ & -18.6$^\circ$ & [-0.20, 0.20] & 2.3 km &  0.40 &  1.17 & 26.17$\pm$0.11 (0.46) & 26.17$\pm$0.11 (0.34) & 26.82$\pm$0.11 (0.41) & 26.83$\pm$0.11 (0.35) \\ 
N1656921161 & 2010-185T07:06:59 & 143.3$^\circ$ & -18.6$^\circ$ & [-0.20, 0.20] & 2.3 km &  0.40 &  1.12 & 26.18$\pm$0.11 (0.46) & 26.16$\pm$0.11 (0.34) & 26.90$\pm$0.11 (0.41) & 26.84$\pm$0.11 (0.37) \\ 
N1656921555 & 2010-185T07:13:33 & 143.4$^\circ$ & -18.7$^\circ$ & [-0.20, 0.20] & 2.3 km &  0.38 &  1.10 & 26.17$\pm$0.11 (0.46) & 26.16$\pm$0.11 (0.33) & 26.92$\pm$0.11 (0.40) & 26.85$\pm$0.11 (0.37) \\ 
N1656921949 & 2010-185T07:20:07 & 143.6$^\circ$ & -18.7$^\circ$ & [-0.20, 0.20] & 2.3 km &  0.40 &  1.17 & 26.16$\pm$0.11 (0.45) & 26.16$\pm$0.11 (0.33) & 26.80$\pm$0.11 (0.42) & 26.87$\pm$0.11 (0.35) \\ 
N1656922343 & 2010-185T07:26:41 & 143.7$^\circ$ & -18.7$^\circ$ & [-0.20, 0.20] & 2.3 km &  0.40 &  1.15 & 26.17$\pm$0.11 (0.45) & 26.17$\pm$0.11 (0.33) & 26.80$\pm$0.11 (0.42) & 26.87$\pm$0.11 (0.35) \\ 
N1656922737 & 2010-185T07:33:15 & 143.8$^\circ$ & -18.7$^\circ$ & [-0.20, 0.20] & 2.3 km &  0.40 &  1.11 & 26.15$\pm$0.11 (0.46) & 26.15$\pm$0.11 (0.34) & 26.88$\pm$0.11 (0.40) & 26.84$\pm$0.11 (0.36) \\ 
N1656923131 & 2010-185T07:39:49 & 144.0$^\circ$ & -18.7$^\circ$ & [-0.20, 0.20] & 2.3 km &  0.39 &  1.15 & 26.14$\pm$0.11 (0.46) & 26.14$\pm$0.11 (0.33) & 26.84$\pm$0.11 (0.41) & 26.87$\pm$0.11 (0.35) \\ 
N1656923525 & 2010-185T07:46:23 & 144.1$^\circ$ & -18.7$^\circ$ & [-0.20, 0.20] & 2.3 km &  0.40 &  1.18 & 26.14$\pm$0.11 (0.46) & 26.14$\pm$0.11 (0.34) & 26.80$\pm$0.11 (0.41) & 26.87$\pm$0.11 (0.35) \\ 
N1656923919 & 2010-185T07:52:57 & 144.3$^\circ$ & -18.7$^\circ$ & [-0.20, 0.20] & 2.3 km &  0.32 &  1.03 & 26.14$\pm$0.11 (0.46) & 26.15$\pm$0.11 (0.33) & 26.87$\pm$0.11 (0.40) & 26.87$\pm$0.11 (0.36) \\ 
N1656924313 & 2010-185T07:59:31 & 144.4$^\circ$ & -18.7$^\circ$ & [-0.20, 0.20] & 2.3 km &  0.39 &  1.15 & 26.12$\pm$0.10 (0.46) & 26.13$\pm$0.10 (0.33) & 26.91$\pm$0.11 (0.41) & 26.86$\pm$0.11 (0.37) \\ 
N1656924707 & 2010-185T08:06:05 & 144.6$^\circ$ & -18.7$^\circ$ & [-0.20, 0.20] & 2.3 km &  0.39 &  1.18 & 26.10$\pm$0.10 (0.45) & 26.12$\pm$0.10 (0.33) & 26.74$\pm$0.11 (0.44) & 26.85$\pm$0.11 (0.34) \\ 
N1656925101 & 2010-185T08:12:39 & 144.7$^\circ$ & -18.7$^\circ$ & [-0.20, 0.20] & 2.2 km &  0.34 &  0.97 & 26.09$\pm$0.10 (0.44) & 26.11$\pm$0.10 (0.32) & 26.91$\pm$0.11 (0.40) & 26.91$\pm$0.11 (0.35) \\ 
N1656925495 & 2010-185T08:19:13 & 144.9$^\circ$ & -18.7$^\circ$ & [-0.20, 0.20] & 2.2 km &  0.40 &  1.14 & 26.08$\pm$0.10 (0.45) & 26.10$\pm$0.10 (0.32) & 26.90$\pm$0.11 (0.41) & 26.86$\pm$0.11 (0.37) \\ 
N1656925889 & 2010-185T08:25:47 & 145.0$^\circ$ & -18.8$^\circ$ & [-0.20, 0.20] & 2.2 km &  0.38 &  1.10 & 26.06$\pm$0.10 (0.44) & 26.09$\pm$0.10 (0.32) & 26.84$\pm$0.11 (0.42) & 26.87$\pm$0.11 (0.35) \\ 
N1656926283 & 2010-185T08:32:21 & 145.2$^\circ$ & -18.8$^\circ$ & [-0.20, 0.20] & 2.2 km &  0.40 &  1.17 & 26.05$\pm$0.10 (0.43) & 26.09$\pm$0.10 (0.32) & 26.74$\pm$0.11 (0.44) & 26.84$\pm$0.11 (0.34) \\ 
N1656926677 & 2010-185T08:38:55 & 145.3$^\circ$ & -18.8$^\circ$ & [-0.20, 0.20] & 2.2 km &  0.38 &  1.13 & 26.06$\pm$0.10 (0.44) & 26.09$\pm$0.10 (0.32) & 26.87$\pm$0.11 (0.42) & 26.85$\pm$0.11 (0.36) \\ 
N1656927071 & 2010-185T08:45:29 & 145.5$^\circ$ & -18.8$^\circ$ & [-0.20, 0.20] & 2.2 km &  0.30 &  0.80 & 26.05$\pm$0.10 (0.44) & 26.09$\pm$0.10 (0.31) & 26.98$\pm$0.11 (0.41) & 26.68$\pm$0.10 (0.40) \\ 
N1656927465 & 2010-185T08:52:03 & 145.7$^\circ$ & -18.8$^\circ$ & [-0.20, 0.20] & 2.2 km &  0.38 &  1.14 & 26.04$\pm$0.10 (0.45) & 26.08$\pm$0.10 (0.32) & 26.88$\pm$0.10 (0.41) & 26.87$\pm$0.10 (0.36) \\ 
N1656927859 & 2010-185T08:58:37 & 145.8$^\circ$ & -18.8$^\circ$ & [-0.20, 0.20] & 2.2 km &  0.32 &  1.02 & 26.05$\pm$0.10 (0.45) & 26.10$\pm$0.10 (0.32) & 26.91$\pm$0.10 (0.42) & 26.84$\pm$0.10 (0.38) \\ 
N1656928253 & 2010-185T09:05:11 & 146.0$^\circ$ & -18.8$^\circ$ & [-0.20, 0.20] & 2.2 km &  0.37 &  1.20 & 26.03$\pm$0.10 (0.45) & 26.08$\pm$0.10 (0.33) & 26.90$\pm$0.10 (0.41) & 26.87$\pm$0.10 (0.36) \\ 
N1656928647 & 2010-185T09:11:45 & 146.1$^\circ$ & -18.8$^\circ$ & [-0.20, 0.20] & 2.2 km &  0.35 &  1.14 & 26.03$\pm$0.10 (0.45) & 26.08$\pm$0.10 (0.33) & 26.83$\pm$0.10 (0.42) & 26.84$\pm$0.10 (0.35) \\ 
N1656929041 & 2010-185T09:18:19 & 146.3$^\circ$ & -18.8$^\circ$ & [-0.20, 0.20] & 2.2 km &  0.34 &  1.01 & 26.02$\pm$0.10 (0.45) & 26.08$\pm$0.10 (0.33) & 26.85$\pm$0.10 (0.42) & 26.83$\pm$0.10 (0.36) \\ 
N1656929435 & 2010-185T09:24:53 & 146.4$^\circ$ & -18.8$^\circ$ & [-0.20, 0.20] & 2.2 km &  0.39 &  1.15 & 26.01$\pm$0.10 (0.44) & 26.06$\pm$0.10 (0.33) & 26.87$\pm$0.10 (0.41) & 26.85$\pm$0.10 (0.36) \\ 
N1656929829 & 2010-185T09:31:27 & 146.6$^\circ$ & -18.8$^\circ$ & [-0.20, 0.20] & 2.1 km &  0.38 &  1.14 & 26.01$\pm$0.10 (0.44) & 26.06$\pm$0.10 (0.32) & 26.89$\pm$0.10 (0.42) & 26.85$\pm$0.10 (0.36) \\ 
N1656930223 & 2010-185T09:38:01 & 146.8$^\circ$ & -18.9$^\circ$ & [-0.20, 0.20] & 2.1 km &  0.38 &  1.10 & 26.01$\pm$0.10 (0.43) & 26.05$\pm$0.10 (0.32) & 26.85$\pm$0.10 (0.42) & 26.84$\pm$0.10 (0.36) \\ 
N1656930617 & 2010-185T09:44:35 & 146.9$^\circ$ & -18.9$^\circ$ & [-0.20, 0.20] & 2.1 km &  0.38 &  1.15 & 26.00$\pm$0.10 (0.42) & 26.06$\pm$0.10 (0.32) & 26.69$\pm$0.10 (0.46) & 26.81$\pm$0.10 (0.34) \\ 
N1656931011 & 2010-185T09:51:09 & 147.1$^\circ$ & -18.9$^\circ$ & [-0.20, 0.20] & 2.1 km &  0.37 &  1.13 & 26.00$\pm$0.10 (0.43) & 26.06$\pm$0.10 (0.32) & 26.89$\pm$0.10 (0.41) & 26.85$\pm$0.10 (0.35) \\ 
N1656931405 & 2010-185T09:57:43 & 147.3$^\circ$ & -18.9$^\circ$ & [-0.20, 0.20] & 2.1 km &  0.37 &  1.19 & 26.03$\pm$0.10 (0.44) & 26.12$\pm$0.10 (0.33) & 26.93$\pm$0.10 (0.41) & 26.85$\pm$0.10 (0.36) \\ 
N1656931799 & 2010-185T10:04:17 & 147.4$^\circ$ & -18.9$^\circ$ & [-0.20, 0.20] & 2.1 km &  0.36 &  1.15 & 26.00$\pm$0.10 (0.43) & 26.07$\pm$0.10 (0.32) & 26.81$\pm$0.10 (0.42) & 26.84$\pm$0.10 (0.34) \\ 
\hline
\multicolumn{6}{|c|}{Average Values} &  0.37 &  1.12 & 26.08 & 26.11 & 26.86 & 26.85 \\ 
\multicolumn{6}{|c|}{Observed Standard Deviations} &  0.03 &  0.07 & 0.06 & 0.04 & 0.05 & 0.03 \\ 
\multicolumn{6}{|c|}{Expected Standard Deviations}  &  &  & 0.11 & 0.11 & 0.11 & 0.11 \\ 
\hline\end{tabular}}
\end{table}

\begin{table}
\caption{Detailed list of images used in this analysis (continued)}
\label{obstab3}
\resizebox{6in}{!}{\begin{tabular}{|c|c|c|c|c|c|c|c|c|c|c|c|}\hline
Image & UTC time & Phase & $B$ & $\cos\phi/\tan B$ & Radial & $A_r$ & $A_z$ & $\Lambda_r$ from $dp_1/dr$ (km) & $\Lambda_r$ from $dp_1/dr(\bar{p}_1)^{-1}$ (km) & $\Lambda_z$ from $p_2$ (km) & $\Lambda_z$ from $p_2/p_1$ (km) \\
 Name &  & Angle &  & values fit & Res. & (km) & (km) & Value$\pm$Error (Width) & Value$\pm$Error (Width) & Value$\pm$Error (Width) & Value$\pm$Error (Width) \\
\hline
N1719548905 & 2012-180T03:36:09 & 155.0$^\circ$ & -18.4$^\circ$ & [-0.20, 0.20] & 2.1 km &  0.43 &  0.86 & 24.19$\pm$0.09 (0.37) & 24.16$\pm$0.09 (0.27) & 24.92$\pm$0.09 (0.38) & 24.95$\pm$0.09 (0.35) \\ 
N1719550225 & 2012-180T03:58:09 & 143.9$^\circ$ & -18.1$^\circ$ & [ 0.10, 0.50] & 2.1 km &  0.51 &  1.12 & 24.24$\pm$0.09 (0.39) & 24.40$\pm$0.09 (0.32) & 25.08$\pm$0.10 (0.38) & 25.05$\pm$0.10 (0.34) \\ 
N1719564625 & 2012-180T07:58:09 & 151.5$^\circ$ & -16.4$^\circ$ & [-0.20, 0.20] & 1.8 km &  0.48 &  1.19 & 24.37$\pm$0.08 (0.37) & 24.45$\pm$0.08 (0.31) & 24.99$\pm$0.08 (0.38) & 25.00$\pm$0.08 (0.33) \\ 
\hline
\multicolumn{6}{|c|}{Average Values} &  0.47 &  1.06 & 24.27 & 24.34 & 25.00 & 25.00 \\ 
\multicolumn{6}{|c|}{Observed Standard Deviations} &  0.04 &  0.17 & 0.10 & 0.15 & 0.08 & 0.05 \\ 
\multicolumn{6}{|c|}{Expected Standard Deviations}  &  &  & 0.09 & 0.09 & 0.09 & 0.09 \\ 
\hline
N1729213871 & 2012-292T00:17:53 & 139.0$^\circ$ &  12.7$^\circ$ & [-0.20, 0.40] & 1.6 km &  0.58 &  1.14 & 24.01$\pm$0.07 (0.36) & 23.97$\pm$0.07 (0.26) & 24.50$\pm$0.07 (0.41) & 24.62$\pm$0.07 (0.36) \\ 
N1729220822 & 2012-292T02:13:44 & 131.2$^\circ$ &  17.6$^\circ$ & [-0.20, 0.20] & 1.5 km &  0.62 &  1.62 & 23.96$\pm$0.06 (0.34) & 23.95$\pm$0.06 (0.26) & 24.60$\pm$0.07 (0.41) & 24.44$\pm$0.07 (0.32) \\ 
N1729221097 & 2012-292T02:18:19 & 130.9$^\circ$ &  17.8$^\circ$ & [-0.20, 0.20] & 1.5 km &  0.63 &  1.81 & 23.96$\pm$0.06 (0.33) & 23.95$\pm$0.06 (0.26) & 24.54$\pm$0.07 (0.45) & 24.39$\pm$0.07 (0.32) \\ 
N1729221372 & 2012-292T02:22:54 & 130.6$^\circ$ &  18.0$^\circ$ & [-0.20, 0.20] & 1.5 km &  0.64 &  1.75 & 23.96$\pm$0.06 (0.33) & 23.96$\pm$0.06 (0.26) & 24.59$\pm$0.07 (0.44) & 24.41$\pm$0.07 (0.33) \\ 
N1729221647 & 2012-292T02:27:29 & 130.2$^\circ$ &  18.2$^\circ$ & [-0.20, 0.20] & 1.5 km &  0.62 &  1.68 & 23.96$\pm$0.06 (0.33) & 23.96$\pm$0.06 (0.26) & 24.60$\pm$0.07 (0.42) & 24.39$\pm$0.06 (0.31) \\ 
N1729221922 & 2012-292T02:32:04 & 129.9$^\circ$ &  18.4$^\circ$ & [-0.15, 0.20] & 1.5 km &  0.62 &  1.68 & 23.96$\pm$0.06 (0.32) & 23.96$\pm$0.06 (0.26) & 24.42$\pm$0.06 (0.45) & 24.32$\pm$0.06 (0.31) \\ 
N1729222197 & 2012-292T02:36:39 & 129.6$^\circ$ &  18.6$^\circ$ & [-0.15, 0.15] & 1.5 km &  0.64 &  1.72 & 23.96$\pm$0.06 (0.32) & 23.97$\pm$0.06 (0.25) & 24.58$\pm$0.07 (0.48) & 24.38$\pm$0.06 (0.33) \\ 
N1729222472 & 2012-292T02:41:14 & 129.3$^\circ$ &  18.8$^\circ$ & [-0.15, 0.15] & 1.5 km &  0.64 &  2.06 & 23.97$\pm$0.06 (0.33) & 23.97$\pm$0.06 (0.26) & 24.61$\pm$0.07 (0.36) & 24.45$\pm$0.06 (0.30) \\ 
N1729222747 & 2012-292T02:45:49 & 128.9$^\circ$ &  19.0$^\circ$ & [-0.15, 0.15] & 1.5 km &  0.68 &  2.05 & 23.96$\pm$0.06 (0.32) & 23.96$\pm$0.06 (0.26) & 24.50$\pm$0.06 (0.47) & 24.35$\pm$0.06 (0.30) \\ 
N1729223022 & 2012-292T02:50:24 & 128.6$^\circ$ &  19.2$^\circ$ & [-0.15, 0.15] & 1.5 km &  0.57 &  1.48 & 23.94$\pm$0.06 (0.34) & 23.96$\pm$0.06 (0.26) & 24.56$\pm$0.07 (0.88) & 24.40$\pm$0.06 (0.30) \\ 
N1729223297 & 2012-292T02:54:59 & 128.3$^\circ$ &  19.4$^\circ$ & [-0.15, 0.15] & 1.5 km &  0.64 &  2.22 & 23.97$\pm$0.06 (0.33) & 23.97$\pm$0.06 (0.26) & 24.49$\pm$0.06 (0.42) & 24.36$\pm$0.06 (0.31) \\ 
\hline
\multicolumn{6}{|c|}{Average Values} &  0.63 &  1.75 & 23.96 & 23.96 & 24.54 & 24.41 \\ 
\multicolumn{6}{|c|}{Observed Standard Deviations} &  0.03 &  0.30 & 0.02 & 0.01 & 0.06 & 0.08 \\ 
\multicolumn{6}{|c|}{Expected Standard Deviations}  &  &  & 0.06 & 0.06 & 0.07 & 0.07 \\ 
\hline
N1731280660 & 2012-315T22:24:09 & 140.6$^\circ$ &  11.8$^\circ$ & [-0.40, 0.40] & 1.6 km &  0.45 &  1.10 & 23.84$\pm$0.07 (0.33) & 23.90$\pm$0.07 (0.24) & 24.45$\pm$0.07 (0.35) & 24.53$\pm$0.07 (0.32) \\ 
N1731280997 & 2012-315T22:29:46 & 140.3$^\circ$ &  12.0$^\circ$ & [-0.40, 0.40] & 1.6 km &  0.44 &  1.15 & 23.84$\pm$0.07 (0.33) & 23.90$\pm$0.07 (0.24) & 24.52$\pm$0.07 (0.35) & 24.56$\pm$0.07 (0.31) \\ 
N1731281334 & 2012-315T22:35:23 & 139.9$^\circ$ &  12.3$^\circ$ & [-0.40, 0.30] & 1.6 km &  0.45 &  1.15 & 23.82$\pm$0.07 (0.33) & 23.88$\pm$0.07 (0.24) & 24.43$\pm$0.07 (0.36) & 24.51$\pm$0.07 (0.32) \\ 
N1731281671 & 2012-315T22:41:00 & 139.6$^\circ$ &  12.5$^\circ$ & [-0.20, 0.30] & 1.6 km &  0.45 &  1.15 & 23.81$\pm$0.07 (0.34) & 23.88$\pm$0.07 (0.24) & 24.47$\pm$0.07 (0.37) & 24.56$\pm$0.07 (0.32) \\ 
N1731282008 & 2012-315T22:46:37 & 139.2$^\circ$ &  12.7$^\circ$ & [-0.40, 0.30] & 1.6 km &  0.42 &  1.10 & 23.80$\pm$0.07 (0.34) & 23.89$\pm$0.07 (0.24) & 24.44$\pm$0.07 (0.38) & 24.54$\pm$0.07 (0.32) \\ 
N1731282345 & 2012-315T22:52:14 & 138.8$^\circ$ &  13.0$^\circ$ & [-0.40, 0.30] & 1.6 km &  0.40 &  1.16 & 23.77$\pm$0.07 (0.33) & 23.85$\pm$0.07 (0.25) & 24.47$\pm$0.07 (0.35) & 24.52$\pm$0.07 (0.32) \\ 
N1731282682 & 2012-315T22:57:51 & 138.5$^\circ$ &  13.2$^\circ$ & [-0.30, 0.30] & 1.6 km &  0.44 &  1.24 & 23.79$\pm$0.07 (0.34) & 23.90$\pm$0.07 (0.24) & 24.35$\pm$0.07 (0.39) & 24.44$\pm$0.07 (0.33) \\ 
N1731283019 & 2012-315T23:03:28 & 138.1$^\circ$ &  13.5$^\circ$ & [-0.30, 0.30] & 1.6 km &  0.43 &  1.23 & 23.78$\pm$0.07 (0.34) & 23.89$\pm$0.07 (0.24) & 24.43$\pm$0.07 (0.38) & 24.50$\pm$0.07 (0.32) \\ 
N1731283356 & 2012-315T23:09:05 & 137.7$^\circ$ &  13.7$^\circ$ & [-0.30, 0.30] & 1.6 km &  0.41 &  1.08 & 23.77$\pm$0.07 (0.34) & 23.87$\pm$0.07 (0.24) & 24.31$\pm$0.07 (0.41) & 24.43$\pm$0.07 (0.34) \\ 
N1731283693 & 2012-315T23:14:42 & 137.4$^\circ$ &  13.9$^\circ$ & [-0.30, 0.30] & 1.6 km &  0.40 &  1.08 & 23.77$\pm$0.07 (0.34) & 23.87$\pm$0.07 (0.25) & 24.47$\pm$0.07 (0.38) & 24.52$\pm$0.07 (0.33) \\ 
N1731284030 & 2012-315T23:20:19 & 137.0$^\circ$ &  14.2$^\circ$ & [-0.30, 0.30] & 1.6 km &  0.38 &  1.02 & 23.76$\pm$0.07 (0.35) & 23.87$\pm$0.07 (0.25) & 24.42$\pm$0.07 (0.41) & 24.48$\pm$0.07 (0.35) \\ 
N1731284367 & 2012-315T23:25:56 & 136.6$^\circ$ &  14.4$^\circ$ & [-0.20, 0.20] & 1.5 km &  0.47 &  1.40 & 23.77$\pm$0.07 (0.34) & 23.90$\pm$0.07 (0.25) & 24.53$\pm$0.07 (0.39) & 24.53$\pm$0.07 (0.31) \\ 
N1731285041 & 2012-315T23:37:10 & 135.9$^\circ$ &  14.9$^\circ$ & [-0.20, 0.20] & 1.5 km &  0.49 &  1.29 & 23.76$\pm$0.06 (0.35) & 23.90$\pm$0.07 (0.25) & 24.41$\pm$0.07 (0.40) & 24.47$\pm$0.07 (0.32) \\ 
N1731285378 & 2012-315T23:42:47 & 135.5$^\circ$ &  15.1$^\circ$ & [-0.20, 0.20] & 1.5 km &  0.45 &  1.16 & 23.76$\pm$0.06 (0.36) & 23.90$\pm$0.07 (0.25) & 24.46$\pm$0.07 (0.41) & 24.52$\pm$0.07 (0.32) \\ 
N1731285715 & 2012-315T23:48:24 & 135.1$^\circ$ &  15.4$^\circ$ & [-0.20, 0.20] & 1.5 km &  0.43 &  1.25 & 23.77$\pm$0.06 (0.36) & 23.91$\pm$0.07 (0.25) & 24.41$\pm$0.07 (0.39) & 24.45$\pm$0.07 (0.33) \\ 
N1731286052 & 2012-315T23:54:01 & 134.7$^\circ$ &  15.6$^\circ$ & [-0.20, 0.20] & 1.5 km &  0.43 &  1.10 & 23.78$\pm$0.06 (0.35) & 23.90$\pm$0.06 (0.26) & 24.53$\pm$0.07 (0.39) & 24.56$\pm$0.07 (0.32) \\ 
N1731286389 & 2012-315T23:59:38 & 134.3$^\circ$ &  15.8$^\circ$ & [-0.20, 0.20] & 1.5 km &  0.40 &  1.06 & 23.77$\pm$0.06 (0.36) & 23.90$\pm$0.06 (0.26) & 24.37$\pm$0.07 (0.43) & 24.47$\pm$0.07 (0.36) \\ 
N1731286726 & 2012-316T00:05:15 & 133.9$^\circ$ &  16.1$^\circ$ & [-0.20, 0.20] & 1.5 km &  0.38 &  1.06 & 23.75$\pm$0.06 (0.36) & 23.87$\pm$0.06 (0.27) & 24.44$\pm$0.07 (0.41) & 24.51$\pm$0.07 (0.35) \\ 
N1731287063 & 2012-316T00:10:52 & 133.6$^\circ$ &  16.3$^\circ$ & [-0.20, 0.20] & 1.5 km &  0.38 &  0.94 & 23.78$\pm$0.06 (0.35) & 23.88$\pm$0.06 (0.27) & 24.42$\pm$0.07 (0.43) & 24.50$\pm$0.07 (0.36) \\ 
N1731287400 & 2012-316T00:16:29 & 133.2$^\circ$ &  16.6$^\circ$ & [-0.15, 0.20] & 1.5 km &  0.45 &  1.07 & 23.83$\pm$0.06 (0.33) & 23.94$\pm$0.06 (0.26) & 24.43$\pm$0.07 (0.35) & 24.47$\pm$0.07 (0.35) \\ 
N1731287737 & 2012-316T00:22:06 & 132.8$^\circ$ &  16.8$^\circ$ & [-0.15, 0.20] & 1.5 km &  0.40 &  1.11 & 23.78$\pm$0.06 (0.35) & 23.88$\pm$0.06 (0.27) & 24.34$\pm$0.07 (0.43) & 24.45$\pm$0.07 (0.32) \\ 
N1731288074 & 2012-316T00:27:43 & 132.4$^\circ$ &  17.0$^\circ$ & [-0.15, 0.20] & 1.5 km &  0.38 &  0.94 & 23.79$\pm$0.06 (0.35) & 23.88$\pm$0.06 (0.28) & 24.49$\pm$0.07 (0.43) & 24.59$\pm$0.07 (0.35) \\ 
N1731288411 & 2012-316T00:33:20 & 132.0$^\circ$ &  17.3$^\circ$ & [-0.20, 0.15] & 1.5 km &  0.37 &  1.03 & 23.79$\pm$0.06 (0.34) & 23.87$\pm$0.06 (0.29) & 24.43$\pm$0.07 (0.34) & 24.46$\pm$0.07 (0.30) \\ 
\hline
\multicolumn{6}{|c|}{Average Values} &  0.42 &  1.12 & 23.79 & 23.89 & 24.44 & 24.50 \\ 
\multicolumn{6}{|c|}{Observed Standard Deviations} &  0.03 &  0.11 & 0.03 & 0.02 & 0.06 & 0.04 \\ 
\multicolumn{6}{|c|}{Expected Standard Deviations}  &  &  & 0.07 & 0.07 & 0.07 & 0.07 \\ 
%
\hline
N1751831892 & 2013-187T19:02:30 & 140.5$^\circ$ &  11.5$^\circ$ & [-0.50, 0.60] & 1.9 km &  0.45 &  1.20 & 23.60$\pm$0.08 (1.97) & 23.48$\pm$0.08 (0.26) & 24.12$\pm$0.08 (0.35) & 24.24$\pm$0.08 (0.27) \\ 
N1751832439 & 2013-187T19:11:37 & 140.1$^\circ$ &  11.9$^\circ$ & [-0.50, 0.60] & 1.9 km &  0.47 &  1.20 & 23.39$\pm$0.08 (0.30) & 23.48$\pm$0.08 (0.26) & 24.10$\pm$0.08 (0.37) & 24.23$\pm$0.08 (0.27) \\ 
N1751832986 & 2013-187T19:20:44 & 139.6$^\circ$ &  12.3$^\circ$ & [-0.40, 0.60] & 1.9 km &  0.46 &  1.21 & 23.62$\pm$0.08 (2.03) & 23.49$\pm$0.08 (0.26) & 24.09$\pm$0.08 (0.40) & 24.25$\pm$0.08 (0.25) \\ 
N1751833533 & 2013-187T19:29:51 & 139.2$^\circ$ &  12.6$^\circ$ & [-0.40, 0.60] & 1.9 km &  0.49 &  1.27 & 23.41$\pm$0.08 (0.30) & 23.51$\pm$0.08 (0.26) & 24.11$\pm$0.08 (0.36) & 24.26$\pm$0.08 (0.26) \\ 
N1751834080 & 2013-187T19:38:58 & 138.7$^\circ$ &  13.0$^\circ$ & [-0.40, 0.50] & 1.9 km &  0.50 &  1.34 & 23.42$\pm$0.08 (0.29) & 23.52$\pm$0.08 (0.26) & 24.09$\pm$0.08 (0.40) & 24.27$\pm$0.08 (0.26) \\ 
N1751834627 & 2013-187T19:48:05 & 138.3$^\circ$ &  13.4$^\circ$ & [-0.40, 0.50] & 1.9 km &  0.53 &  1.48 & 23.45$\pm$0.08 (0.29) & 23.52$\pm$0.08 (0.26) & 24.11$\pm$0.08 (0.38) & 24.28$\pm$0.08 (0.26) \\ 
N1751835174 & 2013-187T19:57:12 & 137.8$^\circ$ &  13.8$^\circ$ & [-0.40, 0.50] & 1.9 km &  0.52 &  1.45 & 23.45$\pm$0.08 (0.29) & 23.52$\pm$0.08 (0.25) & 24.09$\pm$0.08 (0.40) & 24.28$\pm$0.08 (0.26) \\ 
N1751835721 & 2013-187T20:06:19 & 137.3$^\circ$ &  14.2$^\circ$ & [-0.30, 0.40] & 1.9 km &  0.50 &  1.12 & 23.45$\pm$0.08 (0.30) & 23.49$\pm$0.08 (0.25) & 24.11$\pm$0.08 (0.34) & 24.12$\pm$0.08 (0.29) \\ 
N1751836268 & 2013-187T20:15:26 & 136.9$^\circ$ &  14.5$^\circ$ & [-0.30, 0.40] & 1.9 km &  0.48 &  1.08 & 23.44$\pm$0.08 (0.31) & 23.48$\pm$0.08 (0.25) & 24.11$\pm$0.08 (0.36) & 24.16$\pm$0.08 (0.29) \\ 
N1751836815 & 2013-187T20:24:33 & 136.4$^\circ$ &  14.9$^\circ$ & [-0.30, 0.30] & 1.9 km &  0.50 &  1.14 & 23.45$\pm$0.08 (0.31) & 23.49$\pm$0.08 (0.25) & 24.08$\pm$0.08 (0.37) & 24.13$\pm$0.08 (0.29) \\ 
N1751837362 & 2013-187T20:33:40 & 136.0$^\circ$ &  15.3$^\circ$ & [-0.30, 0.30] & 1.9 km &  0.49 &  1.21 & 23.44$\pm$0.08 (0.32) & 23.47$\pm$0.08 (0.25) & 24.08$\pm$0.08 (0.36) & 24.16$\pm$0.08 (0.29) \\ 
N1751838456 & 2013-187T20:51:54 & 135.1$^\circ$ &  16.1$^\circ$ & [-0.30, 0.30] & 1.9 km &  0.48 &  1.12 & 23.44$\pm$0.08 (0.32) & 23.48$\pm$0.08 (0.24) & 24.08$\pm$0.08 (0.41) & 24.14$\pm$0.08 (0.30) \\ 
N1751839003 & 2013-187T21:01:01 & 134.6$^\circ$ &  16.5$^\circ$ & [-0.20, 0.30] & 1.9 km &  0.49 &  1.13 & 23.43$\pm$0.08 (0.32) & 23.48$\pm$0.08 (0.24) & 24.12$\pm$0.08 (0.36) & 24.13$\pm$0.08 (0.31) \\ 
N1751839550 & 2013-187T21:10:08 & 134.1$^\circ$ &  16.8$^\circ$ & [-0.20, 0.30] & 1.9 km &  0.47 &  1.05 & 23.41$\pm$0.08 (0.33) & 23.45$\pm$0.08 (0.25) & 24.10$\pm$0.08 (0.40) & 24.12$\pm$0.08 (0.31) \\ 
N1751840097 & 2013-187T21:19:15 & 133.7$^\circ$ &  17.2$^\circ$ & [-0.20, 0.30] & 1.9 km &  0.45 &  1.10 & 23.40$\pm$0.08 (0.33) & 23.45$\pm$0.08 (0.25) & 24.05$\pm$0.08 (0.42) & 24.03$\pm$0.08 (0.32) \\ 
N1751840644 & 2013-187T21:28:22 & 133.2$^\circ$ &  17.6$^\circ$ & [-0.20, 0.20] & 1.9 km &  0.48 &  1.06 & 23.39$\pm$0.08 (0.33) & 23.44$\pm$0.08 (0.24) & 24.03$\pm$0.08 (0.41) & 24.04$\pm$0.08 (0.35) \\ 
N1751841191 & 2013-187T21:37:29 & 132.7$^\circ$ &  18.0$^\circ$ & [-0.20, 0.20] & 1.9 km &  0.46 &  1.17 & 23.37$\pm$0.08 (0.34) & 23.42$\pm$0.08 (0.25) & 24.06$\pm$0.08 (0.37) & 24.05$\pm$0.08 (0.31) \\ 
\hline
\multicolumn{6}{|c|}{Average Values} &  0.48 &  1.20 & 23.44 & 23.48 & 24.09 & 24.17 \\ 
\multicolumn{6}{|c|}{Observed Standard Deviations} &  0.02 &  0.13 & 0.07 & 0.03 & 0.02 & 0.09 \\ 
\multicolumn{6}{|c|}{Expected Standard Deviations}  &  &  & 0.08 & 0.08 & 0.08 & 0.08 \\ 
\hline\end{tabular}}
\end{table}

\end{document}